\documentclass[twocolumn,nofootinbib,groupaddress,preprintnumbers,floatfix,prd]{revtex4-1}

\usepackage[bookmarks=false,hyperfootnotes=false]{hyperref}
\usepackage{graphicx}
\usepackage{amsmath}
\usepackage{xspace}
\usepackage{xcolor}
\usepackage{float}
\usepackage{comment}
\usepackage{subcaption}
\usepackage{textcomp}
\usepackage{placeins}
\usepackage[normalem]{ulem}

\newcommand{\eq}[1]{eq.~\eqref{eq:#1}}

\newcommand{\eqstwo}[2]{eqs.~\eqref{eq:#1} and \eqref{eq:#2}}

\renewcommand{\sec}[1]{sec.~\ref{sec:#1}}

\newcommand{\subsec}[1]{sec.~\ref{subsec:#1}}

\newcommand{\fig}[1]{fig.~\ref{fig:#1}}

\newcommand{\figs}[2]{figs.~\ref{fig:#1}--\ref{fig:#2}}

\newcommand{\citeref}[1]{\mbox{Ref.~\cite{#1}}}

\newcommand{\citerefs}[1]{\mbox{Refs.~\cite{#1}}}

\newcommand{\ord}[1]{\mathcal{O}(#1)}

\newcommand{\df}{\mathrm{d}}

\newcommand{\as}{\alpha_{\rm s}}

\newcommand{\Tau}{\mathcal{T}}

\newcommand{\GeV}{\ensuremath{\,\mathrm{GeV}}\xspace}
\newcommand{\TeV}{\ensuremath{\,\mathrm{TeV}}\xspace}

\newcommand{\nn}{\nonumber}

\newcommand{\cP}{\mathcal{P}}

\newcommand{\cut}{\mathrm{cut}}

\newcommand{\NS}{\mathrm{NS}}
\newcommand{\FO}{\mathrm{FO}}

\newcommand{\NLO}{\mathrm{NLO}}

\newcommand{\NNLL}{\mathrm{NNLL}'}

\newcommand{\nons}{\mathrm{nons}}

\newcommand{\one}{{(1)}}

\newcommand{\zerojet}{$0$-jet\xspace}
\newcommand{\onejet}{$1$-jet\xspace}
\newcommand{\twojet}{$2$-jet\xspace}

\newcommand{\zeroonejet}{$0$/$1$-jet\xspace}
\newcommand{\onetwojet}{$1$/$2$-jet\xspace}
\newcommand{\zerojettiness}{$0$-jettiness\xspace}
\newcommand{\onejettiness}{$1$-jettiness\xspace}

\newcommand{\Njettiness}{$N$-jettiness\xspace}

\newcommand{\lqcd}{\Lambda_\mathrm{QCD}}
\newcommand{\dsigMC}{\df\sigma^\textsc{mc}}
\newcommand{\obs}{X}
\newcommand{\taucut}{\Tau_0^{\cut}}

\newcommand{\geneva}{\textsc{Geneva}\xspace}

\newcommand{\powheg}{\textsc{Powheg}\xspace}
\newcommand{\minlo}{\textsc{MiNLO}\xspace}

\newcommand{\pythia}{\textsc{Pythia}\xspace}
\newcommand{\pythiaEight}{\textsc{Pythia8}\xspace}

\newcommand{\Matrix}{\textsc{Matrix}\xspace}
\newcommand{\openloops}{\textsc{OpenLoops}\xspace}

\newcommand{\LHAPDF}{\textsc{Lhapdf}\xspace}
\newcommand{\FastJet}{\textsc{FastJet}\xspace}

\allowdisplaybreaks[1]

\newcommand{\rescaleoneplot}{0.8\columnwidth}
\newcommand{\rescaletwoplots}{0.41\textwidth}
\newcommand{\hspacebetweentwoplots}{1em}
\newcommand{\rescalethreeplots}{0.32\textwidth}
\newcommand{\hspacebetweenthreeplots}{\fill}
\newcommand{\vspacebetweentwoplots}{2ex}
\newcommand{\spacebeforefigurecaption}{-1ex}
\begin{document}


\title{Higgsstrahlung at NNLL$'+$NNLO Matched to Parton Showers in GENEVA}

\author{Simone Alioli}
\affiliation{Universit\`{a} degli Studi di Milano-Bicocca \& INFN, Piazza della Scienza 3, Milano 20126, Italia\vspace{0.5ex}}

\author{Alessandro Broggio}
\affiliation{Universit\`{a} degli Studi di Milano-Bicocca \& INFN, Piazza della Scienza 3, Milano 20126, Italia\vspace{0.5ex}}

\author{Stefan Kallweit}
\affiliation{Universit\`{a} degli Studi di Milano-Bicocca \& INFN, Piazza della Scienza 3, Milano 20126, Italia\vspace{0.5ex}}

\author{Matthew A.~Lim}
\affiliation{Universit\`{a} degli Studi di Milano-Bicocca \& INFN, Piazza della Scienza 3, Milano 20126, Italia\vspace{0.5ex}}

\author{Luca Rottoli}
\affiliation{Universit\`{a} degli Studi di Milano-Bicocca \& INFN, Piazza della Scienza 3, Milano 20126, Italia\vspace{0.5ex}}

\date{\today}

\begin{abstract}

We present results for the Higgsstrahlung process within the
\geneva Monte Carlo framework. We combine the fully differential
NNLO calculation with the higher-order resummation in the \mbox{\zerojettiness}
resolution variable (beam thrust). The resulting parton-level events
are further showered and hadronised by \pythiaEight.
The beam thrust resummation is carried out to NNLL$'$ accuracy,
which consistently incorporates all singular virtual and real NNLO
corrections. It thus provides a natural perturbative connection
between the NNLO calculation and the parton shower regime, including a
systematic assessment of perturbative uncertainties. In this way,
observables which are inclusive over the additional radiation are correct to NNLO, while the description
of \mbox{\zerojet-like} resummation variables is improved beyond the parton shower approximation. We provide
predictions for the 13\TeV LHC.

\end{abstract}

\maketitle


\section{Introduction}
\label{sec:intro}

The study of the properties of the Higgs boson forms an important part of the experimental programme at the LHC. Though first observed in final states with two photons, the dominant decay mode of the Higgs boson is in fact to a pair of bottom quarks which has only recently been measured in associated production with a vector boson~\cite{Sirunyan:2018kst,Aaboud:2018zhk}. This class of \textit{Higgsstrahlung} processes is, therefore, an interesting subject of study. In particular, it allows one to probe couplings of the Higgs boson to electroweak vector bosons as well as to heavy quarks. It is also a particularly clean channel, since the leptonic signature arising from the bosonic decay can be efficiently distinguished above the QCD background.

At tree level the calculation involves a relatively simple extension of the Drell--Yan process in which a Higgs boson is emitted from the vector boson. In the case of $ZH$ production, the gluon-initiated channel in which the Higgs boson couples to a top-quark loop becomes accessible beginning at $\ord{\as^2}$. This provides a finite contribution to the next-to-next-to-leading order~(NNLO) cross section and is particularly sensitive to the presence of New Physics. Other top-quark loop mediated contributions, while in principle present, have been shown to provide only a percent level contribution to the total cross section~\cite{Brein:2011vx} and are often neglected in differential calculations, being separately finite and gauge invariant.

Fixed-order~(FO) calculations at NNLO have been available for some time, both at the inclusive~\cite{Brein:2003wg,Brein:2011vx,Brein:2012ne} and fully differential level~\cite{Ferrera:2011bk,Ferrera:2013yga,Ferrera:2014lca}. More recently, differential calculations including the effects of top-quark loops~\cite{Campbell:2016jau} and including the decay of the Higgs boson to a $b\bar{b}$ pair at NNLO~\cite{Ferrera:2017zex,Caola:2017xuq,Gauld:2019yng} have appeared.  Analytic resummations of threshold and jet-veto logarithms have been extensively discussed in \citerefs{Dawson:2012gs,Shao:2013uba,Li:2014ria,Harlander:2014wda}. Next-to-leading order~(NLO) electroweak~(EW) corrections have been presented in \citerefs{Denner:2011id,Denner:2014cla}.

There has also been considerable progress in the development of Monte Carlo event generators. NLO calculations matched to parton showers for Higgsstrahlung have been available in the \powheg framework~\cite{Luisoni:2013kna} for quite some time. More recently merged NLO calculations of $VH$ and $VH+$jet have appeared~\cite{Luisoni:2013kna}, utilising the \minlo method~\cite{Hamilton:2012np,Hamilton:2012rf}. Of the available methods which can attain NNLO accuracy matched to parton shower~\cite{Hamilton:2012rf,Hoeche:2014aia,Alioli:2015toa,Monni:2019whf}, to date only the \minlo approach has been applied to the Higgsstrahlung case~\cite{Astill:2016hpa,Astill:2018ivh}.
Merged NLO samples of $VH$ and $VH+$jet including NLO EW corrections matched to parton showers also produced using the \minlo method have been presented in \citeref{Granata:2017iod}.

In this paper, we present results for the Higgsstrahlung processes \mbox{$pp\to ZH\to \ell^+\ell^- H$} and \mbox{$pp\to W^\pm H \to \ell^\pm\nu_\ell H$} at NNLO matched to the \pythia8 parton shower according to the \geneva method~\cite{Alioli:2012fc,Alioli:2013hqa,Alioli:2015toa,Alioli:2016wqt}, assuming an on-shell Higgs boson. The fully differential Higgsstrahlung calculation at NNLO is improved with next-to-next-to-leading logarithmic (NNLL$'$) resummation of \zerojettiness~(beam thrust) and subsequently showered while maintaining NNLO accuracy at FO for the underlying process.

The content of the paper is organised as follows. In \sec{TheoreticalFramework} we present a review of the \geneva framework in some detail, in order to provide background for the uninitiated reader. We discuss the implementation of the Higgsstrahlung processes and highlight details specific to this case. Next, in \sec{Validation}, we describe the validation of our results against existing NNLO predictions. In \sec{Results} we then present our predictions for various distributions at partonic, showered and hadronised levels. Finally, we make some general comments on the outcome of the work and present potential future directions in \sec{Conclusions}.

\section{Theoretical Framework}
\label{sec:TheoreticalFramework}

The theoretical framework of our calculation is based on the \geneva method,
which has been developed in \citerefs{Alioli:2012fc,Alioli:2013hqa,Alioli:2015toa,Alioli:2016wqt}. While we report in the following
all the necessary ingredients for the specific Higgsstrahlung process under investigation, we refer the interested readers to the original papers where the derivations of the results used here are presented in greater detail.

\subsection{General Setup}
\label{subsec:setup}
The \geneva framework is based on the definition of physical and infrared~(IR)-finite events, generated at a given perturbative accuracy and obtained from both FO and resummed calculations. This is achieved by translating IR-divergent final states with $M$ partons into IR-finite final states with $N$ partonic jets (where $M \geq N$) for which the divergences cancel on an event-by-event basis. The translation is performed using an $N$-jet resolution variable $\Tau_N$ which partitions the phase space into regions with different numbers of resolved emissions in the final state. For example, the \geneva \zerojet cross section $\dsigMC_0$ receives contributions not only from $0$-parton events, but also from $1$-parton events where the additional emission is unresolved, i.e.\ below the $\Tau_0^\cut$  value of the \zerojet resolution, and from $2$-parton events where both additional emissions are unresolved. The partitioning of the phase space is achieved by defining cuts on the $\Tau_0$ and $\Tau_1$ resolution parameters and separating regimes as follows:\footnote{We exploit the notation in \eq{NNLOevents} to highlight the  dependence of the $\dsigMC_i$ cross section on the resolution parameters. When an argument contains a single term, e.g. $\Tau_N^\cut$, it means that the corresponding quantity has been integrated over up to the value of the argument. An argument \mbox{$\Tau_N > \Tau_N^\cut$} implies instead that the corresponding cross section is still differential in the relevant resolution variable for values larger than the cutoff.}
\begin{align}
\label{eq:NNLOevents}
  \text{$\Phi_0$ events: }
& \qquad \frac{\dsigMC_0}{\df\Phi_0}(\Tau_0^\cut)
\,,\nn\\
\text{$\Phi_1$ events: }
& \qquad
\frac{\dsigMC_{1}}{\df\Phi_{1}}(\Tau_0 > \Tau_0^\cut; \Tau_{1}^\cut)
\,,
\\
\text{$\Phi_2$ events: }
& \qquad
\frac{\dsigMC_{\ge 2}}{\df\Phi_{2}}(\Tau_0 > \Tau_0^\cut, \Tau_{1} > \Tau_{1}^\cut)
\,. \nn
\end{align}
In this way the cross section over the entire phase space is divided into exclusive \zerojet and \onejet cross sections and an inclusive \twojet cross section.
The partonic jet bins thus defined are rather different from those an experimentalist might define using a usual jet algorithm: their definition depends on an IR-safe phase space map $\Phi_N(\Phi_M)$ which projects an $M$-body onto an $N$-body phase space and ensures the individual IR-finiteness of the resulting MC cross sections $\dsigMC_i$.

Using the events in \eq{NNLOevents}, the total cross section for an observable $X$ is then given by 
\begin{align}
\sigma(\obs)
&= \int\!\df\Phi_0\, \frac{\dsigMC_0}{\df\Phi_0}(\Tau_0^\cut)\, M_\obs(\Phi_0) \\ & \quad
+ \int\!\df\Phi_{1}\, \frac{\dsigMC_{1}}{\df\Phi_{1}}(\Tau_0 > \Tau_0^\cut; \Tau_{1}^\cut)\, M_\obs(\Phi_{1})
\nn\\ & \quad
+ \int\!\df\Phi_{2}\, \frac{\dsigMC_{\ge 2}}{\df\Phi_{2}}(\Tau_0 > \Tau_0^\cut, \Tau_{1} > \Tau_{1}^\cut)\, M_\obs(\Phi_{2})
\nn\,,
\end{align}
where $M_X(\Phi_N)$ is the measurement function that computes the observable $X$ for the $N$-parton final state $\Phi_N$. Despite appearances, this object is not identical to the result one would obtain in a standard FO calculation since the unresolved emissions are assigned to the projected phase space points $\Phi_N(\Phi_M)$ rather than to the exact points $\Phi_M$. However, provided that the definitions of the $\dsigMC_{i}$ cross sections correctly capture all the singular contributions at the given order, the nonsingular difference due to the projection vanishes in the limit \mbox{$\Tau_N^\cut \to 0$}. We are therefore motivated to choose as small a value for $\Tau_N^\cut$ as possible. We should then expect to encounter large logarithms of $\Tau_N$ and $\Tau_N^\cut$ as we take this limit, and, to prevent the convergence of our perturbation theory being spoiled, these terms must be resummed.

The attentive reader may have noticed that the requirement of projectability from \mbox{$\Phi_M \to \Phi_N$} allows for nonsingular (non-projectable) events to be assigned to the higher multiplicity jet bins despite their value for the $\Tau_N$ resolution variable being below the $\Tau_N^\cut$ cutoff. A simple example can be seen in $\Phi_1$ configurations \mbox{$q g \to Z H q $}: when the direction of the outgoing quark is collinear to that of the incoming quark (and therefore anti-collinear to the incoming gluon)  the resulting $\Phi_0$ projection would require a  \mbox{$gg\to ZH$} tree-level configuration which does not exist. These events are therefore classified as $\Phi_1$ and, lacking any collinear or soft enhancement due to their nonsingular nature, they are assigned the corresponding tree-level cross section.

\subsubsection{\zeroonejet separation}
\label{subsec:01jetsep}

The separation between the \zerojet and \onejet regimes is determined by a \zerojet resolution variable $\Tau_0$. In \geneva, resummation of this variable is performed at NNLL$'$ accuracy, therefore including all contributions singular in $\Tau_0$ to $\ord{\as^2}$.\footnote{The inclusion of terms of the form $\as^2 \delta(\Tau_0)$ determines the difference between NNLL$'$ and NNLL accuracy.} The exclusive \zerojet and inclusive \onejet cross sections can then be written as 
\begin{align}
\frac{\dsigMC_0}{\df\Phi_0}(\Tau_0^\cut)
&= \frac{\df\sigma^{\rm NNLL'}}{\df\Phi_0}(\Tau_0^\cut)
+ \frac{\df\sigma_0^{\rm nons}}{\df\Phi_0}(\Tau_0^\cut)
\,,
\label{eq:0master}
\end{align}
\begin{align}
\frac{\dsigMC_{\geq 1}}{\df\Phi_{1}}(\Tau_0 > \Tau_0^\cut)
&= \frac{\df\sigma^{\rm NNLL'}}{\df\Phi_{0}\df \Tau_0} \, \cP(\Phi_1) \theta\left(\Tau_0 > \Tau_0^\cut\right)
\label{eq:1incmaster}
\\&\quad\nn
+ \frac{\df\sigma_{\ge 1}^{\rm nons}}{\df\Phi_1}(\Tau_0 > \Tau_0^\cut)
\,,
\end{align}
where the nonsingular terms contain at worst integrable singularities. In the case of the inclusive \onejet cross section, it is necessary to extend the dependence of $\df\sigma^{\rm NNLL'}$ from $\Phi_0$ to $\Phi_1$ by including the differential dependence on the radiation phase space, parameterised in terms of $\Tau_0$ and two other variables. This has been done by considering the resummed differential spectrum in $\Tau_0$, \mbox{$\df\sigma^{\rm NNLL'}/\df\Tau_0\df\Phi_0$}, and introducing a normalised splitting function $\cP(\Phi_1)$ to account for the dependence on the two remaining variables. These could be for example the fractional energy $z$ of one daughter in the splitting and an azimuthal angle $\varphi$. In order not to spoil the normalisation for each point in the $\Tau_0$ spectrum, the splitting function must satisfy
\begin{align}
\label{eq:cPnorm}
\int \! \frac{\df\Phi_1}{\df \Phi_{0} \df \Tau_0} \, \cP(\Phi_1) = 1
\,.\end{align}
Since we wish to obtain overall NNLO accuracy, we must have $\dsigMC_0$ and $\dsigMC_{\geq 1}$ at NNLO$_0$ and NLO$_1$, respectively, which determines the nonsingular matching contributions to be
\begin{align}
&\frac{\df\sigma_0^\nons}{\df\Phi_{0}}(\Tau_0^\cut)
\label{eq:0nons}
\\
& \quad
= \frac{\df\sigma_0^{{\rm NNLO_0}}}{\df\Phi_{0}}(\Tau_0^\cut)
- \biggl[\frac{\df\sigma^{\rm NNLL'}}{\df\Phi_{0}}(\Tau_0^\cut) \biggr]_{\rm NNLO_0}
\,, \nn
\end{align}
\begin{align}
&\frac{\df\sigma_{\ge 1}^\nons}{\df\Phi_{1}}(\Tau_0 > \Tau_0^\cut)
\label{eq:1nons}
  \\ \nn & \quad
= \frac{\df\sigma_{\ge 1}^{{\rm NLO_1}}}{\df\Phi_{1}}(\Tau_0 > \Tau_0^\cut)   \\ \nn &\quad \quad 
-  \biggl[\frac{\df\sigma^{\rm NNLL'}}{\df\Phi_0 \df \Tau_0}\cP(\Phi_1) \biggr]_{\rm NLO_1} \!\!\!\theta\left(\Tau_0 > \Tau_0^\cut\right)
\,.
\end{align}
The terms in square brackets are the FO expansions to $\ord{ \as^2}$ of the resummed cumulant and spectrum. Inserting the expressions for the FO cross sections, we obtain
\begin{align}
\frac{\dsigMC_0}{\df\Phi_0}(\Tau_0^\cut)
&= \frac{\df\sigma^{\rm NNLL'}}{\df\Phi_0}(\Tau_0^\cut)
\label{eq:0masterful}
 \\ & \quad
- \biggl[\frac{\df\sigma^{\rm NNLL'}}{\df\Phi_{0}}(\Tau_0^\cut) \biggr]_{\rm NNLO_0}
\nn \\& \quad
+ (B_0 + V_0 + W_0)(\Phi_0) \nn \\ & \quad
+ \int \! \frac{\df \Phi_1}{\df \Phi_0}\, (B_1 + V_1)(\Phi_1)\, \theta\left(\Tau_0(\Phi_1) < \Tau_0^\cut\right)
\nn \\ & \quad
+ \int \! \frac{\df \Phi_2}{\df \Phi_0}\, B_2 (\Phi_2)\, \theta\left(\Tau_0(\Phi_2) < \Tau_0^\cut\right)
\,,\nn
\end{align}

\begin{align}
\frac{\dsigMC_{\geq 1}}{\df\Phi_{1}}(\Tau_0 > \Tau_0^\cut)
&= \frac{\df\sigma^{\rm NNLL'}}{\df\Phi_{0}\df \Tau_0} \, \cP(\Phi_1) \ \theta\left(\Tau_0 > \Tau_0^\cut\right)
\label{eq:1masterful}
\\&\quad 
-  \biggl[\frac{\df\sigma^{\rm NNLL'}}{\df\Phi_0 \df \Tau_0}\cP(\Phi_1) \biggr]_{\rm NLO_1} \hspace*{-2em}\theta\left(\Tau_0 > \Tau_0^\cut\right)
\nn\\&\quad
+(B_1 + V_1)(\Phi_1)\,\theta\left(\Tau_0(\Phi_1) > \Tau_0^\cut\right)
\nn \\ & \quad
+ \int\!\!\frac{\df\Phi_2}{\df\Phi_1^\Tau}\, B_2(\Phi_2)\,\theta\left(\Tau_0(\Phi_2) \!>\! \Tau_0^\cut\right)\nn
\,,
\end{align}
where $B_M$ contains the $M$-parton tree-level contributions, $V_M$ the $M$-parton one-loop contributions, and $W_0$ the two-loop contribution. We have also introduced the shorthand notation
\begin{align}
\label{eq:dPhiRatio}
\frac{\df \Phi_{M}}{\df \Phi_N} = \df \Phi_{M} \, \delta[ \Phi_N - \Phi_N(\Phi_M) ]
\,.\end{align}
Because the resummed contribution is differential in $\Tau_0$, particular care is needed when integrating the FO $2$-parton contribution $B_2$ of the inclusive \onejet cross section in \eq{1masterful}.
The associated radiation phase space must be parameterised specifically by $\Tau_0$ and two other arbitrary variables, e.g.\ \mbox{$\df\Phi_1=\df\Phi_0\df\Tau_0\df z_1\df z_2$} where the $z_i$ might for example be $\{z,\varphi\}$. The projection \mbox{$\df\Phi_2/\df\Phi_1^\Tau$}, which implicitly defines $\Phi_1^\Tau$, must therefore use a map which preserves the value of $\Tau_0$,
\begin{equation} \label{eq:Tau0map}
\Tau_0(\Phi_1^\Tau(\Phi_2)) = \Tau_0(\Phi_2)
\,,\end{equation}
such that the pointwise singular $\Tau_0$ dependence is alike among all terms in \eq{1masterful} and cancellation of these singular terms is guaranteed.
The projection used is defined as
\begin{equation} \label{eq:Phi1TauProj}
\frac{\df\Phi_2}{\df\Phi_1^\Tau} \equiv \df\Phi_2\,\delta[\Phi_1 - \Phi_1^\Tau(\Phi_2)]\,\Theta^\Tau(\Phi_2)
\,,\end{equation}
where \mbox{$\Theta^\Tau(\Phi_2)$} defines the region of $\Phi_2$ that can be projected onto the physical $\Phi_1$ phase space via the IR-safe  map $\Phi_1^\Tau(\Phi_2)$. Only this projectable region of $\Phi_2$ is included in \mbox{$\df\sigma_{\geq 1}^\nons/\df\Phi_1$}, while the remainder will be included in the nonsingular $\Phi_2$ events below.

\vspace{1em}
\subsubsection{\onetwojet separation}
\label{subsec:12jetsep}
The separation of the inclusive \onejet cross section into an exclusive \onejet cross section and an inclusive \twojet cross section proceeds in analogy to the \zeroonejet case, with the relevant resolution variable now $\Tau_1$ and the requirement on the resummation accuracy relaxed to NLL.
We write
\begin{widetext}
\begin{align}
\frac{\dsigMC_{1}}{\df\Phi_{1}} (\Tau_0 > \Tau_0^\cut; \Tau_{1}^\cut)
&= \frac{\df\sigma_1^{\rm NLL}}{\df\Phi_{1}}(\Tau_0 > \Tau_0^\cut; \Tau_{1}^\cut)
+ \frac{\df\sigma_1^\nons}{\df \Phi_{1}}(\Tau_0 > \Tau_0^\cut; \Tau_{1}^\cut)
\,,  \label{eq:1master}
\\
\frac{\dsigMC_{\geq 2}}{\df\Phi_{2}} (\Tau_0 > \Tau_0^\cut, \Tau_{1}>\Tau_{1}^\cut)
&= \frac{\df\sigma^{\rm NLL}_{\geq 2}}{\df\Phi_2}\left(\Tau_0 > \Tau_0^\cut , \Tau_{1} > \Tau_{1}^\cut \right)
+ \frac{\df \sigma^\nons_{\geq 2}}{\df \Phi_2}(\Tau_0 > \Tau_0^\cut, \Tau_1 > \Tau_1^\cut)
\,,
\label{eq:2master}
\end{align}
where now, in contrast to the \zeroonejet case, it is sufficient to consider contributions only up to NLL to ensure that the matching terms, which determine
the FO accuracy, are free from singular logarithmic enhancements in $\Tau_1$.

For NNLO accuracy, $\dsigMC_1$ and $\dsigMC_{\geq 2}$ must be correct to NLO$_1$ and LO$_2$ respectively, and so the nonsingular matching contributions are (c.f.\ \eqstwo{0nons}{1nons})
\begin{align}
\label{eq:1match}
\frac{\df\sigma_1^\nons}{\df\Phi_{1}}(\Tau_0>\Tau_0^\cut;\Tau_1^\cut) &=\frac{\df\sigma_1^{{\rm NLO_1}}}{\df\Phi_{1}}(\Tau_0>\Tau_0^\cut;\Tau_1^\cut) - \biggl[\frac{\df\sigma_1^{\rm NLL}}{\df\Phi_{1}}(\Tau_0>\Tau_0^\cut;\Tau_1^\cut) \biggr]_{\rm NLO_1}
\,,
\\
\label{eq:2match}
\frac{\df\sigma_{\ge 2}^\nons}{\df\Phi_{2}}(\Tau_0 > \Tau_0^\cut,\Tau_1 > \Tau_1^\cut) &= \frac{\df\sigma_{\ge 2}^{{\rm LO_2}}}{\df\Phi_{2}}(\Tau_0 > \Tau_0^\cut,\Tau_1 > \Tau_1^\cut)
-  \biggl[\frac{\df\sigma_{\geq 2}^{\rm NLL}}{\df\Phi_2} \left(\Tau_0 > \Tau_0^\cut, \Tau_1 > \Tau_1^\cut\right)  \biggr]_{\rm LO_2}
\,.
\end{align}
At NLL, the resummed contributions take the form
\begin{align} \label{eq:Tau1resum}
\frac{\df\sigma_1^{\rm NLL}}{\df\Phi_{1}}(\Tau_0 > \Tau_0^\cut; \Tau_{1}^\cut) &= \frac{\df\sigma_{\geq 1}^C}{\df\Phi_1} \, U_1(\Phi_1, \Tau_1^\cut)\, \theta\left(\Tau_0 > \Tau_0^\cut\right)
\,,
\end{align}
\begin{align} \label{eq:Tau1resum2}
\frac{\df\sigma^{\rm NLL}_{\geq 2}}{\df\Phi_2}(\Tau_0 > \Tau_0^\cut, \Tau_1 > \Tau_1^\cut) &=
\frac{\df\sigma_{\geq 1}^C}{\df\Phi_1}\, U_1'(\Phi_1, \Tau_1)\, \theta\left(\Tau_0 > \Tau_0^\cut\right) \Big\vert_{\Phi_1 = \Phi_1^\Tau(\Phi_2)} \!\! \cP(\Phi_2) \, \theta\left(\Tau_1 > \Tau_1^\cut\right)\,.
\end{align}
\end{widetext}
We now discuss separately each contribution appearing in the equations \eqstwo{Tau1resum}{Tau1resum2} above. In order to understand the r\^ole of $\df\sigma_{\geq 1}^C$, it is perhaps clearer if we temporarily neglect the $\Tau_0$ resummation which was constructed at NNLL$'$ accuracy in the previous subsection.
In this approximation,  $\df\sigma_{\geq 1}^C$  is a proxy for $\df\sigma^{\rm FO}_{\geq 1}$ in the limit that \mbox{$\Tau_1 \to 0$}.
We choose it to have a similar form  to the FO contribution to \eq{1masterful} but with the full double-real matrix element $B_2$ replaced with its singular approximant $C_2(\Phi_2)$ and with a different projection, \mbox{$\df\Phi_2/\df\Phi_1^C \equiv \df\Phi_2\,\delta[\Phi_1 - \Phi_1^C(\Phi_2)]$}. 
The expanded form of  $\df\sigma_{\geq 1}^C$ at NLO$_1$ is therefore given by

\begin{align} \label{eq:NLO1singular}
\biggl[\frac{\df\sigma^{C}_{\geq 1}}{\df\Phi_{1}}\biggr]_{\NLO_1}
&= (B_1 + V_1)(\Phi_1) + \int\!\frac{\df\Phi_2}{\df \Phi_1^C}\, C_2(\Phi_2)\nn
\\&\equiv (B_1 + V_1^C)(\Phi_1)
\,.\end{align}

The $C_2$ term acts as a standard NLO subtraction that reproduces the pointwise singular behaviour of $B_2$ -- in practice, we have implemented the FKS subtractions~\cite{Frixione:1995ms}.

\mbox{$U_1(\Phi_1, \Tau_1^\cut)$} denotes the Sudakov factor that resums the dependence on $\Tau_1^\cut$ to NLL accuracy and \mbox{$U_1'(\Phi_1, \Tau_1)$} denotes its derivative with respect to $\Tau_1^\cut$.
At this order, it is given by~\cite{Stewart:2010tn,Pietrulewicz:2016nwo}
\begin{widetext} 
\begin{align}
 U_1(\Phi_1, \Tau_1^\cut) & = \frac{U}{\Gamma\bigg(1+2(2 C_F + C_A) \left[\eta_\Gamma^{\rm NLL} \left(\mu_S,\mu_H\right)-\eta_\Gamma^{\rm NLL} \left(\mu_J,\mu_H\right)\right]\bigg)} 
\end{align}
with $\Gamma$ the Euler gamma function and 
\begin{align}
\ln U =& \  2 (2 C_F + C_A)\Bigg[ 
2 K_\Gamma^{\rm NLL}\left(\mu_J, \mu_H\right)  -  K_\Gamma^{\rm NLL}\left(\mu_S,\mu_H\right) \Bigg]
  \\ & +
2 C_F \Bigg[ - \eta_\Gamma^{\rm NLL}\left(\mu_J, \mu_H\right) \ln \left(\frac{w_q w_{\bar q}}{\mu_H^2}\right)  
  + \eta_\Gamma^{\rm NLL} \left(\mu_S,\mu_H\right)\ln \left(\frac{w_q w_{\bar q}}{s_{q\bar q}}\right)\Bigg] \nn \\
&
 + 
C_A \Bigg[ - \eta_\Gamma^{\rm NLL} \left(\mu_J,\mu_H\right) \ln \left(\frac{w_g^2}{\mu_H^2}\right) + \eta_\Gamma^{\rm NLL} \left(\mu_S,\mu_H\right) \ln \left(\frac{w_g^2 s_{q \bar q}}{s_{qg}s_{\bar qg}}\right)\Bigg]
 \nn \\
 & + 
K_\gamma^{\rm NLL}\left(\mu_J, \mu_H\right) - 2 \gamma_{\rm E} (2 C_F + C_A)\left[\eta_\Gamma^{\rm NLL} \left(\mu_S,\mu_H\right)-\eta_\Gamma^{\rm NLL} \left(\mu_J,\mu_H\right)\right] \nn
\,.\end{align}
The functions appearing in the above are defined as
\begin{align}
K_\Gamma^{\rm NLL}(\mu_1, \mu_2) & = -\frac{\Gamma_0}{4\beta_0^2} \biggl[ \frac{4\pi}{\as(\mu_1)}\left( 1 - \frac{1}{r} - \ln r\right) + \left(\frac{\Gamma_1}{\Gamma_0} - \frac{\beta_1}{\beta_0}\right)\left( 1 - r + \ln r\right) + \frac{\beta_1}{2\beta_0}  \ln^2 r \biggr]\,,
\end{align}

\begin{equation}
\eta_\Gamma^{\rm NLL}(\mu_1, \mu_2) = -\frac{1}{2} \frac{\Gamma_0}{\beta_0} \biggl[ \ln r  + \frac{\as(\mu_1)}{4\pi}\left(\frac{\Gamma_1}{\Gamma_0} - \frac{\beta_1}{\beta_0}\right)(r-1)\biggr]\,, \qquad
K_\gamma^{\rm NLL}(\mu_1, \mu_2) = -\frac{1}{2} \frac{\gamma_0}{\beta_0} \ln r
\,,\end{equation}
\end{widetext}
with \mbox{$r = \as(\mu_2) / \as(\mu_1)$} and the dependence on $\Tau_1^\cut$ appears via the dependence on the scales
\begin{align}
\mu_S = \Tau_1^\cut\,, \qquad \mu_H = \Tau_1^{\max}\,, \qquad \mu_J = \sqrt{\mu_S \, \mu_H}
\,.\end{align}
Here, $\Tau_1^{\max}$ is the value at which the $\Tau_1$ resummation is turned off, which is chosen near the maximum kinematically allowed value of $\Tau_1$ for a given phase space point $\Phi_1$.
The cusp and non-cusp anomalous dimensions entering the above expressions are well known,
\begin{align}
\Gamma_0 & = 4
\,,\qquad
\Gamma_1 = 4\biggl[\Bigr(\frac{67}{9} - \frac{\pi^2}{3}\Bigl) C_A - \frac{20}{9} T_F n_f \biggr]
\,,\nn\\
\gamma_0 & = 12 C_F + 2 \beta_0
\,, \qquad
\beta_0 = \frac{11}{3} C_A - \frac{4}{3} T_F n_f
\,,\end{align}
while the kinematical terms are
\begin{align}
& s_{ab} = p_a^- p_b^+\,, \quad  s_{a1} = p_a^- p_1^+\,, \quad s_{b1} = p_a^+ p_1^-
\nn\\ \nn
& w_a = p_a^- e^{-Y_{VH}}\,, \quad  w_b = p_b^+ e^{Y_{VH}}\,,\nn\\
& w_1 = p_1^+ e^{Y_{VH}} + p_1^- e^{-Y_{VH}}
\,,\end{align}
where $p_a$, $p_b$, and $p_1$ are the massless four-momenta of the $\Phi_1$ phase space point, and \mbox{${p^+ = p^0 - p^3}$},  \mbox{${p^- = p^0 + p^3}$}. The assignment of $p_a$, $p_b$, and $p_1$ to $p_q$, $p_{\bar q}$, and $p_g$ is according to the flavour structure of $\Phi_1$. For example, for a \mbox{$q \bar q \to H Z g$} flavour structure we have \mbox{$p_q = p_a$}, \mbox{$p_{\bar q} = p_b$} and \mbox{$p_g = p_1$}.

Inserting now \eqstwo{Tau1resum}{NLO1singular} into \eq{1match}, we find that the matching term for the exclusive \onejet cross section is given by
\begin{widetext}
\begin{align} \label{eq:Phi1Nons}
\frac{\df\sigma_1^\nons}{\df\Phi_1}(\Tau_0 > \Tau_0^\cut; \Tau_1^\cut)
&= \int\!\biggl[\frac{\df\Phi_{2}}{\df\Phi_1^\Tau}\,B_{2}(\Phi_2)\, \theta\left(\Tau_0(\Phi_2) > \Tau_0^\cut\right)\,\theta(\Tau_{1} < \Tau_1^\cut)
- \frac{\df\Phi_2}{\df \Phi_1^C}\, C_{2}(\Phi_{2})\, \theta(\Tau_0 > \Tau_0^\cut) \biggr]
\nn \\ & \quad
- B_1(\Phi_1)\, U_1^\one(\Phi_1, \Tau_1^\cut)\, \theta(\Tau_0 > \Tau_0^\cut)
\,,
\end{align}
while analogously in the \twojet case we obtain
\begin{align}
\label{eq:Phi2Nons}
\frac{\df\sigma_{\geq 2}^\nons}{\df\Phi_2}(\Tau_0 > \Tau_0^\cut, \Tau_1 > \Tau_1^\cut)
&= \bigl\{ B_2(\Phi_2)\,[1 - \Theta^\Tau(\Phi_2)\,\theta(\Tau_1 < \Tau_1^\cut)]
\\ & \quad\nn
- B_1(\Phi_1^\Tau)\,U_1^{\one\prime}(\Phi_1^\Tau, \Tau_1)\,\cP(\Phi_2)\, \theta(\Tau_1 > \Tau_1^\cut)
\bigr\}\, \theta\left(\Tau_0(\Phi_2) > \Tau_0^\cut\right)
\,.
\end{align}
\end{widetext}
 In these expressions, $U^{(1)}_1$ and $U^{(1)\prime}_1$ are the $\mathcal{O}(\as)$ terms in the expansions of the objects $U_1$ and $U_1'$, which cancel the logarithmic terms in $\Tau_1$ in the $B_2$ pieces. It is worth noticing that the contributions in  \eqstwo{Phi1Nons}{Phi2Nons} are actually nonsingular at $\ord{\as}$, despite the accuracy of the $\Tau_1$ resummation being only NLL and not NLL$'$. This is due to the fact that we have included the full $\ord{\as}$ virtual, soft and collinear contributions in \eq{NLO1singular}.
\newpage
Thus far, we have discussed the construction of an additive NLO$_1+$NLL$_{\Tau_1}$ matching but have neglected to include the $\Tau_0$ resummation which we constructed at NNLL$'$ in the previous subsection.
 In order to include this correctly we must ensure that the integral of the $\dsigMC_1$ and $\dsigMC_{\geq2}$ cross section reproduces the $\Tau_0$-resummed result for the inclusive \onejet MC cross section $\dsigMC_{\geq1}$ in \eq{1masterful}. That is,
\begin{widetext}
\begin{align} \label{eq:MC1plusMC2cond}
\frac{\dsigMC_{\geq 1}}{\df\Phi_1}(\Tau_0 > \Tau_0^\cut)
&= \frac{\dsigMC_{1}}{\df\Phi_{1}} (\Tau_0 > \Tau_0^\cut; \Tau_{1}^\cut)
+ \int\!\frac{\df\Phi_{2}}{\df\Phi_{1}^\Tau}\, \frac{\dsigMC_{\ge 2}}{\df\Phi_{2}} (\Tau_0 > \Tau_0^\cut, \Tau_{1} > \Tau_{1}^\cut)
 \\ \nn
&= \frac{\df\sigma_{\geq 1}^C}{\df\Phi_1}\, \theta(\Tau_0 > \Tau_0^\cut)
+ \int \! \biggl[
\frac{\df \Phi_2}{\df \Phi_1^\Tau}\, B_2(\Phi_2)\, \theta\left(\Tau_0(\Phi_2) > \Tau_0^\cut\right)
- \frac{\df\Phi_2}{\df\Phi_1^C}\, C_2(\Phi_2)\, \theta(\Tau_0 > \Tau_0^\cut) \biggr]
\,,\end{align}
where we have used the identity (noting that $U_1(\Phi_1, \Tau_1^{\max}) \equiv 1$ and  c.f. \eq{cPnorm})
\begin{align}
\label{eq:unitarity}
U_1(\Phi_1, \Tau_1^\cut) + \int \! \frac{\df \Phi_2}{\df \Phi_1^\Tau}\, U_1'(\Phi_1, \Tau_1)\, \cP(\Phi_2)\,
\theta(\Tau_1 > \Tau_1^\cut) = 1
\,.\end{align}
Inserting the expression for $\dsigMC_{\geq 1}$ in \eq{1masterful} into \eq{MC1plusMC2cond}, we obtain the result for \mbox{$\df\sigma_{\geq 1}^C$} beyond NLO$_1$
\begin{align} \label{eq:dsiggeq1MC}
\frac{\df\sigma_{\geq 1}^C}{\df\Phi_1}
&= \frac{\df \sigma^{\rm NNLL'}}{\df\Phi_0\df\Tau_0}\cP(\Phi_1)
+ (B_1 + V_1^C)(\Phi_1)
- \biggl[\frac{\df\sigma^{\rm NNLL'}}{\df\Phi_0\df\Tau_0}\cP(\Phi_1)\,\biggr]_{\NLO_1}
\end{align}
and thus the full expressions for the exclusive \onejet and inclusive \twojet cross sections,
\begin{align} \label{eq:1masterfull}
\frac{\dsigMC_{1}}{\df\Phi_{1}} (\Tau_0 > \Tau_0^\cut; \Tau_{1}^\cut)
&= \Bigg\{ \frac{\df\sigma^{\rm NNLL'}}{\df\Phi_0\df\Tau_0}\cP(\Phi_1)
+ (B_1 + V_1^C)(\Phi_1)
- \biggl[\frac{\df\sigma^{\rm NNLL'}}{\df\Phi_0\df\Tau_0}\cP(\Phi_1)\,\biggr]_{\NLO_1}\Bigg\}
 \\ & \quad
\times U_1(\Phi_1, \Tau_1^\cut)\, \theta(\Tau_0 > \Tau_0^\cut)
\nn \\ & \quad
+\int\!\biggl[\frac{\df\Phi_{2}}{\df\Phi_1^\Tau}\,B_{2}(\Phi_2)\, \theta\left(\Tau_0(\Phi_2) > \Tau_0^\cut\right)\,\theta(\Tau_{1} < \Tau_1^\cut)
- \frac{\df\Phi_2}{\df \Phi_1^C}\, C_{2}(\Phi_{2})\, \theta(\Tau_0 > \Tau_0^\cut) \biggr]
\nn \\ & \quad
- B_1(\Phi_1)\, U_1^\one(\Phi_1, \Tau_1^\cut)\, \theta(\Tau_0 > \Tau_0^\cut)
\,,\nn \end{align}
\begin{align} \label{eq:2masterful}
\frac{\dsigMC_{\geq 2}}{\df\Phi_{2}} (\Tau_0 > \Tau_0^\cut, \Tau_{1}>\Tau_{1}^\cut)
&=\Bigg\{ \frac{\df\sigma^{\rm NNLL'}}{\df\Phi_0\df\Tau_0}\cP(\Phi_1)
+ (B_1 + V_1^C)(\Phi_1)
- \biggl[\frac{\df\sigma^{\rm NNLL'}}{\df\Phi_0\df\Tau_0}\cP(\Phi_1)\,\biggr]_{\NLO_1}\Bigg\}
 \\ & \quad
\times U_1'(\Phi_1, \Tau_1)\, \theta(\Tau_0 > \Tau_0^\cut) \Big\vert_{\Phi_1 = \Phi_1^\Tau(\Phi_2)} \!\! \cP(\Phi_2) \, \theta(\Tau_1 > \Tau_1^\cut)
\nn \\ & \quad
+\bigl\{ B_2(\Phi_2)\,[1 - \Theta^\Tau(\Phi_2)\,\theta(\Tau_1 < \Tau_1^\cut)]
\nn \\ & \quad
- B_1(\Phi_1^\Tau)\,U_1^{\one\prime}(\Phi_1^\Tau, \Tau_1)\,\cP(\Phi_2)\, \theta(\Tau_1 > \Tau_1^\cut)
\bigr\}\, \theta\left(\Tau_0(\Phi_2) > \Tau_0^\cut\right) \nn
\,.\end{align}
These contain the differential $\Tau_0$ resummation via $\df\sigma^{\rm NNLL'}$ and completely define the fully differential jet cross sections.
\end{widetext}
\subsection{Implementation of the $\boldsymbol{VH}$ process in G{\scriptsize ENEVA}}
\label{subsec:details}

\subsubsection{Choice of the jet resolution variables}
\label{subsec:TauNDefs}

We use \Njettiness~\cite{Stewart:2010tn} as our $N$-jet resolution variable, defined as

\begin{align} \label{eq:TauNdef}
\Tau_N = \sum_k \min \Bigl\{ \hat q_a \cdot p_k, \hat q_b \cdot p_k, \hat q_1 \cdot p_k, \ldots , \hat q_N \cdot p_k \Bigr\}
\,,\end{align}

where the sum over $k$ runs over all coloured final-state particles and where \mbox{$\hat q_i = n_i = (1, \vec n_i)$} are light-like reference vectors along the jet and beam directions. While the reference vectors which lie along the beam directions are the same for any $N$ so that we can choose $\vec n_a = \hat z$ and $\vec n_b = -\hat z$, for values of $N\geq 1$ the definition of the reference vector along the jet direction depends on a clustering metric. We refer the interested reader to~\cite{Alioli:2015toa} for details.

\Njettiness quantifies the degree to which the final state is $N$-jet-like for a given $N$, and has the useful property that \mbox{$\Tau_N=0$} in the limit that a configuration is composed of exactly $N$ partons. It can be used to cluster the final state into $N$-jet and beam regions in an IR-safe manner without resorting to any additional clustering algorithms. Crucially for our purposes, both its singular structure and resummation are known to the requisite accuracy.

For production of any colour singlet at NNLL$'$, the two resolution variables \zerojettiness $\Tau_0$ and \onejettiness $\Tau_1$ are needed to partition the phase space. Since this construction is identical to what has been previously done for the Drell--Yan process~\cite{Alioli:2015toa} we refer the reader to the discussion therein.

\subsubsection{The $\Tau_0$ spectrum at NNLL$'$ from SCET}
\label{subsec:SCETResummation}

The all-orders parton-level factorisation theorem for \zerojettiness is given by~\cite{Stewart:2009yx, Stewart:2010pd}
\begin{align}
\label{eq:Tau0Factorization}
\frac{\df \sigma^{\rm SCET}}{\df \Phi_0 \df \Tau_0}
&=
\sum_{ij} \frac{\df\sigma_{ij}^B}{\df\Phi_0} H_{ij} (Q^2, \mu) \int\!\df t_a\, \df t_b \, B_i (t_a, x_a, \mu)
\nn \\ & \quad \times
B_j (t_b, x_b, \mu) \, S\Bigl(\Tau_0 - \frac{t_a + t_b}{Q}, \mu \Bigr)
\,\end{align}
where \mbox{$\df\sigma_{ij}^B/\df\Phi_0$} is the Born cross section for the \mbox{$ij \to ZH \to \ell^+ \ell^- H$} or \mbox{$ij \to W^\pm H \to \ell^\pm\nu_\ell H$} hard scattering. The hard function $H_{ij}(Q)$ contains the corresponding Born and virtual squared matrix elements, and the sum runs over all possible $q\bar q$ pairs \mbox{$ij = \{u\bar u, \bar u u, d\bar d, \bar d d, \ldots\}$}. The $B_i(t, x)$ are inclusive (anti)quark beam functions~\cite{Stewart:2009yx}, with partonic virtualities $t_{a,b}$ and momentum fractions $x_{a,b}$  given in terms of the total rapidity $Y_{VH}$ and invariant mass \mbox{$Q = M_{VH}$} of the $VH$ final state as well as the hadronic centre-of-mass energy $E_{\rm cm}$ by
\begin{equation}
x_a = \frac{Q}{E_{\rm cm}}\, e^{Y_{VH}}
\,, \qquad x_b = \frac{Q}{E_{\rm cm}}\, e^{-Y_{VH}}
\,.\end{equation}
The beam functions are computed perturbatively in terms of standard PDFs $f_j$; schematically \mbox{$B_i = \sum_j \mathcal{I}_{ij}\otimes f_j$} where the $\mathcal{I}_{ij}$ are perturbative coefficients. Finally, $S(k)$ is the quark hemisphere soft function for beam thrust~\cite{Berger:2010xi}.

The preceding~\eq{Tau0Factorization} is derived using Soft Collinear Effective Theory~(SCET)~\cite{Bauer:2000ew, Bauer:2000yr, Bauer:2001ct, Bauer:2001yt, Bauer:2002nz,Beneke:2002ni,Beneke:2002ph}. We note that each of $H$, $B$, $S$ depends only on a single characteristic scale. The consequence is that the perturbative expansions of these constituent functions do not feature any large logarithms when a suitable choice of scale is made, viz.
\begin{align}
\label{eq:canonicalScales}
\mu_H = Q \,, \quad \mu_B = \sqrt{Q \Tau_0}\,, \quad \mu_S = \Tau_0
\,.\end{align}
In \eq{Tau0Factorization}, however, all ingredients must be evaluated at an arbitrary common scale $\mu$, whose dependence exactly cancels between the different functions at the appropriate order. We achieve this by using the renormalisation group evolution in the effective theory to evolve each function from its own scale to $\mu$. We thus obtain the resummed $\Tau_0$ spectrum used in \eq{1nons}:
\begin{align} \label{eq:resummedspectrum}
\frac{\df \sigma^{\rm NNLL'}}{\df \Phi_0 \df \Tau_0}
&= \sum_{ij} \frac{\df \sigma_{ij}^B}{\df \Phi_0} H_{ij}(Q^2, \mu_H)\, U_H(\mu_H, \mu)
\nn\\
& \quad 
\otimes \bigl[ B_i (x_a, \mu_B) \otimes U_B(\mu_B, \mu) \bigr]
\nn\\
& \quad
\otimes \bigl[ B_j (x_b, \mu_B) \otimes U_B(\mu_B, \mu) \bigr]
\nn\\
& \quad
\otimes \bigl[ S(\mu_S) \otimes U_S(\mu_S, \mu) \bigr]
\,,\end{align} 
where the large logarithmic terms arising from the ratios of scales have been resummed by the renormalisation group evolution (RGE) factors \mbox{$U_X(\mu_X, \mu)$}. At NNLL$'$ accuracy, the boundary conditions for the evolution of the functions are required at 2-loop order, while the evolution kernel itself must be inserted at 3(2)-loop order in the cusp (non-cusp) anomalous dimensions. It suffices to say that all required expressions are known~\cite{Idilbi:2006dg, Becher:2006mr, Stewart:2010qs, Monni:2011gb, Kelley:2011ng, Hornig:2011iu, Gaunt:2014xga, Gaunt:2014cfa, Kang:2015moa, Gaunt:2015pea}; they are in fact mostly identical to those used for Drell--Yan production.

\begin{figure}[t]
  \begin{subfigure}[b]{\columnwidth}
    \includegraphics[width=0.7\columnwidth]{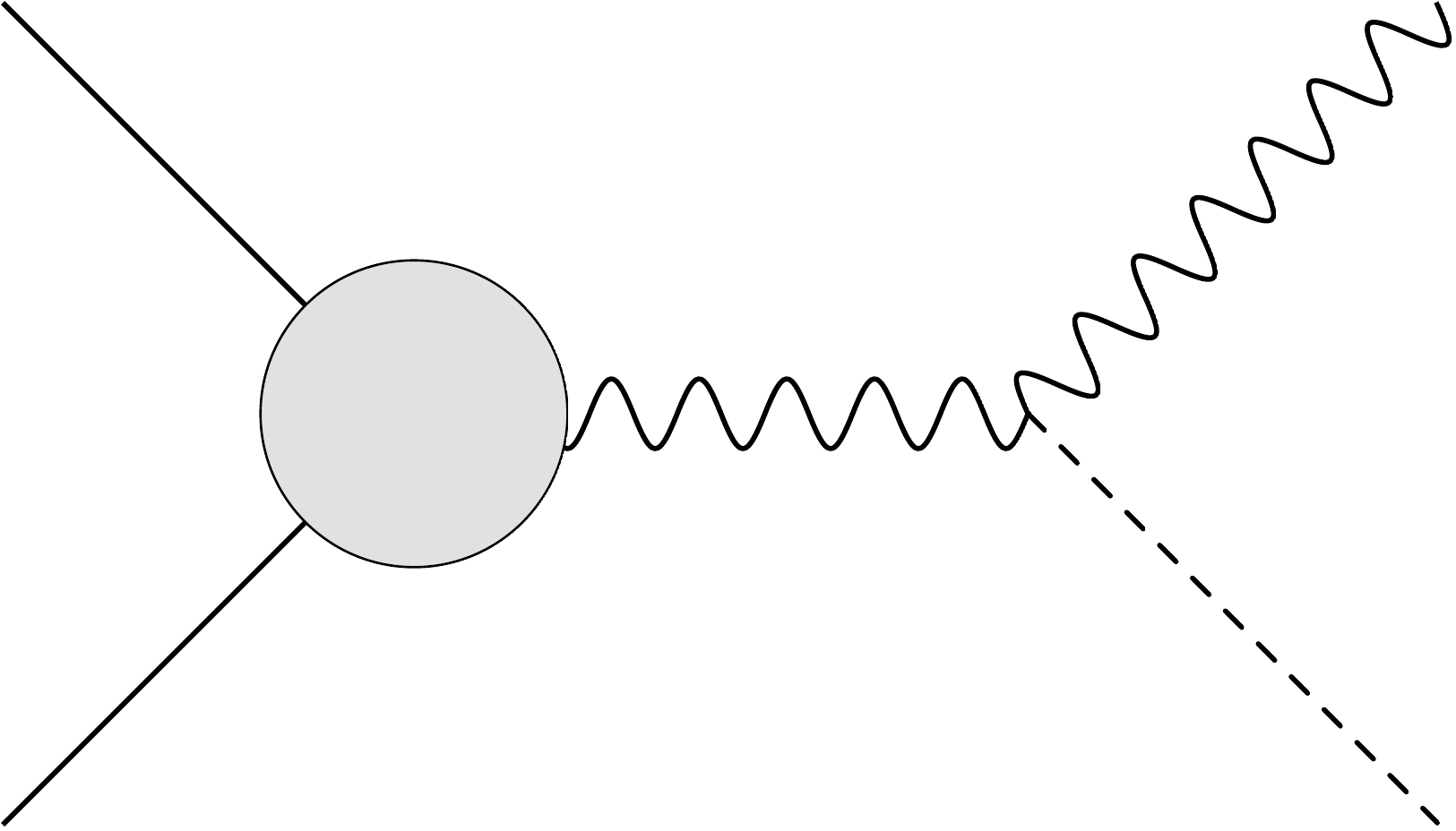}%
    \caption{DY-like contribution}
    \label{fig:DYlike}
  \end{subfigure}
  \par\medskip
  \begin{subfigure}[b]{\columnwidth}
    \includegraphics[width=0.7\columnwidth]{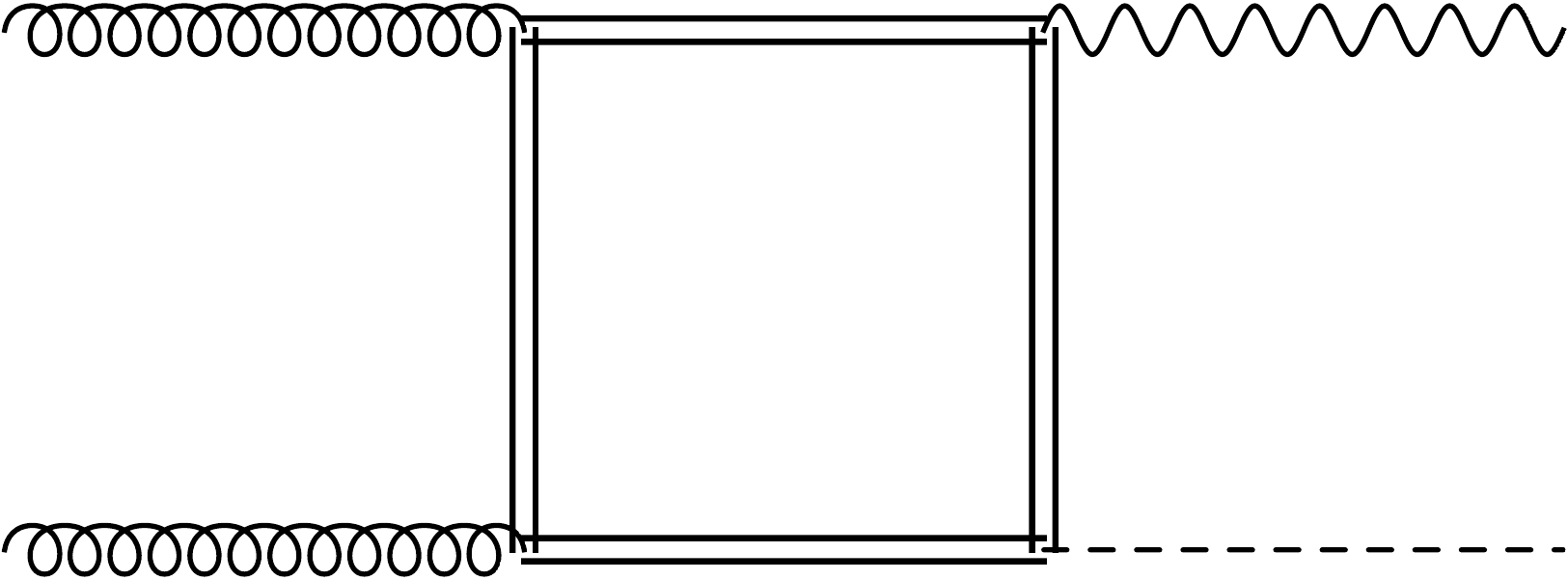}%
    \caption{Top-quark loop contribution}
    \label{fig:ggZH}
    \end{subfigure}
  \vspace{\spacebeforefigurecaption}
  \caption{Classes of diagrams contributing to $VH$ production.}
\label{fig:diagrams}
\end{figure}

The only exception is the hard function, for which many of the relevant Feynman diagrams are still closely related to those appearing in the Drell--Yan case. In particular, for the class of diagrams illustrated in \fig{DYlike} the loop corrections are identical and differences are due solely to the emission of the Higgs boson from the final-state vector boson. Indeed, writing the hard function as the product of a hadronic and an electroweak tensor,
\begin{equation}
\label{eq:FactorisedHF}
H_{ij}=\mathcal{H}_{ij}^\mu(p_i,p_j,Q)\mathcal{W}_\mu(Q,p_V,p_H)
\end{equation}
it is apparent that the hadronic tensor $\mathcal{H}_{ij}^\mu$ (where the indices $i,j$ run over the allowed partonic flavours)  is identical in the Higgsstrahlung and Drell--Yan cases for the class of diagrams in \fig{DYlike} and only the electroweak tensor is modified. Since in \citeref{Alioli:2015toa} the hard function is implemented as a factor multiplying the squared Born amplitude, we may use that same factor here and simply replace the tree level piece with the appropriate Higgsstrahlung contribution. In the case of $ZH$ production at NNLO, however, a second class of diagrams appears, depicted in \fig{ggZH}. These gluon-initiated contributions are finite and enter at $\ord{ \as^2}$; due to the dominance of the gluon PDF at the LHC it is important that they are included, as they have an $\ord{10\%}$ effect on the total cross section~\cite{Astill:2018ivh}. As their contribution is purely nonsingular in nature, they do not affect the implementation of the resummation and can be included separately.

In the present calculation we neglect 2-loop contributions involving top quarks (which are separately finite and gauge-invariant) as their effect has been shown to contribute to the total cross section only at $\mathcal{O}(1\%)$~\cite{Brein:2011vx} and their exact form remains unknown. However, we include real--virtual corrections in which the Higgs boson couples to a top-quark loop with exact top-quark mass dependence.

\subsubsection{Scale choices and profile scales}
\label{subsec:Scales}
\begin{figure}[t]
  \includegraphics[width=\rescaleoneplot]{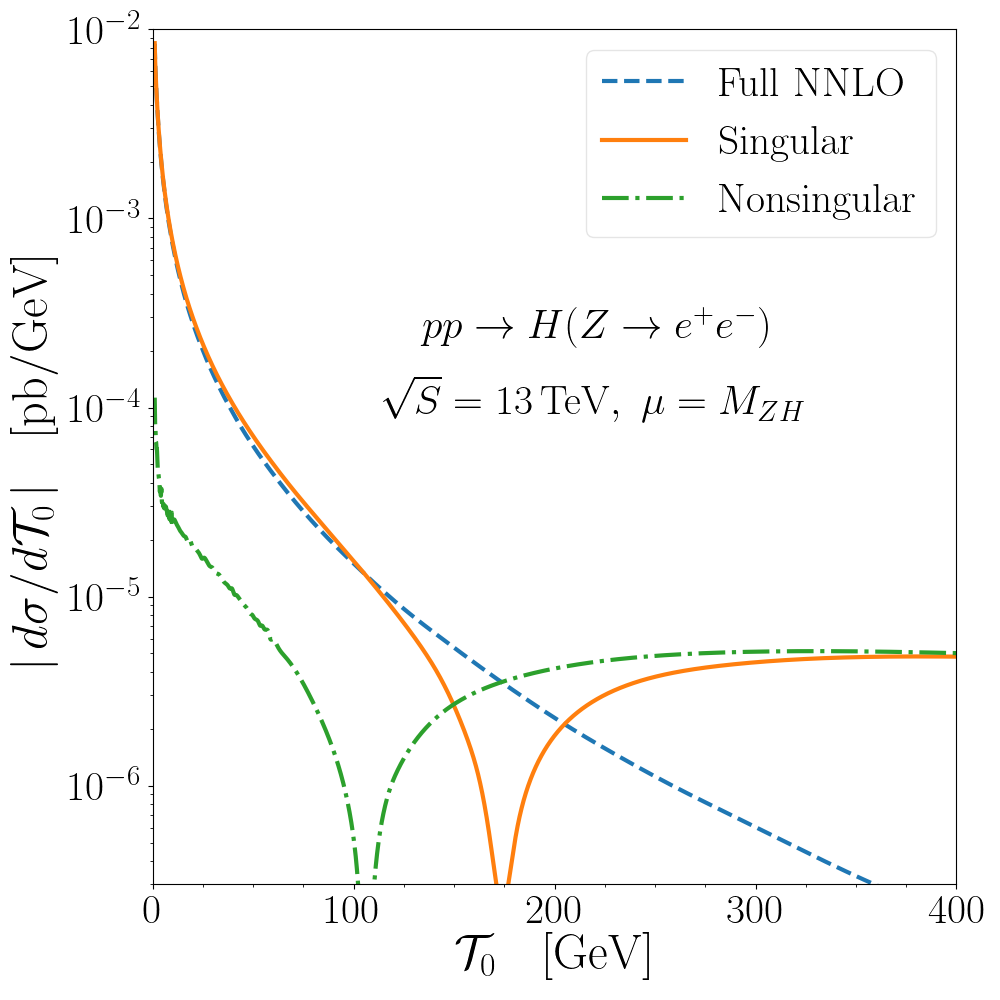}%
  \caption{The absolute value of the singular and nonsingular contributions to the $ZH$ cross section as a function of $\Tau_0$.}
  \label{fig:nonsingplot}
\end{figure}

The purpose of the resummation is to correctly account for the effects of the logarithms of $\Tau_0/Q$ when such terms are large, i.e.\ at small values of $\Tau_0$, where again \mbox{$Q\sim M_{VH}$} denotes a hard scale. For larger values of $\Tau_0$, however, the logarithmic terms are modest or small in size and do not spoil the convergence of the perturbative series. Moreover, keeping the resummation in this region is in fact inapposite: continuing to resum logarithms in this regime would spoil the cancellation between singular and nonsingular terms, preventing one from obtaining the correct FO result. It is therefore important to switch off the resummation in the calculation before this occurs. In order to determine the relevant value of $\Tau_0$, it is instructive to consider the relative sizes of the singular and nonsingular contributions as functions of $\Tau_0$. We show the absolute values of these contributions in \fig{nonsingplot}. We see that these become similar in magnitude at \mbox{$\Tau_0\approx 150\GeV$} and therefore turn off the resummation around this point.

\begin{figure*}[ht!]
  \begin{subfigure}[b]{\rescalethreeplots}
    \includegraphics[width=\textwidth]{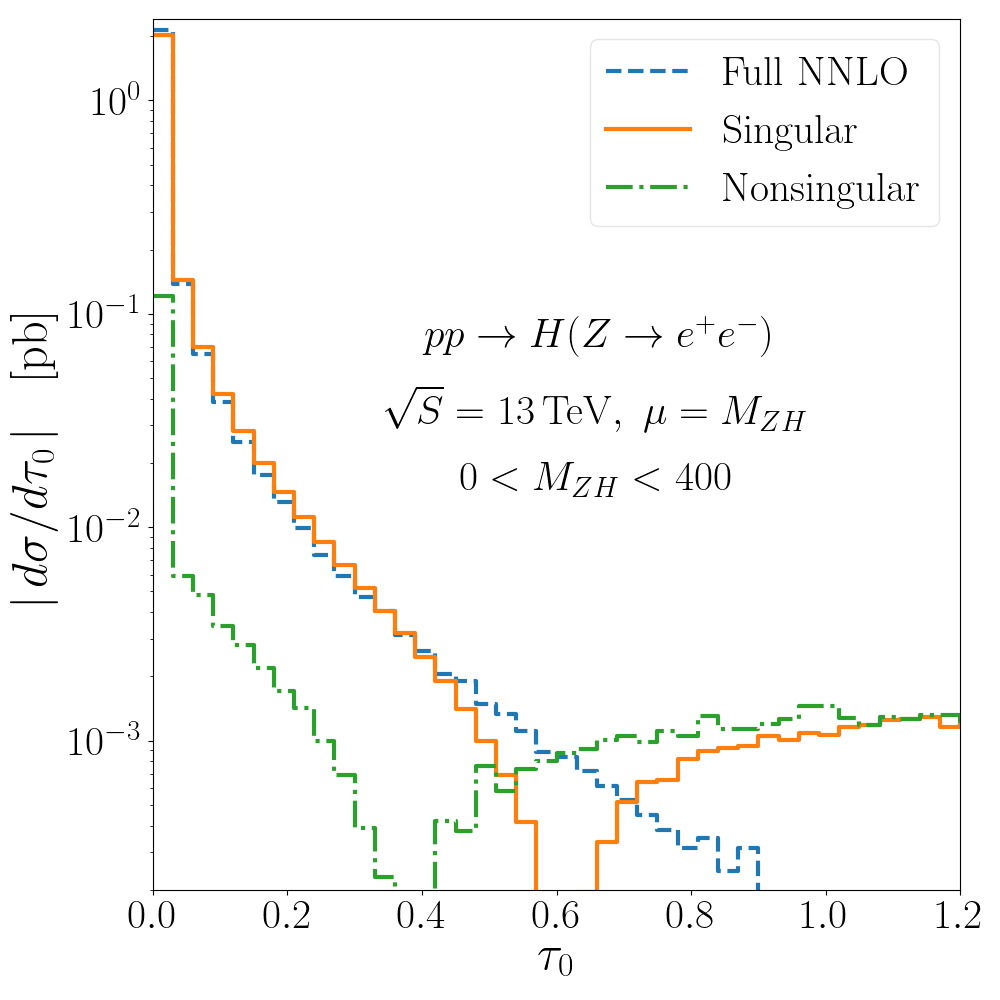}%
    \label{fig:tau0MBin1}
  \end{subfigure}
  \hspace*{\hspacebetweenthreeplots}
  \begin{subfigure}[b]{\rescalethreeplots}
    \includegraphics[width=\textwidth]{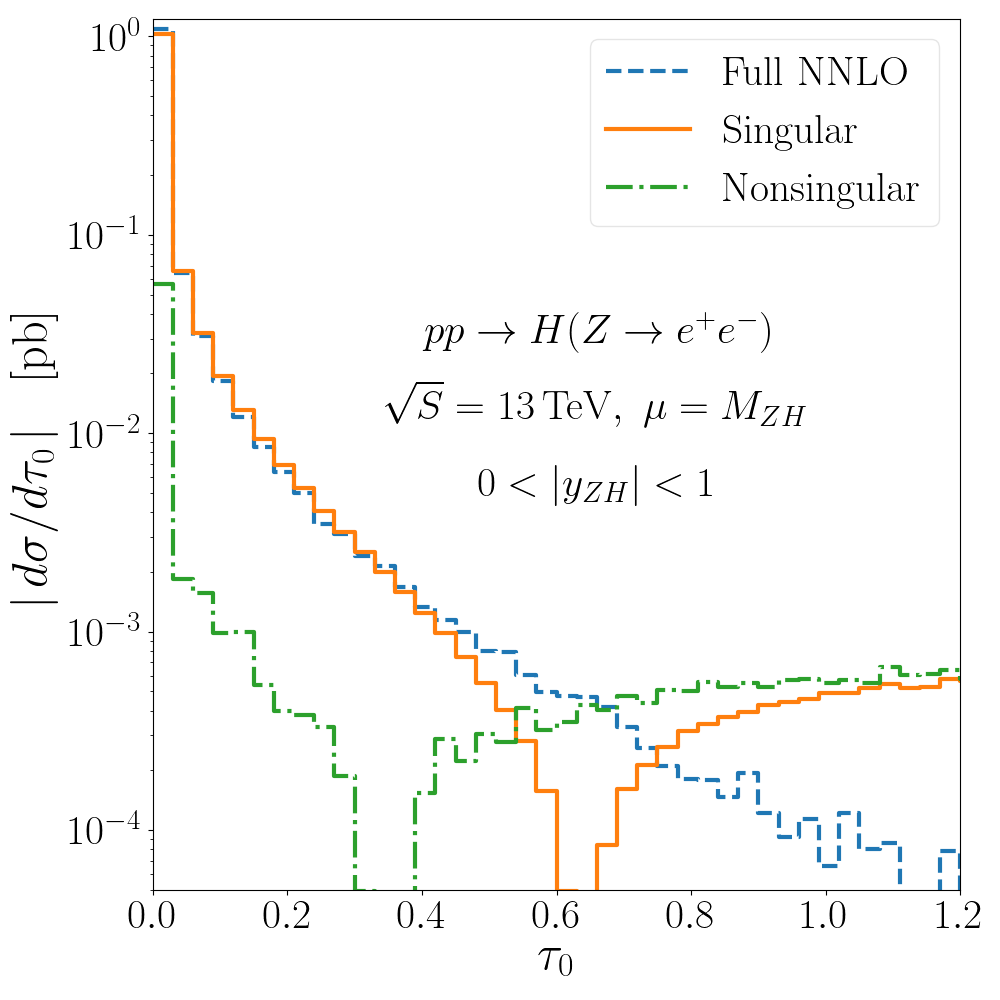}%
    \label{fig:tau0MBin2}
  \end{subfigure}
  \hspace*{\hspacebetweenthreeplots}
  \begin{subfigure}[b]{\rescalethreeplots}
    \includegraphics[width=\textwidth]{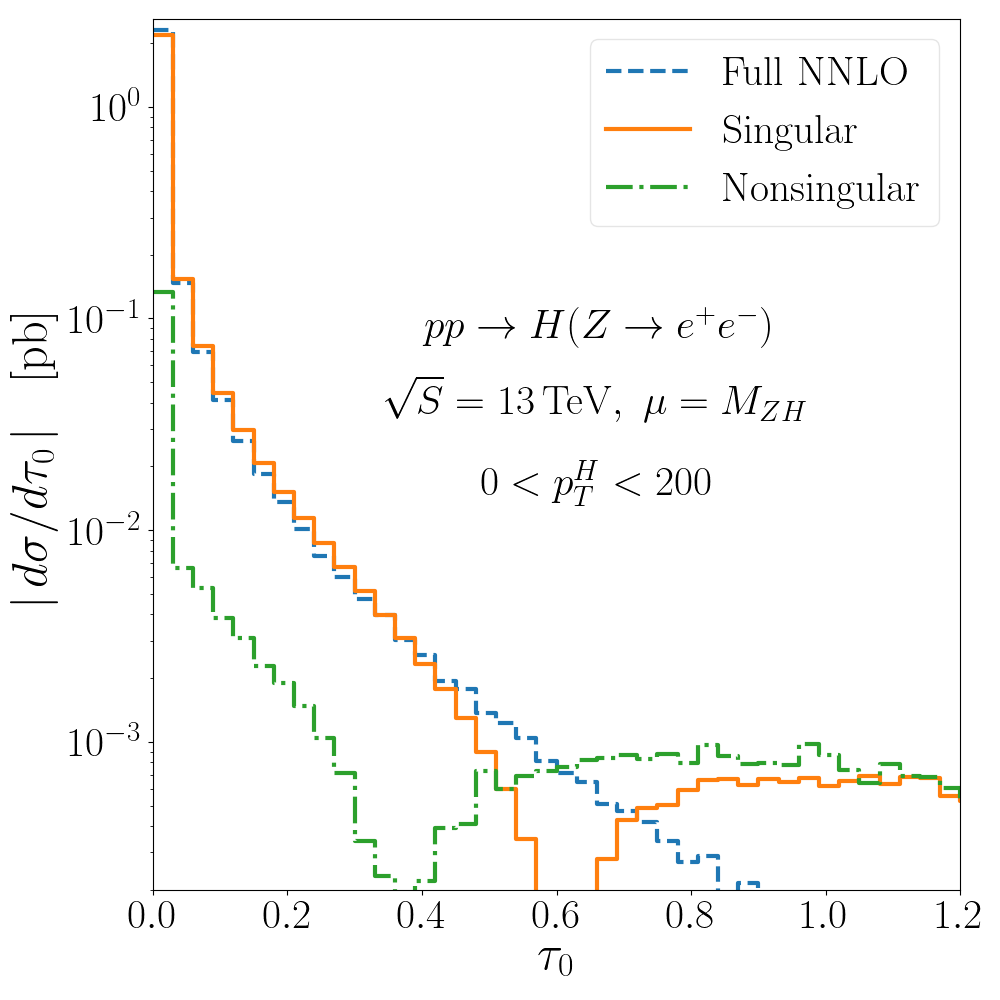}%
    \label{fig:tau0pTBin1}
  \end{subfigure}
  \\[\vspacebetweentwoplots]
  \begin{subfigure}[b]{\rescalethreeplots}
    \includegraphics[width=\textwidth]{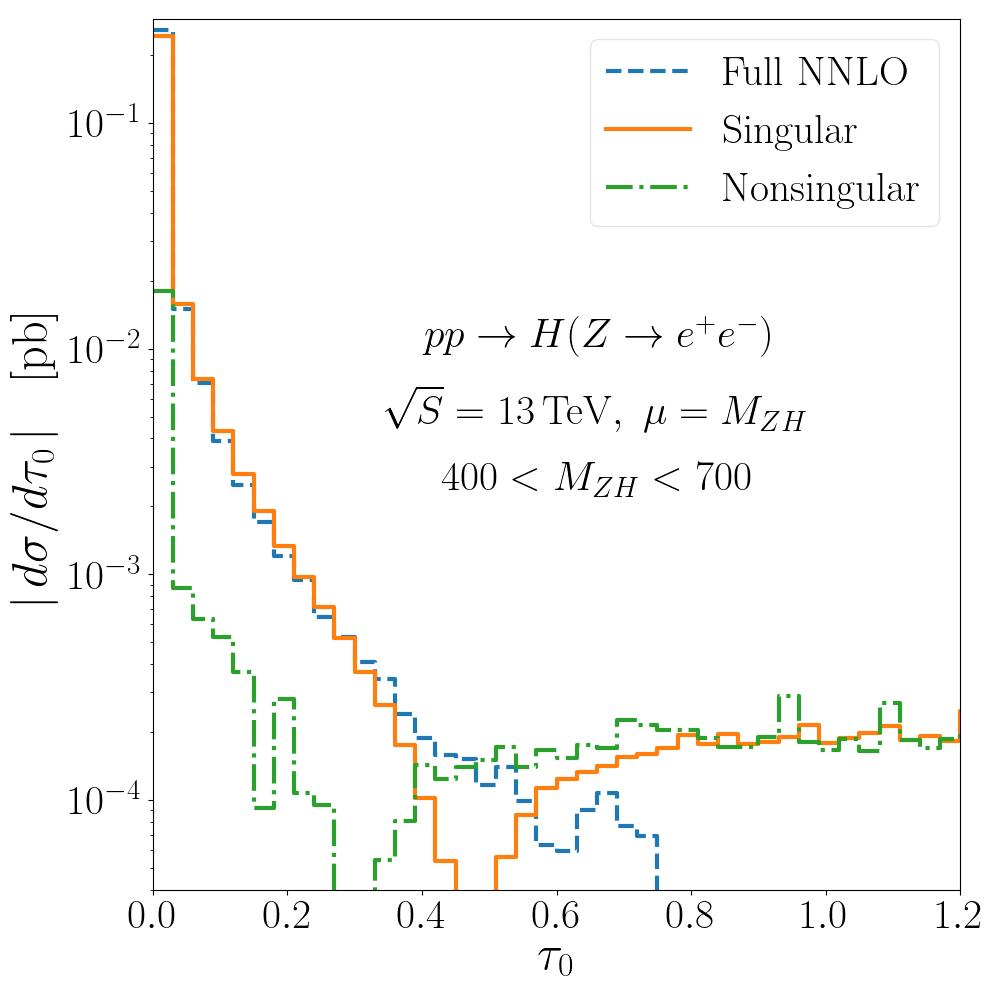}%
    \label{fig:tau0pTBin2}
  \end{subfigure}
  \hspace*{\hspacebetweenthreeplots}
  \begin{subfigure}[b]{\rescalethreeplots}
    \includegraphics[width=\textwidth]{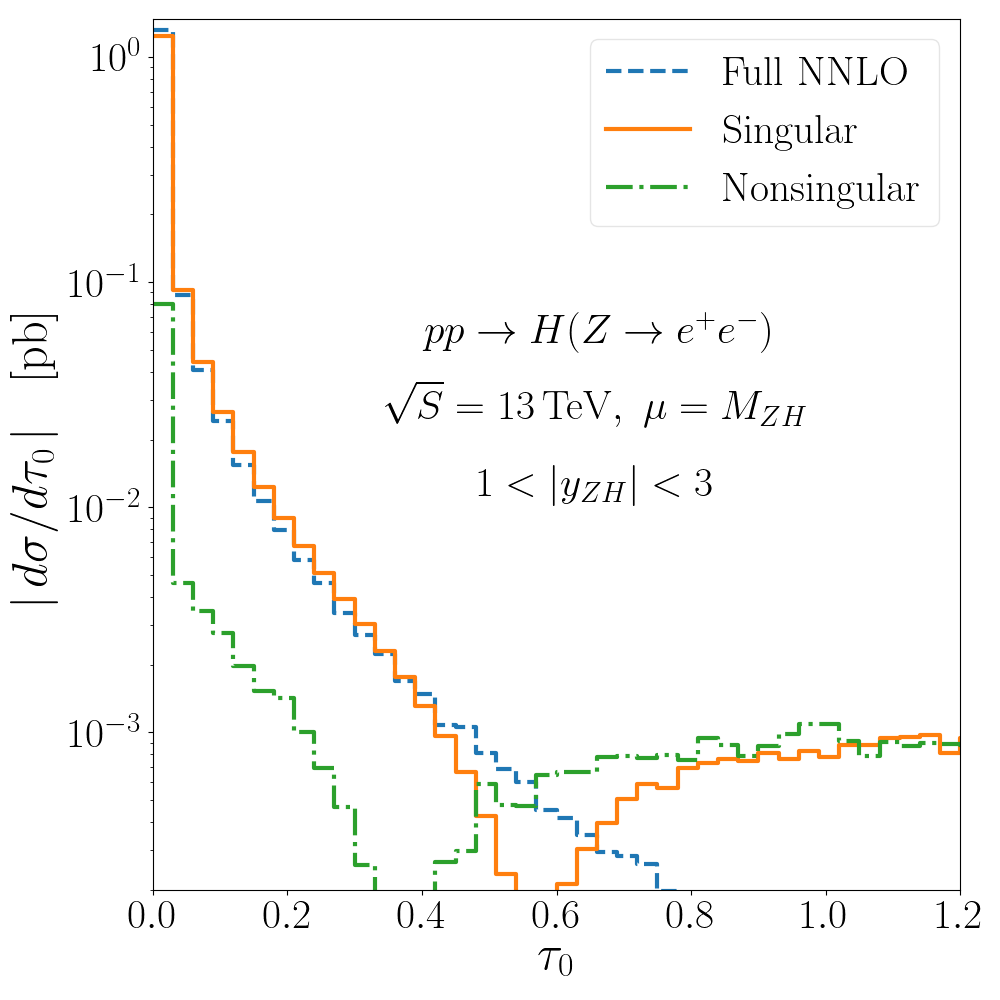}%
    \label{fig:tau0yBin1}
  \end{subfigure}
  \hspace*{\hspacebetweenthreeplots}
  \begin{subfigure}[b]{\rescalethreeplots}
    \includegraphics[width=\textwidth]{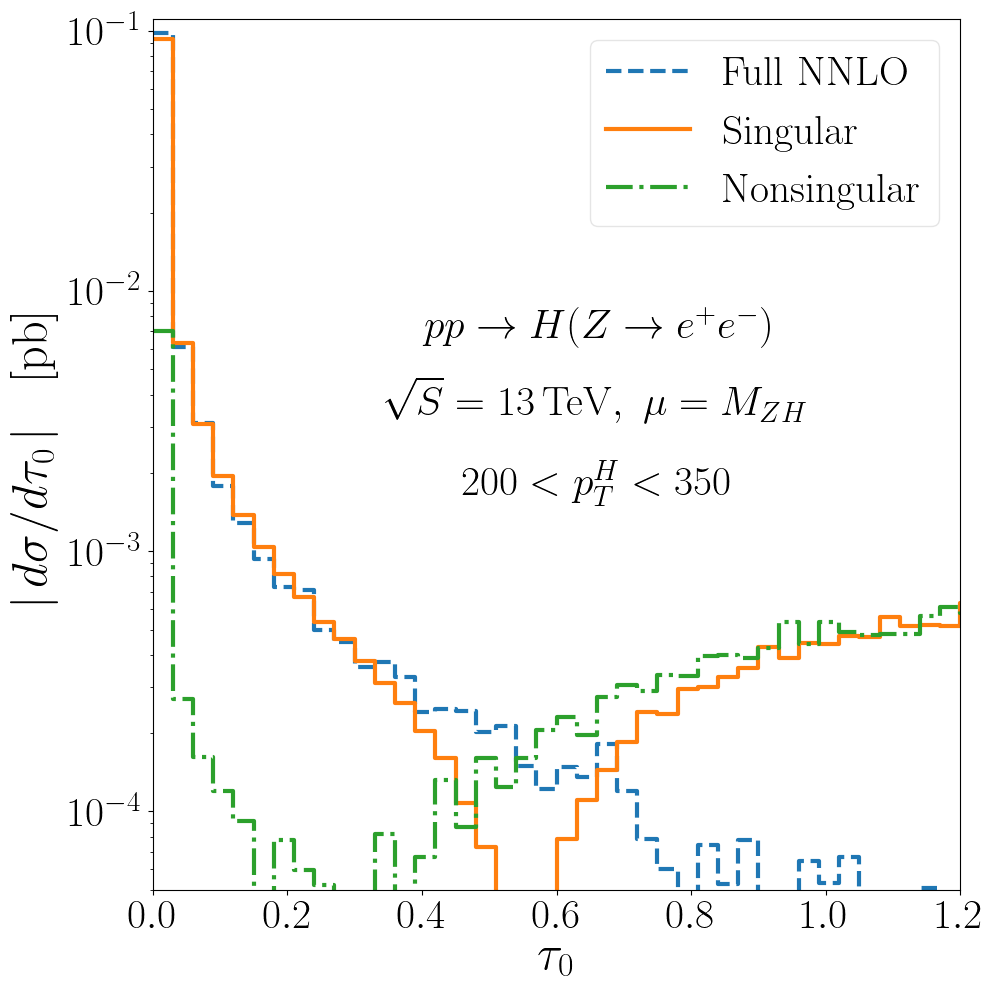}%
    \label{fig:tau0yBin2}
  \end{subfigure}
  \vspace{\spacebeforefigurecaption}
  \caption{The absolute values of the singular and nonsingular contributions to the $ZH$ cross section from bins of various distributions as a function of \mbox{$\tau_0=\Tau_0/M_{ZH}$}.}
\label{fig:nonsingbinned}
\end{figure*}

While this consideration holds for the total spectrum, integrated over all the possible Born-like kinematics, one may wonder exactly what the kinematical dependence on the coefficients of the logarithms is and whether this dependence might affect where the resummation needs to be turned off. We consider this possibility in
\fig{nonsingbinned}, where we plot the singular and nonsingular spectra as functions  of \mbox{$\tau_0 = \Tau_0/M_{VH}$}. The figure shows that the location of the crossing point between the spectra exhibits  modest dependence on the exact kinematic regime. We may therefore switch off the resummation in a similar manner for all classes of event.


Since the resummation is achieved via RGE running, it is sufficient
to set all resummation scales to a common nonsingular scale, viz.\ \mbox{$\mu_\NS = \mu_S = \mu_B = \mu_H$}, to stop the evolution. In order to ensure a natural transition between the resummation and FO regimes, we make use of profile scales $\mu_B(\Tau_0)$ and $\mu_S(\Tau_0)$  which are constructed to interpolate smoothly from the characteristic scales to $\mu_\NS$~\cite{Ligeti:2008ac, Abbate:2010xh}. We have
\begin{align} \label{eq:centralscale} 
\mu_H &= \mu_\NS
\,,  \\
\mu_S(\Tau_0) & = \mu_\NS f_{\rm run}(\Tau_0/Q)
\,, \nn \\
\mu_B(\Tau_0) &=  \mu_\NS \sqrt{f_{\rm run}(\Tau_0/Q)}
\nn\,,\end{align}
where the common profile function $f_{\rm run}(x)$ is as in \citeref{Stewart:2013faa},
\begin{align}
f_{\rm run}(x) &=
\begin{cases} x_0 \bigl[1+ (x/x_0)^2/4 \bigr] & x \le 2x_0\,,
 \\ x & 2x_0 \le x \le x_1\,,
 \\ x + \frac{(2-x_2-x_3)(x-x_1)^2}{2(x_2-x_1)(x_3-x_1)} & x_1 \le x \le x_2\,,
 \\  1 - \frac{(2-x_1-x_2)(x-x_3)^2}{2(x_3-x_1)(x_3-x_2)} & x_2 \le x \le x_3\,,
 \\ 1 & x_3 \le x\,.
\end{cases}
\label{eq:frun}
\,.\end{align}
This form has strict canonical scaling (cf. \eq{canonicalScales}) below $x_1$ and switches off the resummation above $x_3$. Considering \fig{nonsingplot}, we are led to the choice of parameters
\begin{align} \label{eq:TauBprofile}
x_0 = 2.5\GeV/Q\,, \quad
\{x_1,x_2, x_3\} = \{0.2, 0.45, 0.7\}
\,.\end{align}

In the resummation region, the nonsingular scale $\mu_\NS$ must be chosen of the same order as the hard scale $Q$ for the inclusive Higgsstrahlung. In the FO region, it can instead be left free to match
any arbitrary fixed or dynamic scale $\mu_\FO$. The transition is achieved by imposing that \mbox{$Q =M_{VH}$} for values of $\tau_0$
up to $x_3$ and smoothly interpolating the $\mu_\NS$ value between $Q$
and $\mu_\FO$ above that threshold.

We may estimate the uncertainties associated with the resummed and FO calculations by varying the profile scales. In the FO case, we adopt the usual prescription of varying $\mu_\NS$ up and down by a factor of 2 and taking the maximal absolute deviation from the central value as a measure of the uncertainty. This preserves everywhere the ratios between the various scales $\mu_H$, $\mu_B$ and $\mu_S$ and so the arguments of the logarithms which are resummed by the RGE factors are unaffected. In the resummed case, we vary independently the profile scales for $\mu_B$ and $\mu_S$ about their central profiles while keeping \mbox{$\mu_H=\mu_\NS$} fixed. In this way the arguments of the resummed logarithms are varied in order to estimate the size of higher-order corrections in the resummed series while maintaining the scale hierarchy \mbox{$\mu_\NS\sim \mu_H \gg \mu_B \sim \sqrt{\mu_H\mu_S} \gg \mu_S$}. More details on the specifics of this prescription may be found in \citeref{Gangal:2014qda}. In addition, we include two more profiles where we vary all $x_i$ transition points by $\pm0.05$ simultaneously while keeping all the scales at their central values. We thus obtain six profile variations in total and take the maximal absolute deviation in the result from the central value as the resummation uncertainty. The total perturbative uncertainty is then obtained as the quadrature sum of the resummation and FO uncertainties.

Hitherto we have considered only the effect of the scale choice on the $\Tau_0$ spectrum, \mbox{$\df\sigma^{\NNLL} / \df\Tau_0$}, and not the effect on its integral over $\Tau_0$, the cumulant $\df\sigma^{\NNLL}(\Tau_0^{\cut})$. Indeed, while summing over the events distributed according to the $\Tau_0$ spectrum, for example for the calculation of the inclusive total cross section, one performs exactly this integration.
However, since the profile scales have a functional dependence on $\Tau_0$, the integral of the spectrum is not exactly equal to the cumulant evaluated at the highest scale.
Choosing canonical scaling for the cumulant as a function of $\Tau_0^\cut$ like so,
\begin{align}
\mu_H = Q\,, \quad \mu_B = \sqrt{Q \Tau_0^\cut}\,, \quad \mu_S = \Tau_0^\cut
\,,\end{align}
one would obtain
\begin{align}
\label{eq:problem}
 &  \int_{0}^{\Tau_0^{\max}} \frac{\df\sigma^{\NNLL}}{\df\Phi_0 \df\Tau_0}(\mu(\Tau_0))\df\Tau_0 =\\& \nn\qquad \frac{\df\sigma^{\NNLL}}{\df\Phi_0}(\Tau_0^{\max},\mu(\Tau_0^{\max}))+ \mathcal{O}(\rm N^3LL)\,,
\end{align}
where ${\Tau_0^{\max}}$ is the upper kinematical limit, so that the difference is due to terms of higher order. However, these terms could be numerically relevant, especially if one aims at reproducing the exact FO inclusive cross section.

In order to remedy this, we may add a term to the spectrum such that the inclusive FO cross section is correctly recovered. The term added must satisfy the following properties:
\begin{itemize}
    \item The integral of the modified spectrum must recover the FO cross section;
    \item The term must contribute only in the region of $\Tau_0$ where the missing $\rm N^3LL$ terms are sizeable and must vanish elsewhere (especially in the FO region at large $\Tau_0$);
    \item The term must be of the same order as the missing terms so that the NNLL$'$ \;accuracy of the spectrum is not spoiled when the term is added. 
\end{itemize}
\begin{figure*}[ht!]
  \begin{subfigure}[b]{\rescalethreeplots}
    \includegraphics[width=\textwidth]{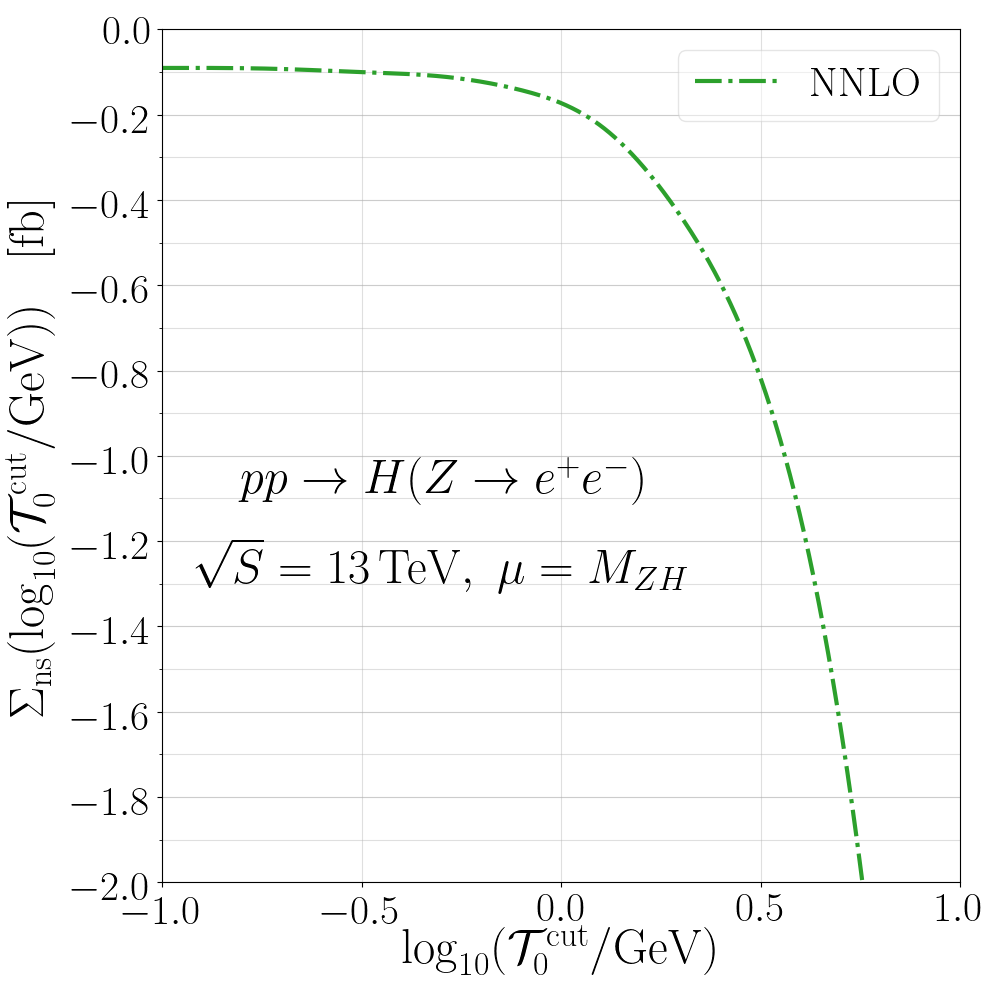}%
    \caption{}
    \label{fig:TauZerocum}
  \end{subfigure}
  \hspace*{\hspacebetweenthreeplots}
  \begin{subfigure}[b]{\rescalethreeplots}
    \includegraphics[width=\textwidth]{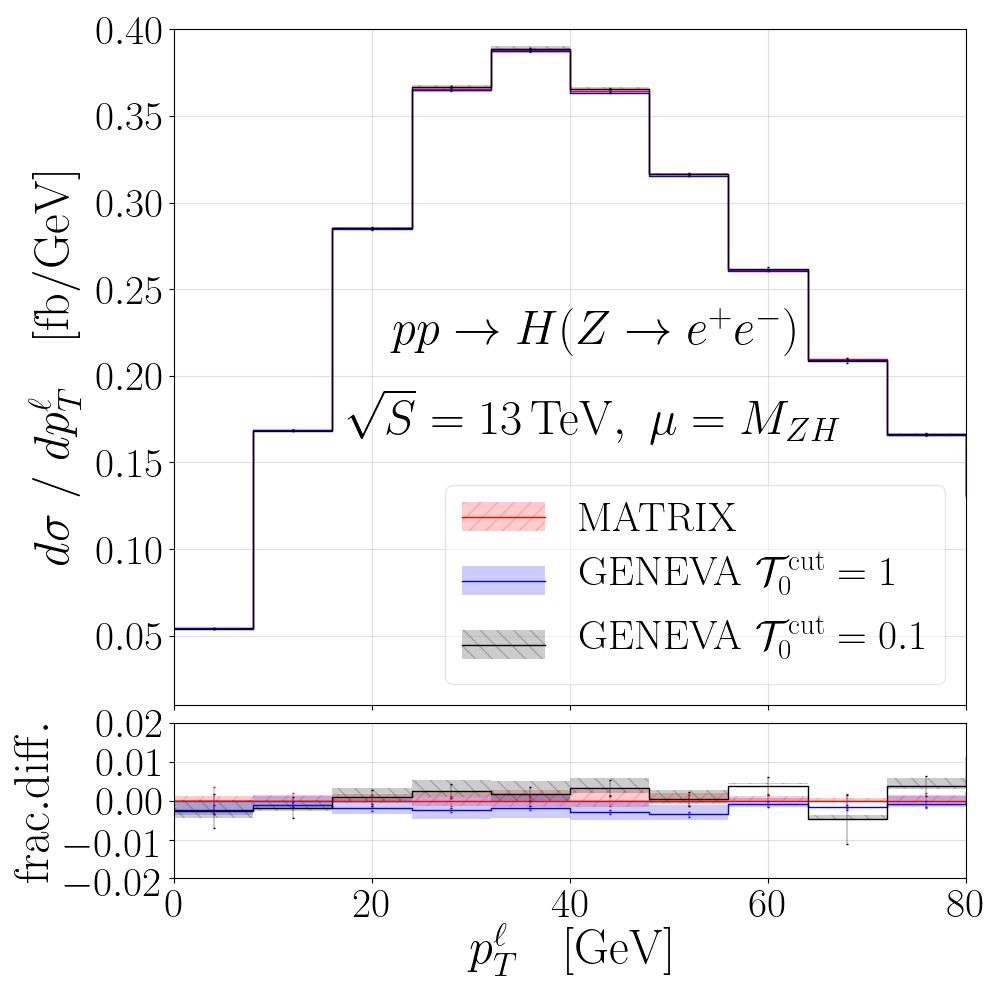}%
    \caption{}
    \label{fig:lpTvaryTaucut}
  \end{subfigure}
  \hspace*{\hspacebetweenthreeplots}
  \begin{subfigure}[b]{\rescalethreeplots}
    \includegraphics[width=\textwidth]{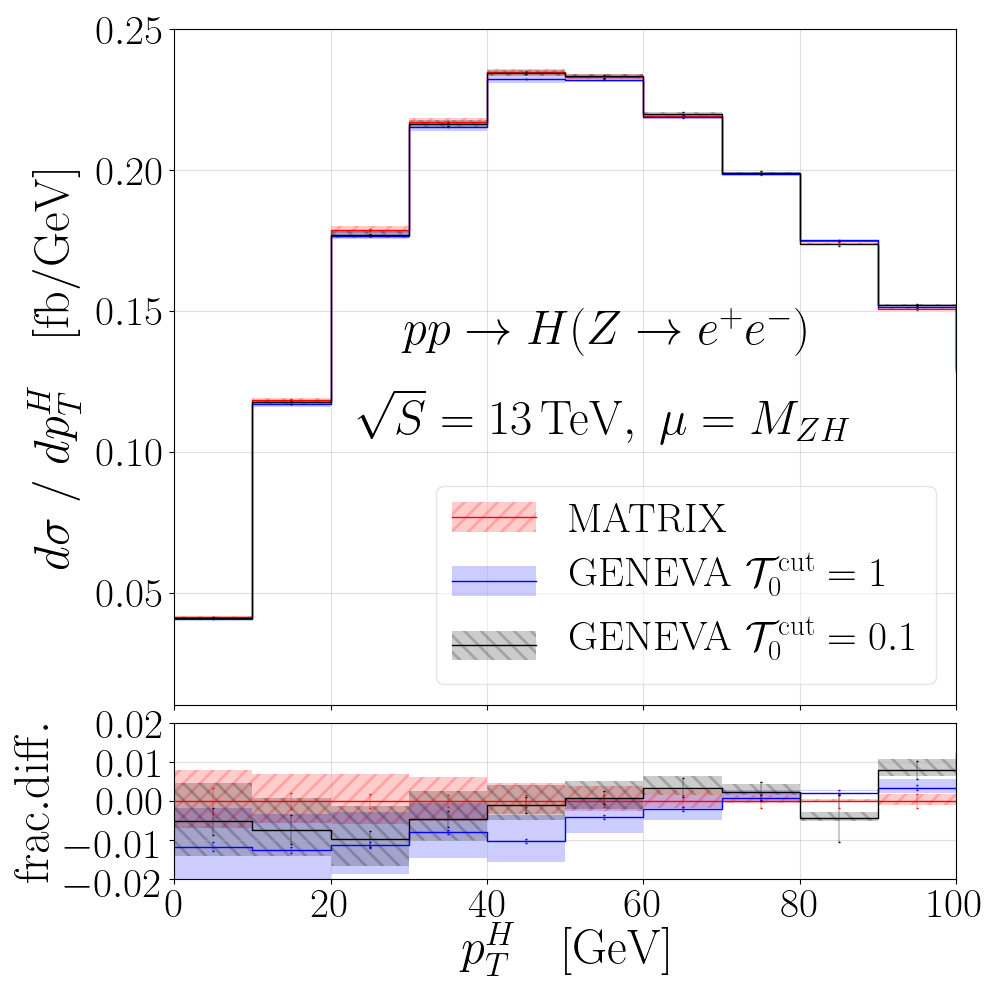}%
    \caption{}
    \label{fig:HpTvaryTaucut}
  \end{subfigure}
  \vspace{\spacebeforefigurecaption}
  \caption{The neglected nonsingular contribution to the $\Tau_0$ cumulant at NNLO (left) and its impact on the hardest lepton (centre) and Higgs boson (right) $p_T$ spectra.}
\end{figure*}

We therefore add the term 
\begin{equation}
\label{eq:term}
\kappa(\Tau_0) \left[ \frac{\df}{\df \Tau_0} \frac{\df \sigma^{\rm NNLL'}}{\df \Phi_0}(\Tau_0, \mu_h(\Tau_0))
 - \frac{\df \sigma^{\rm NNLL'}}{\df \Phi_0 \df \Tau_0} (\mu_h(\Tau_0)) \right],
\end{equation}
where $\kappa(\Tau_0)$ and $\mu_h(\Tau_0)$ are smooth functions. First, we note that the term is, by construction, of higher order and thus our third criterion is satisfied. Second, we note that in the FO region where \mbox{$\mu_h(\Tau_0)=Q$} the difference between the pieces in brackets is zero, as the scales are constant here -- the term therefore vanishes. We may also choose $\kappa(\Tau_0)$ to tend to zero in this region to reduce further the size of this additional contribution before exact cancellation is reached, and choose the profile scale $\mu_h(\Tau_0)$ to reach $Q$ at a lower value of $\Tau_0$ than the rest of the calculation. All this ensures that the accuracy of the tail of the spectrum is not spoiled by the addition of \eq{term} and instead restricts it to act in the region where \mbox{$\Tau_0\sim\taucut\ll Q$}. Our second criterion is therefore satisfied. Third, we may tune \mbox{$\kappa(\Tau_0\to 0)$} such that, upon integration of the sum of the spectrum and \eq{term}, the correct inclusive cross section is recovered. We thus satisfy our first criterion. In fact, we can perform this tuning for each FO scale variation separately. When taking either the central value of $\mu(\Tau_0)$ or any of the resummation variations, we simply take $\mu_h(\Tau_0)$ as before. When we take the FO up and down variations for $\mu(\Tau_0)$, however, we now take \mbox{$\mu_h^{\rm up}(\Tau_0)=2\mu_h(\Tau_0)$} or \mbox{$\mu_h^{\rm down}(\Tau_0)=1/2\mu_h(\Tau_0)$} and readjust the value of \mbox{$\kappa(\Tau_0\to 0)$} such that the value of the inclusive cross section is correctly recovered. In this way we obtain the correct FO scale variations for the inclusive cross section (or indeed any other inclusive quantity).

\subsubsection{Power-suppressed corrections to the nonsingular cumulant}
\label{subsec:Powerterms}
The \zerojet cross section in \eq{0masterful} is NNLO accurate and fully differential in the $\Phi_0$ phase space.
In order to regulate the IR divergences that appear in 4 dimensions in the intermediate stages of the calculation, however, one needs a subtraction method up to NNLO which is also fully differential in $\Phi_0$. While fully general local subtractions are available at NLO~\cite{Frixione:1995ms,Catani:1996jh,Catani:1996vz}, local NNLO general subtraction methods are still in their infancy~\cite{Caola:2017dug,Czakon:2010td,Czakon:2014oma,GehrmannDeRidder:2005cm,Magnea:2018hab,Ferrera:2017zex}.
Even when provided with a NNLO subtraction, the \geneva predictions at $\ord{\as^2}$ stemming from \eq{0masterful}  would be exact for the total cross section but might be correct only up to power corrections in $\Tau_0^{\rm cut}$ for observables dependent on the $\Phi_0$ kinematics. This is a consequence of the projective map used in the definition of the $\Phi_0$ events.\footnote{To be precise, any $\Phi_0$ variable which is left unchanged by the $\Phi_0(\Phi_1)$ and $\Phi_0(\Phi_2)$ mappings, e.g.\ the $M_{VH}$ invariant mass, will achieve NNLO accuracy in our implementation of \eq{0masterful}. However, it is not possible to avoid power corrections for all $\Phi_0$ variables simultaneously.} For this reason, one can avoid the necessity of a NNLO subtraction which is fully differential in $\Phi_0$ and replace the formula for the \zerojet cross section with
\begin{align}
\frac{\widetilde{\dsigMC_0}}{\df\Phi_0}(\Tau_0^\cut)
&= \frac{\df\sigma^{\rm NNLL'}}{\df\Phi_0}(\Tau_0^\cut)
- \biggl[\frac{\df\sigma^{\rm NNLL'}}{\df\Phi_{0}}(\Tau_0^\cut) \biggr]_{\rm NLO_0}
\nn \\& \quad
+ B_0(\Phi_0)  + V_0(\Phi_0)
\label{eq:0tilde}
 \\ & \quad
+ \int \! \frac{\df \Phi_1}{\df \Phi_0}\, B_1 (\Phi_1)\, \theta\left(\Tau_0(\Phi_1) < \Tau_0^\cut\right)
\,,\nn
\end{align}
where only a local NLO subtraction and the expansion of the resummation at $\ord{\as}$ are now required.
The formula above assumes that there is an exact cancellation of both the singular and nonsingular contributions
at $\ord{\as^2}$ between the FO and the resummed-expanded terms.
In reality these terms differ by a nonsingular contribution, which can be written as
\begin{align} \label{eq:Sigmanons}
\frac{\df\sigma_0^\nons}{\df\Phi_{0}}(\Tau_0^\cut)
&= \bigl[ \as f_1(\Tau_0^\cut, \Phi_0) + \as^2 f_2(\Tau_0^\cut, \Phi_0) \bigr]\Tau_0^\cut
\end{align}
where the functions \mbox{$f_i(\Tau_0^{\rm cut},\Phi_0)$} are at worst logarithmically divergent in the small $\Tau_0^{\rm cut}$ limit. While we include the NLO term proportional to \mbox{$f_1(\Tau_0^\cut, \Phi_0)$} in \eq{0tilde}, we neglect the \mbox{$f_2(\Tau_0^\cut, \Phi_0)$} piece. The size of this neglected term as a function of the cut is shown in \fig{TauZerocum} for $ZH$ production. We note that at the default value at which we run, \mbox{$\Tau_0^{\rm cut}=1\GeV$}, the observed effect is rather small in magnitude at just under 0.2 fb ($\sim 0.7\%$ of the total cross section). In order to correct for this discrepancy and obtain the correct NNLO inclusive cross section, we may simply rescale the weights of the $\Phi_0$ events in such a way that the total cross section thus obtained matches the result of an independent NNLO calculation. In practice, we use cross sections obtained from the \Matrix program~\cite{Grazzini:2017mhc} to obtain the reweighting factors for the central scale and its FO variations. By following such a procedure, we are able to include the effects of the $f_2$ term in \eq{Sigmanons} on the total cross section that would have been present had we implemented \eq{0masterful} literally. Since neither \eq{0masterful} nor our approach in \eq{0tilde} achieve the exact $\ord{\as^2}$ $\Phi_0$ dependence of all observables, our approximation does not inherently limit the accuracy of our predictions.

In nearly all spectra, the lack of the correct $\ord{\as^2}$ $\Phi_0$ dependence does not produce striking differences when compared with \Matrix. We therefore conclude that our approximation holds rather well. As an example, we show the transverse-momentum distribution of the hardest lepton produced for different values of
$\Tau_0^{\rm cut}$ in \fig{lpTvaryTaucut}. We observe similar behaviour in many other distributions. In one exceptional case, however, we find a mild effect on a distribution, namely the Higgs boson transverse momentum. In \fig{HpTvaryTaucut}, one can see a difference of $\mathcal{O}(1\%)$ between the \geneva and \Matrix results in the first few bins. This difference is halved when the $\Tau_0^{\rm cut}$ is reduced to 0.1\GeV and is restricted in range compared to the results using the higher cut. Nonetheless, throughout this work we have continued to use a value \mbox{$\Tau_0^{\rm cut}=1\GeV$} for reasons of improved numerical stability (the reader may note the increase in the sizes of the statistical errors associated with the \mbox{$\Tau_0^{\rm cut}=0.1\GeV$} calculation, for example). We find that any discrepancies caused by this choice are consistently small and would most likely fall within PDF uncertainties.

\subsubsection{NLO$_1$ calculation and phase space mapping}

We remark that the approximation described above is applied only to the \zerojet bin. The $\Phi_1$ and $\Phi_2$ events are produced instead following the full forms in \eqstwo{1masterfull}{2masterful}, which employ a proper \zerojettiness subtraction at NNLO combined with the local NLO$_1$ subtraction for the $VHj$ process.
In practice we use the FKS subtraction, but with a specific choice for the mapping employed in splittings and projections. Since the FO terms appearing in \eqstwo{1masterfull}{2masterful} are differential in $\Tau_0$, the mapping used in the NLO$_1$ calculation must preserve the value of $\Tau_0$ during such splittings or projections, see e.g.\ \eq{Phi1TauProj}.
Such a mapping was devised for the Drell--Yan production in \citeref{Alioli:2015toa} and is discussed briefly therein. For this work, we performed only minor modifications in order to accommodate the additional final-state Higgs boson.

\subsubsection{Interface to the parton shower}
\label{subsec:PSmatching}

Since the shower interface is mostly identical for Drell--Yan production and Higgsstrahlung, we provide only a brief recap of its main features here, referring the interested reader to section 3 of \citeref{Alioli:2015toa} for a more detailed discussion.

The partonic jet cross sections $\dsigMC_0$, $\dsigMC_1$ and $\dsigMC_{\ge 2}$ each include contributions from higher multiplicity phase space points, but only in those cases where \mbox{$\Tau_N(\Phi_M)<\Tau_N^\cut$}. In order to make the calculation fully differential in the higher multiplicities, a parton shower is interfaced which adds radiation to each jet cross section in a manner which does not alter the value or accuracy of the integrated cross sections but still produces a fully exclusive final state, i.e.\ in a unitary and recursive fashion. The effect of this is to restore the emissions in $\dsigMC_0$ and $\dsigMC_1$ which were integrated over when the jet cross sections were constructed, as well as to add extra final-state partons to the inclusive $\dsigMC_{\ge 2}$.

In order to simplify the treatment, we consider a shower strongly ordered in $\Tau_N$, such that \mbox{$\Tau_0(\Phi_1)\gg\Tau_1(\Phi_2)\gg\dots$} In practice, no current parton shower program uses $\Tau_N$ as an evolution variable in the way that we have here, choosing instead e.g.\ the transverse momentum of an emission. However, if one were to take the output of a shower ordered in say transverse momentum, one could recluster the partons using the \Njettiness metric in order to obtain a splitting history that was ordered in $\Tau_N$ and equivalent at LL order.

The requirement of the preservation of the accuracy of the jet cross section after applying the shower on a phase space point $\Phi_N$ sets constraints on the point $\Phi_{N+1}$ reached after each emission added by the shower. These constraints are different for the different partonic multiplicities of the events before the shower.

For the cases in which the showered events originate from $\Phi_0$ events, the main constraint is that the integral of the cross section below the $\Tau_0^\cut$ (which is NNLL$'$+NNLO accurate) must not be modified. The emissions generated by the shower must in this case satisfy \mbox{$\Tau_0(\Phi_N)<\Tau_0^\cut$}, so that they recover the events which were integrated over in the construction of the \zerojet exclusive cross section and add events with more emissions below the cut.
In case of a single shower emission we require also that the resulting $\Phi_1$ point is projectable onto $\Phi_0$, as these are the only configurations at this order which are included in \eq{0tilde} (see also \eq{dPhiRatio}).
Both these conditions can be implemented with a careful choice of the starting scale of the shower. The preservation of the cross section below the cut is then ensured by the unitarity of the shower evolution. In practice, we allow for a
tiny spillover up to $5\%$ above $\taucut$ in order to smoothen the transition.

\begin{figure*}[t]
  \begin{subfigure}[b]{\rescalethreeplots}
    \includegraphics[width=\textwidth]{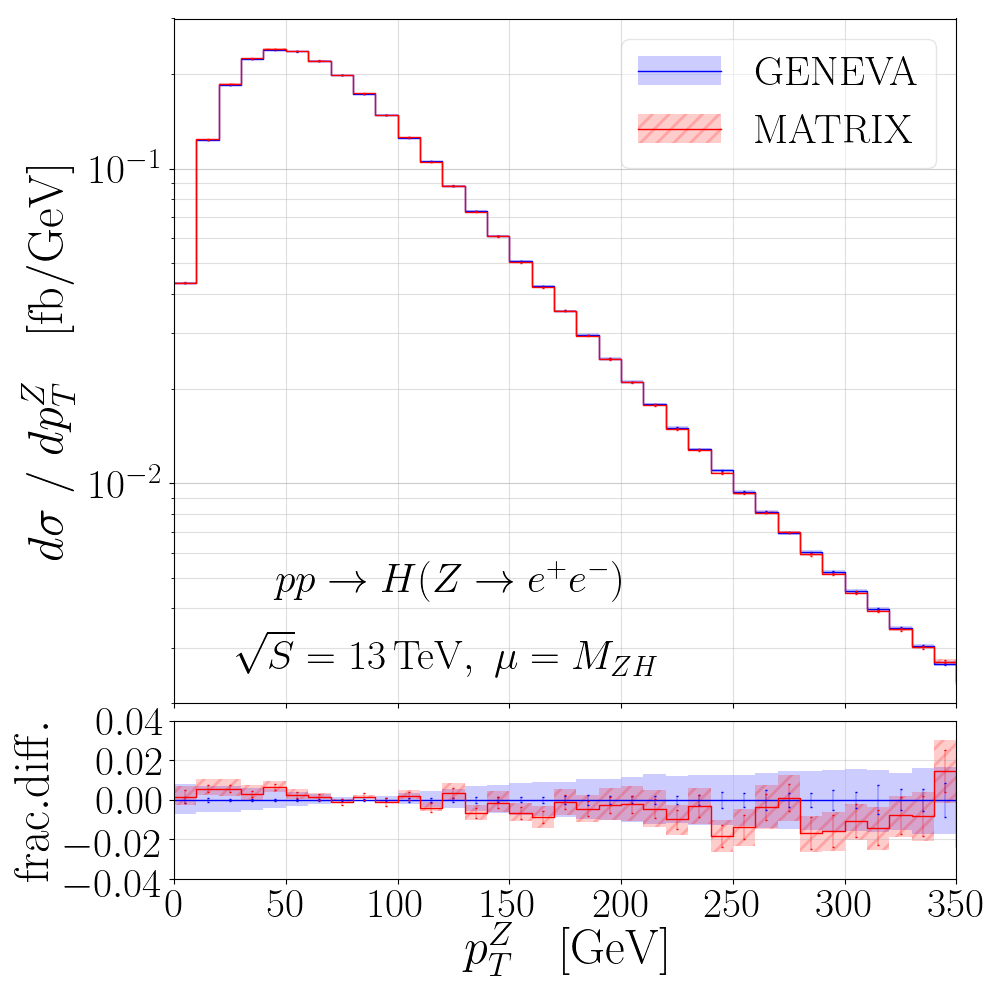}%
  \end{subfigure}
  \hspace*{\hspacebetweenthreeplots}
  \begin{subfigure}[b]{\rescalethreeplots}
    \includegraphics[width=\textwidth]{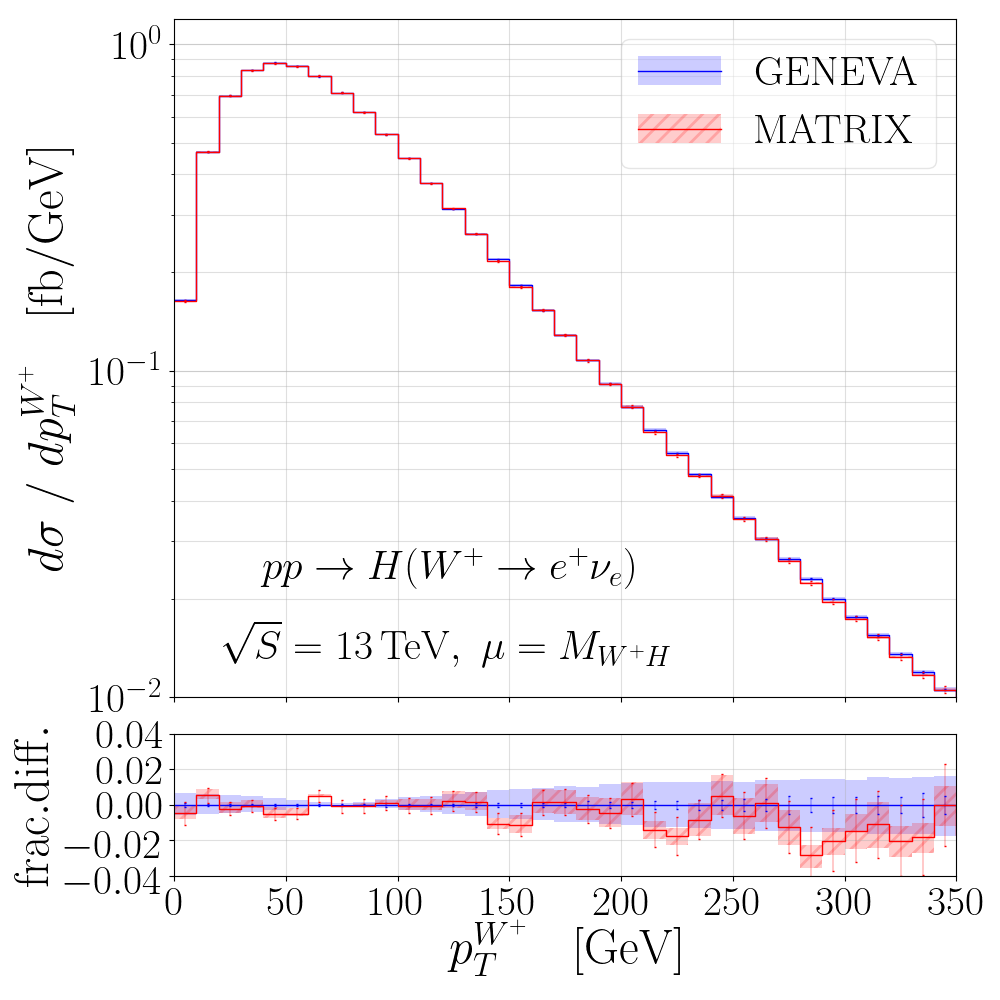}%
  \end{subfigure}
  \hspace*{\hspacebetweenthreeplots}
  \begin{subfigure}[b]{\rescalethreeplots}
    \includegraphics[width=\textwidth]{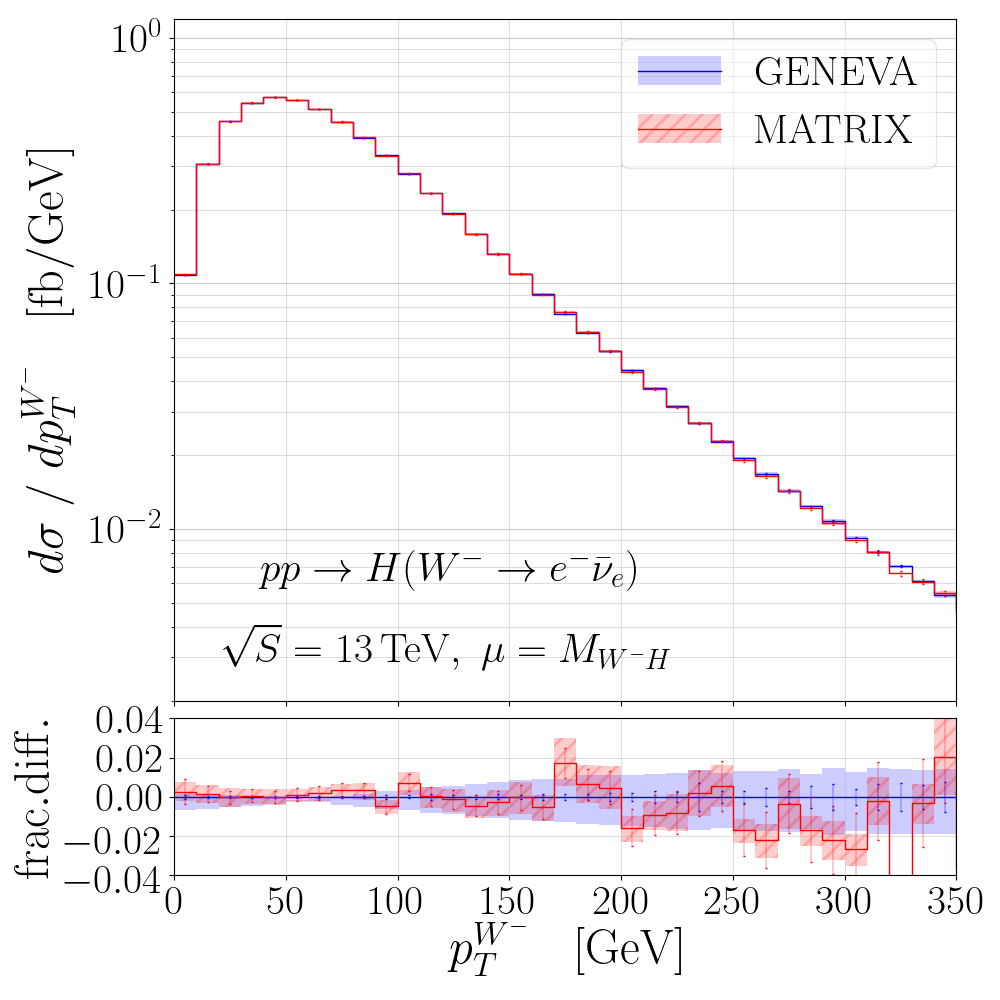}%
  \end{subfigure}
  \vspace{\spacebeforefigurecaption}
  \caption{Vector boson transverse-momentum spectra in $VH$ production: $ZH$ (left), $W^+H$ (centre), $W^-H$ (right) }
\label{fig:VpTMatrix}
\end{figure*}

Shower events originating from $\Phi_1$ and $\Phi_2$ events require instead more care. The reason is that we must now preserve the NNLL$'$+NNLO accuracy of the $\Tau_0$ spectrum.
This means that the $\Phi_2$ points produced after the first emission must be projectable onto $\Phi_1$ using the $\Tau_0$-preserving map mentioned earlier and discussed in detail in \citeref{Alioli:2015toa}. These constraints are most simply implemented by performing the first emissions in \geneva (using the analytic form of the NLL Sudakov factor and phase space maps) and only thereafter letting the shower act as usual, subject to the single restriction \mbox{$\Tau_2(\Phi_N)\leq\Tau_1(\Phi_2)$}. Since it can be shown that the shower acting on the resulting $\Phi_2$ events alters the accuracy of the $\Tau_0$ distribution only beyond NNLL$'$~\cite{Alioli:2015toa}, in practice we apply this procedure only to the $\Phi_1$ events. We find that
\begin{align}
&\frac{\mathrm{d} \sigma_{1}^{\mathrm{MC}}}{\df\Phi_{1}} (\Tau_0 > \Tau_0^\cut, \Tau_1^\cut, \Lambda_1)\\
\nn&\quad= \frac{\dsigMC_{1}}{\df\Phi_{1}} (\Tau_0 > \Tau_0^\cut, \Tau_1^\cut) \, U_1(\Tau_1^\cut, \Lambda_1),
\end{align}
\begin{align}
&\frac{\mathrm{d} \sigma_{\geq 2}^{\mathrm{MC}}}{\df\Phi_{2}} (\Tau_0 > \Tau_0^\cut, \Tau_1^\cut,\Tau_{1}>\Lambda_1)\\\nn
&\quad= \frac{\dsigMC_{\geq 2}}{\df\Phi_{2}} (\Tau_0 > \Tau_0^\cut, \Tau_{1}>\Tau_{1}^\cut) 
\\\nn
& \qquad
+ \frac{\df}{\df \Tau_1} \, \frac{\mathrm{d} \sigma^{\mathrm{MC}}_{1}}{\df\Phi_{1}} (\Tau_0 > \Tau_0^\cut, \Tau_1^\cut, \Tau_1) \,\\
\nn& \qquad \qquad\times\cP(\Phi_2) \,  \theta( \Lambda_1<\Tau_1<\Tau_1^{\max})
\,.
\end{align}

By choosing \mbox{$\Lambda_1\sim \Lambda_{\mathrm{QCD}}$}, the Sudakov factor \mbox{$U_1(\Tau_1^\cut,\Lambda_1)$}  becomes vanishingly small and we can relax the shower conditions on the \onejet contributions. The showered events therefore originate from either $\dsigMC_0$ or $\dsigMC_{\geq 2}$. 

The main differences which occur with respect to the Drell--Yan case are twofold: the decay of the Higgs boson and the choice of the starting scale for the gluon-fusion contributions.
In our study, we have hitherto considered a stable Higgs boson. As the Higgs boson is a scalar with a small width over mass ratio, it is a legitimate strategy to separate the production process from the decay in the narrow width approximation. As long as we limit ourselves to a leading-order description of the decay process, it could be handled entirely by the parton shower program. In the following, we have chosen to work with a stable Higgs boson, in order to simplify the analysis routines, but we might as well have let \pythiaEight decay it in order to achieve a more realistic description of the final state.
The gluon-fusion contribution could be considered as giving rise to $\Phi_0^{gg}$ events which are different in nature to the standard $\Phi_0$ events entering the \geneva formula. As explained in \subsec{SCETResummation}, the reason is that the gluon-fusion contributions are entirely nonsingular and are therefore merely added at FO, lacking any resummed contribution. Since they enter at $\ord{\as^2}$, any emission added by the shower would contribute beyond the claimed NNLO accuracy. We have, therefore, a greater freedom in the choice of the shower starting scale, which we set at the kinematic limit determined by the available centre-of-mass energy.

\section{Validation}
\label{sec:Validation}
\begin{figure*}[t]
  \begin{subfigure}[b]{\rescaletwoplots}
    \includegraphics[width=\textwidth]{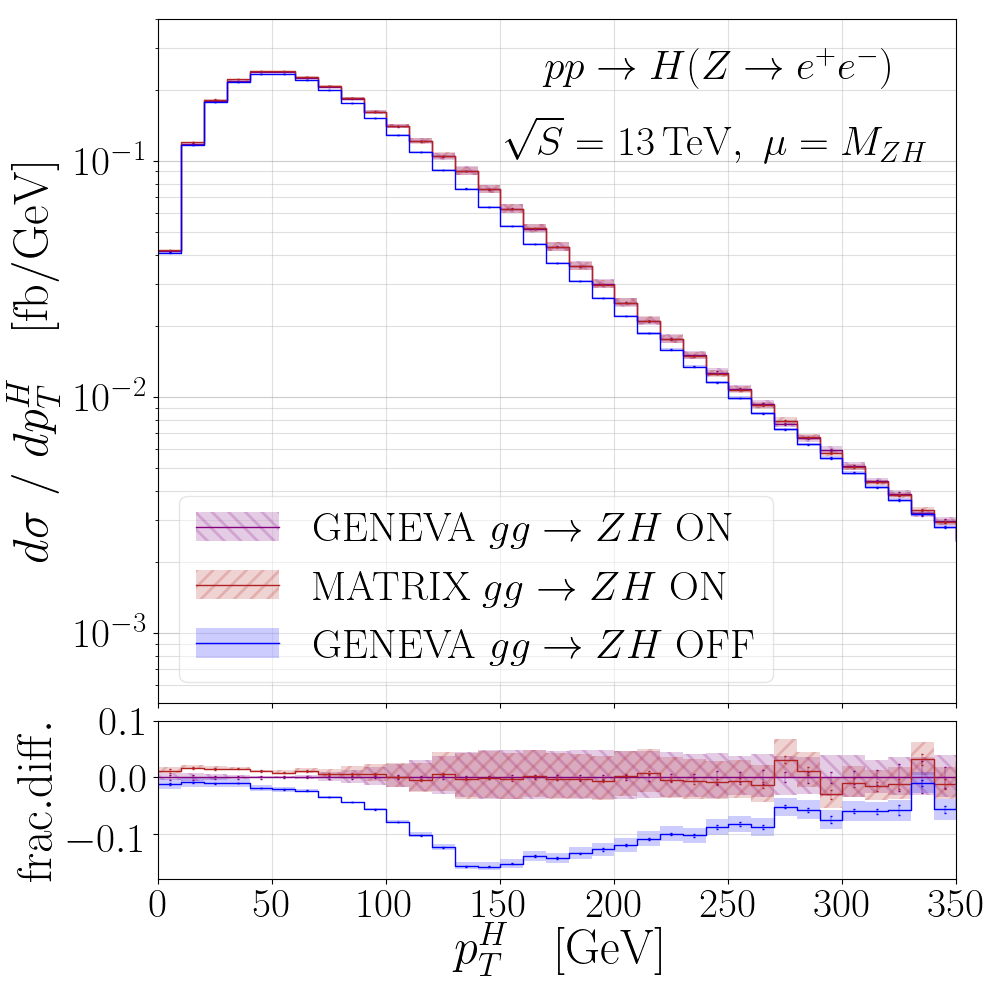}%
    \label{fig:HpTFOgg}
  \end{subfigure}
  \hspace*{\hspacebetweentwoplots}
  \begin{subfigure}[b]{\rescaletwoplots}
    \includegraphics[width=\textwidth]{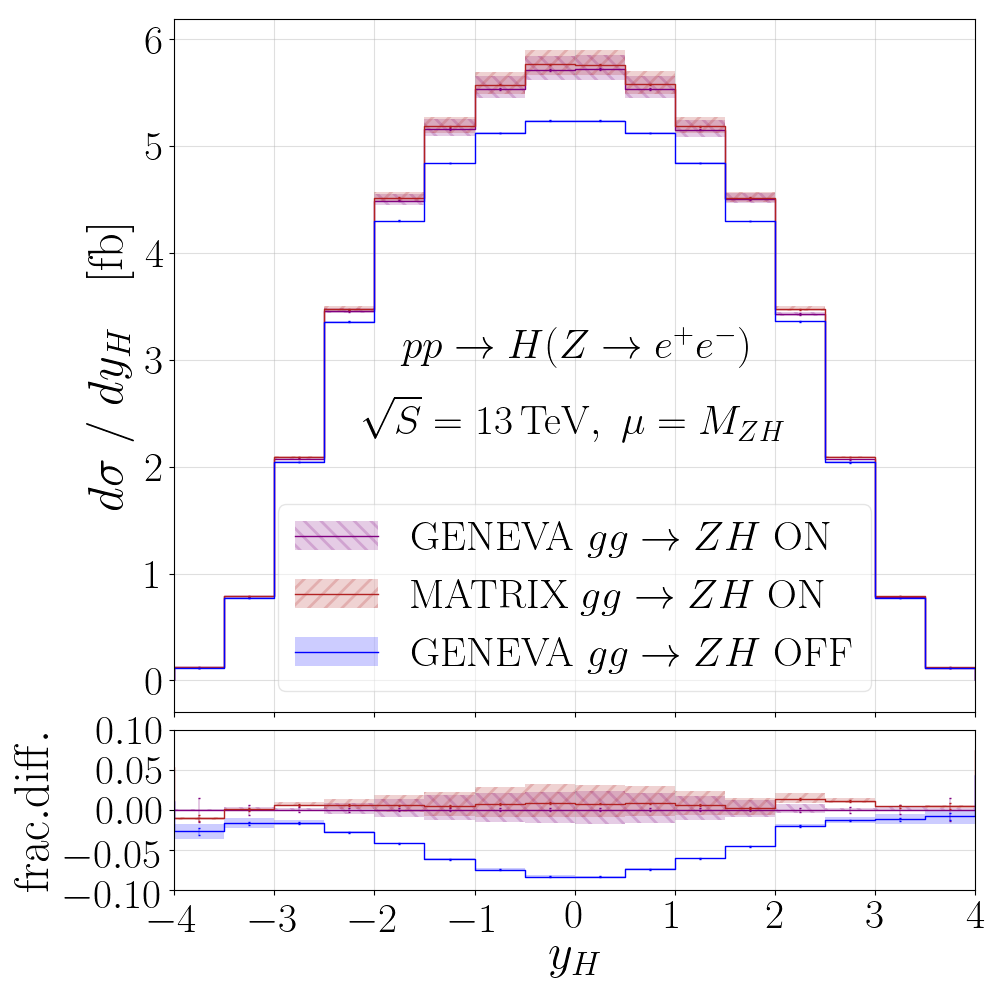}%
    \label{fig:HrapFOgg}
  \end{subfigure}
  \vspace{\spacebeforefigurecaption}
  \caption{Comparison of distributions including the $gg$ contribution between \Matrix and \geneva.}
\label{fig:matrixgg}
\end{figure*}

In order to validate the FO accuracy of our results we have compared with a
custom version of the \Matrix code~\cite{Grazzini:2017mhc} which implements the Higgsstrahlung
process. This program calculates cross sections at NNLO accuracy in QCD through
its fully general implementation of the $q_{\rm{T}}$-subtraction
formalism~\cite{Catani:2007vq,Bozzi:2005wk,Catani:2013tia}, in combination with
the dipole-subtraction formalism~\cite{Catani:1996jh,Catani:1996vz} to deal with
NLO-like singularities.
To eliminate the dependence on the slicing parameter \mbox{$r_\cut\equiv q^{\cut}_{T,VH}/M_{VH}$}
we numerically approach the limit $r_\cut\to0$ by the extrapolation procedure
of \citeref{Grazzini:2017mhc} not only for inclusive cross
sections, but also on a bin-wise level for distributions,
as introduced in \citeref{Catani:2019hip}.
The uncertainties associated with
this extrapolation procedure are combined with the statistical uncertainties to provide
the overall numerical error of the predictions that is shown in \figs{VpTMatrix}{ZHMatrix2} and \ref{fig:matrixgg}.

\begin{figure*}[t]
  \begin{subfigure}[b]{\rescaletwoplots}
    \includegraphics[width=\textwidth]{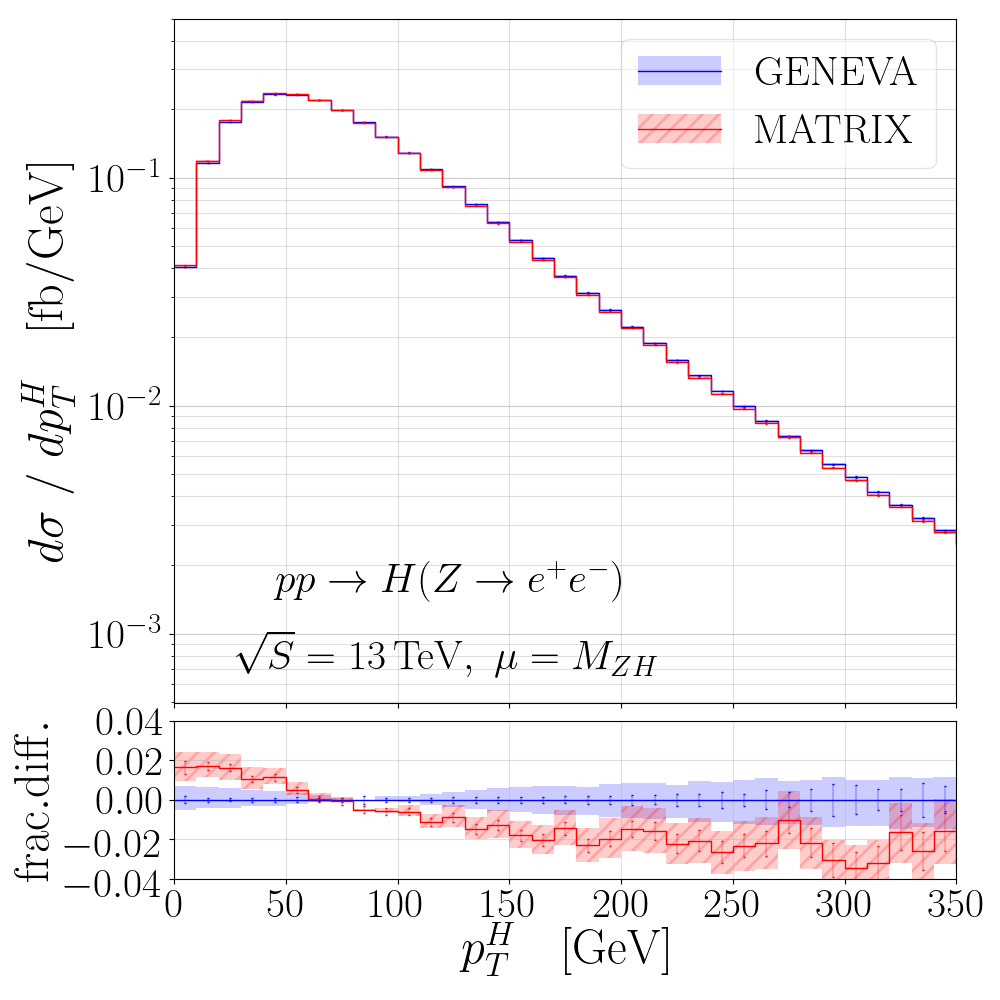}%
  \end{subfigure}
  \hspace*{\hspacebetweentwoplots}
  \begin{subfigure}[b]{\rescaletwoplots}
    \includegraphics[width=\textwidth]{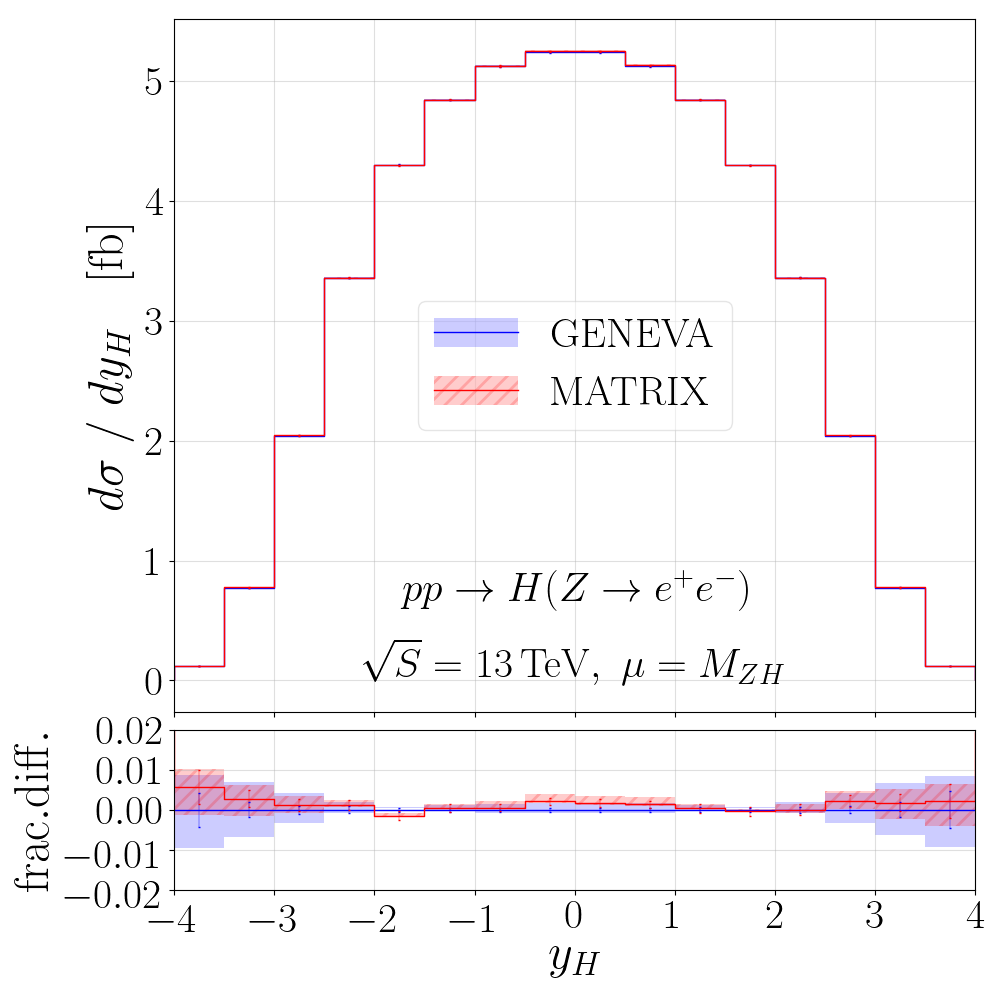}%
  \end{subfigure}
  \vspace{\spacebeforefigurecaption}
  \caption{Transverse momentum (left) and rapidity (right) of the Higgs boson in $ZH$ production.}
  \label{fig:ZHMatrix1}
\end{figure*}
\begin{figure*}[t]
  \begin{subfigure}[b]{\rescaletwoplots}
    \includegraphics[width=\textwidth]{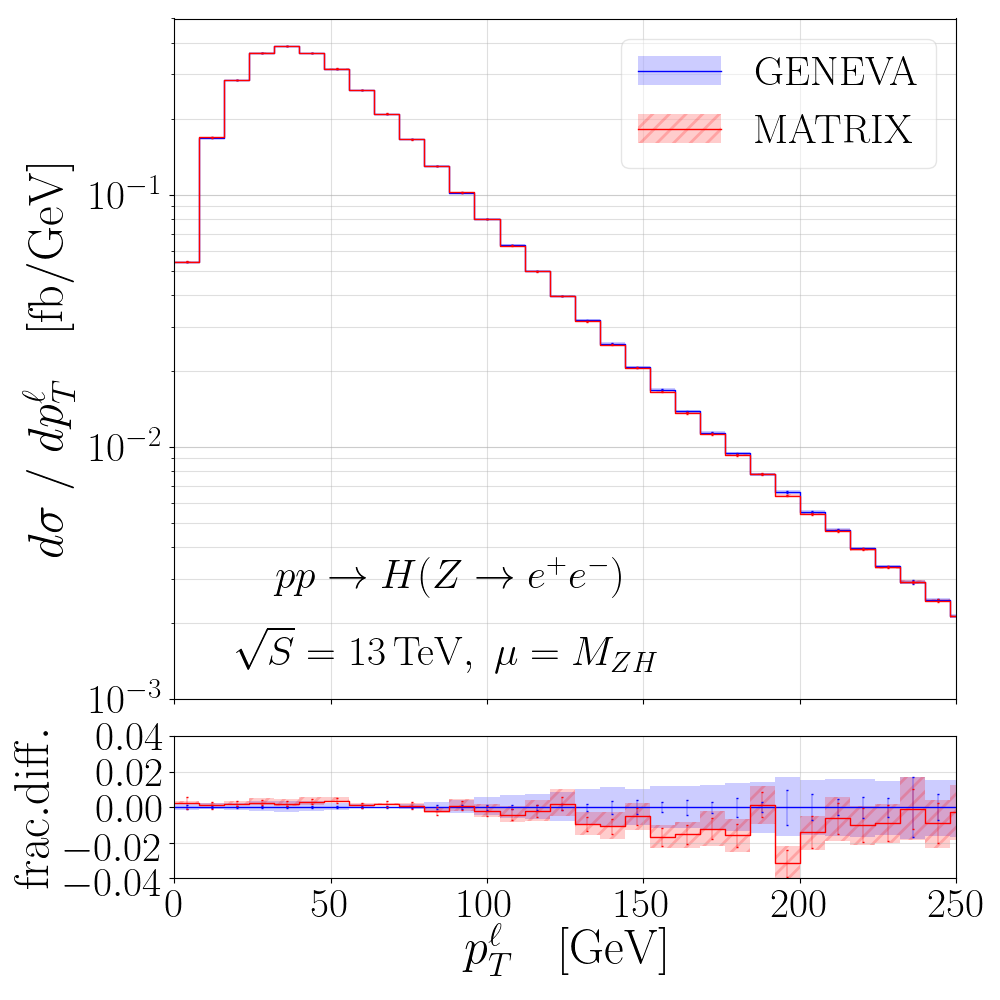}%
  \end{subfigure}
  \hspace*{\hspacebetweentwoplots}
  \begin{subfigure}[b]{\rescaletwoplots}
    \includegraphics[width=\textwidth]{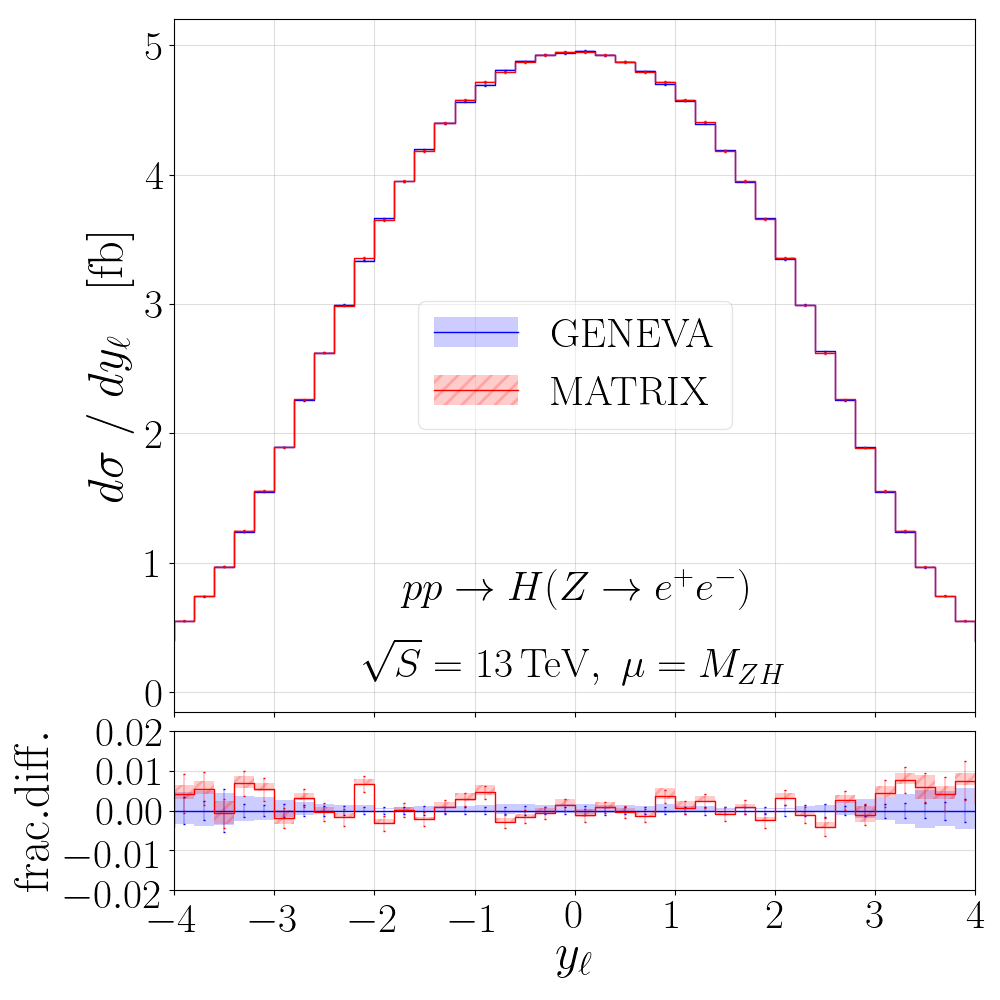}%
  \end{subfigure}
  \vspace{\spacebeforefigurecaption}
  \caption{Transverse momentum (left) and rapidity (right) of the hardest lepton in $ZH$ production.}
\label{fig:ZHMatrix2}
\end{figure*}

Our calculations are performed for $pp$ collisions at the 13\TeV LHC using the \texttt{PDF4LHC15\_nnlo\_100} PDF sets~\cite{Butterworth:2015oua} available via the \LHAPDF interface~\cite{Buckley:2014ana}. In the case of $ZH$ production we restrict the invariant mass of the lepton pair to lie within the range \mbox{$50\GeV<m_{\ell^+\ell^-}<150\GeV$}; we place a much looser \mbox{$1\GeV<m_{\ell\nu}<13\TeV$} restriction in the case of $W^{\pm}H$ production. We use the following values of SM parameters:
\begin{eqnarray}
  &&M_Z =  91.1876\GeV,\quad \Gamma_Z =  2.4952\GeV\nn,\\&&M_W=80.398\GeV, \quad \Gamma_W=2.1054\GeV, \\&& \sin^2 \theta_W^{\rm eff} =  0.2226459 ,\quad   \alpha^{-1}(M_Z) = 132.338
 \,, \nn
\end{eqnarray}
and use also the values of the CKM matrix elements appearing in \citeref{Tanabashi:2018oca}. We set $\taucut$ and $\Tau_1^\cut$ at 1\GeV and adopt the dynamical scale choice \mbox{$\mu_{\rm FO}=M_{VH}$} for the FO scale in this section (except where explicitly specified otherwise).

In \fig{VpTMatrix} we show the transverse-momentum spectra of the produced vector boson for each of the three possible $VH$ processes, excluding for the moment the gluon-initiated channel in $ZH$ production. The red hashed band associated with the \Matrix curve reflects the error obtained from a simultaneous variation of the renormalisation and factorisation scales around the central scale (3-point variation), while the blue band associated with the \geneva prediction has been obtained by following the procedure detailed in \subsec{Scales}. We observe good agreement for the central value  in each case. The scale variation band in the \geneva case becomes slightly larger in the hard region at large transverse momentum because the mechanism for recovering the exact NNLO cross sections and scale variations discussed in \subsec{Scales} becomes less effective, being based on the total cross section and not on differential distributions.

\begin{figure*}[t]
  \begin{subfigure}[b]{\rescaletwoplots}
    \includegraphics[width=\textwidth]{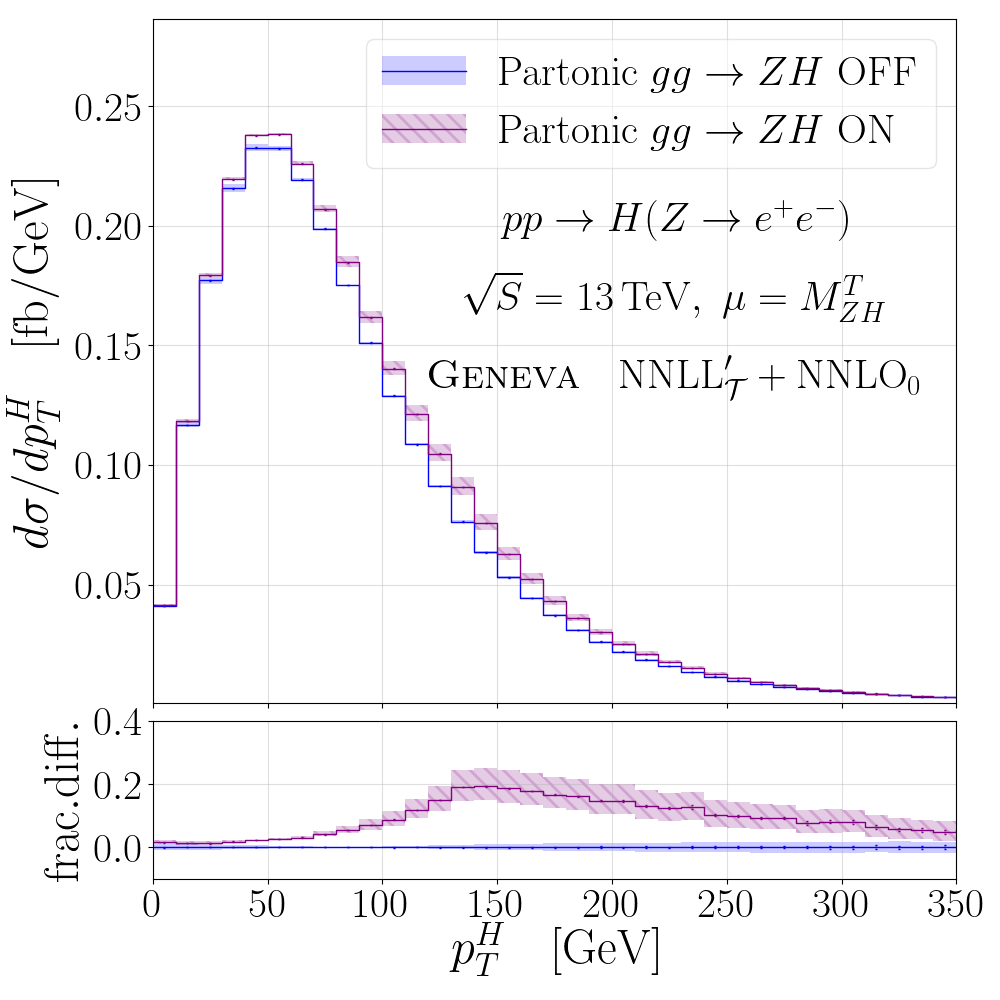}%
    \label{fig:HpTggeff}
  \end{subfigure}
  \hspace*{\hspacebetweentwoplots}
  \begin{subfigure}[b]{\rescaletwoplots}
    \includegraphics[width=\textwidth]{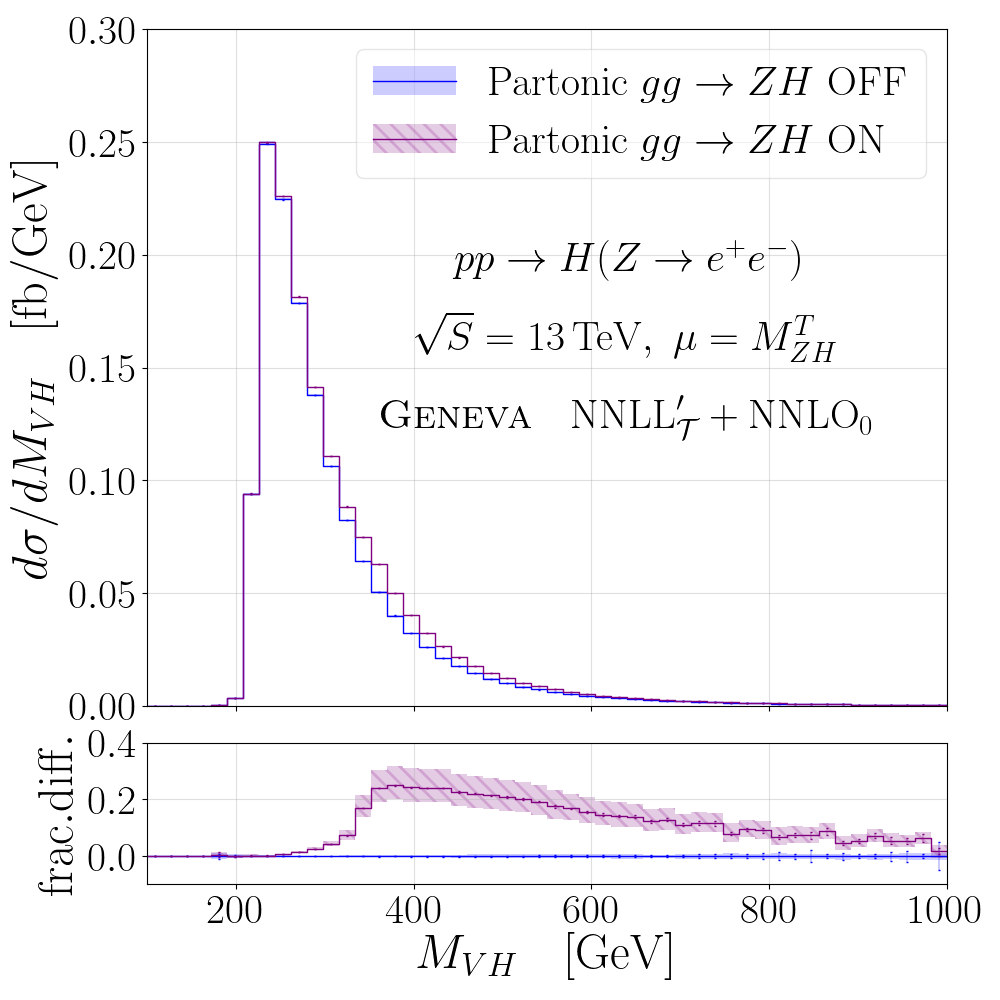}%
    \label{fig:VHmassggeff}
  \end{subfigure}
  \vspace{\spacebeforefigurecaption}
  \caption{Impact of the inclusion of the gluon-fusion contributions to $ZH$ production at the partonic level in \geneva.}
\label{fig:ggeff}
\end{figure*}

\begin{figure*}[t]
  \begin{subfigure}[b]{\rescalethreeplots}
    \includegraphics[width=\textwidth]{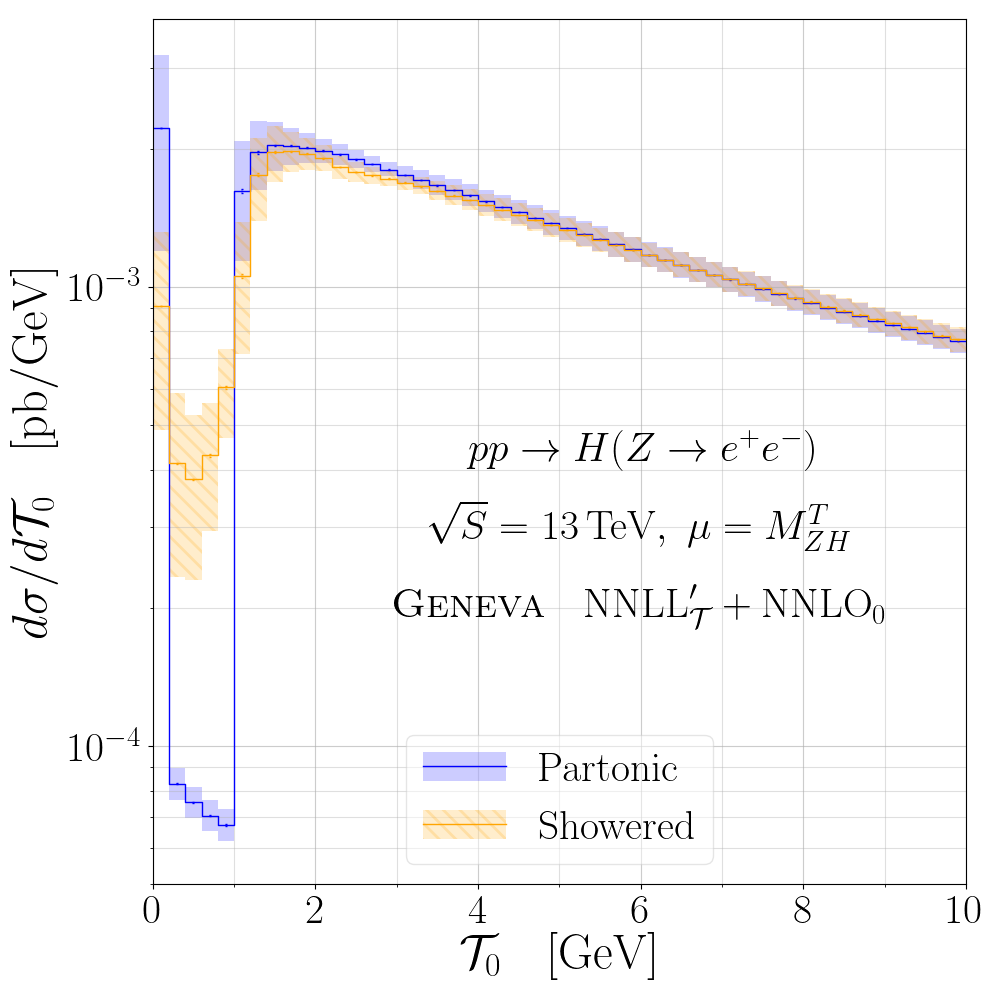}%
    \label{fig:Tau0peak}
  \end{subfigure}
  \hspace*{\hspacebetweenthreeplots}
  \begin{subfigure}[b]{\rescalethreeplots}
    \includegraphics[width=\textwidth]{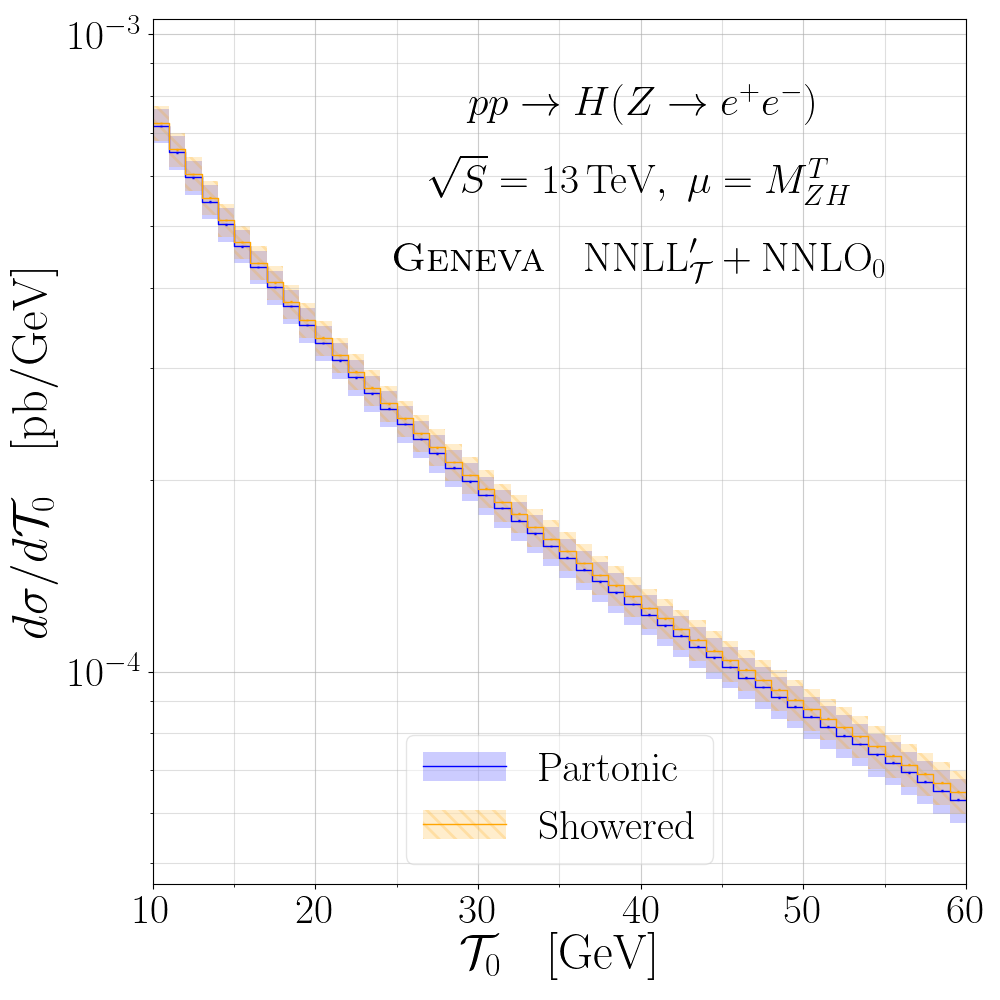}%
    \label{fig:Tau0trans}
  \end{subfigure}
  \hspace*{\hspacebetweenthreeplots}
  \begin{subfigure}[b]{\rescalethreeplots}
    \includegraphics[width=\textwidth]{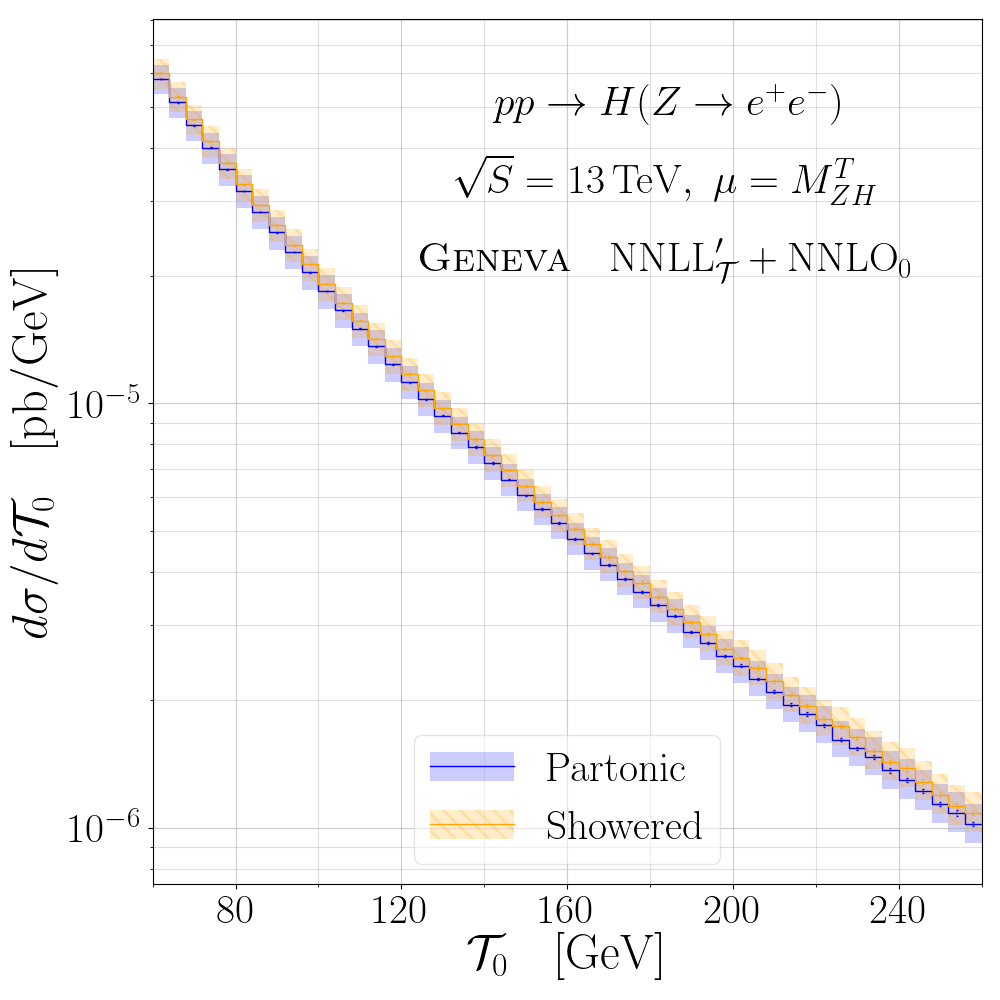}%
    \label{fig:Tau0tail}
  \end{subfigure}
  \vspace{\spacebeforefigurecaption}
  \caption{Validation of the $\Tau_0$ spectrum in \geneva. The partonic NNLL$'$+NNLO $\Tau_0$ resummation is compared to the showered results, before the addition of non-perturbative effects.}
\label{fig:tau0validation}
\end{figure*}
\begin{figure*}[t]
  \begin{subfigure}[b]{\rescalethreeplots}
    \includegraphics[width=\textwidth]{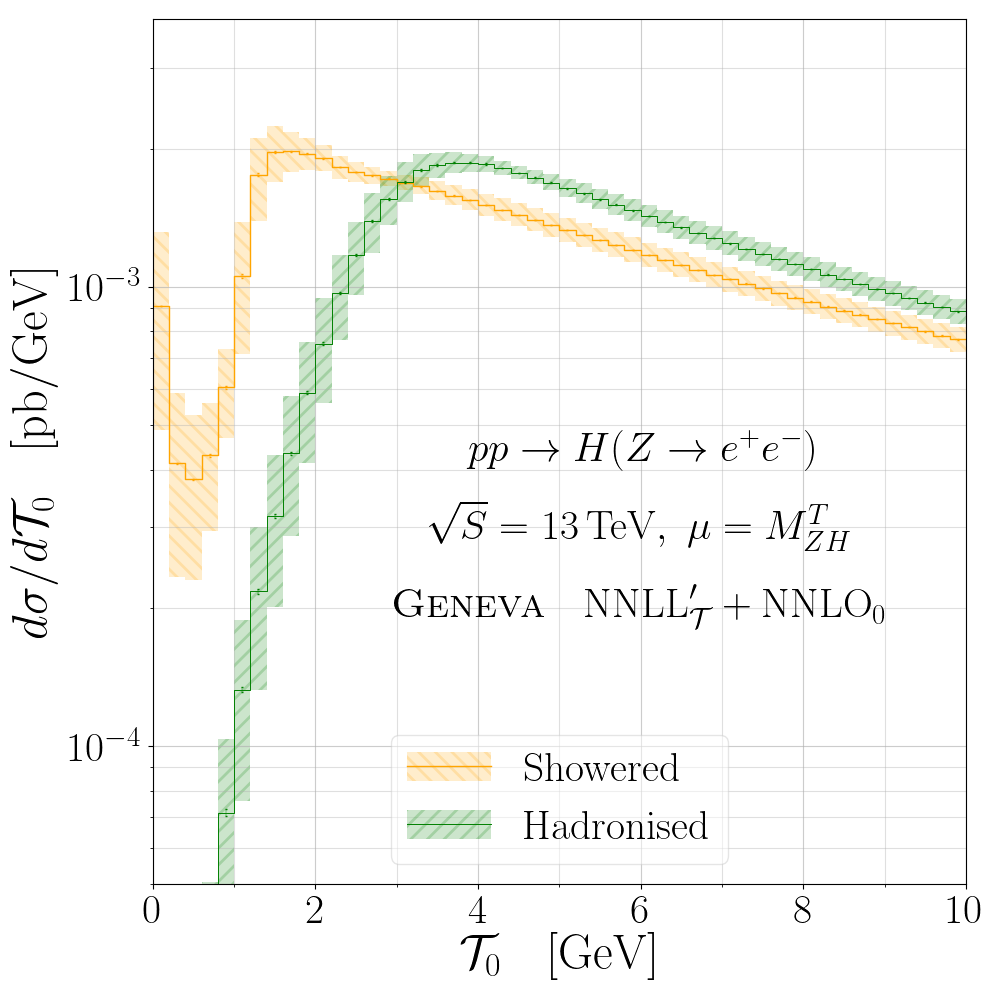}%
    \label{fig:Tau0peakhad}
  \end{subfigure}
  \hspace*{\hspacebetweenthreeplots}
  \begin{subfigure}[b]{\rescalethreeplots}
    \includegraphics[width=\textwidth]{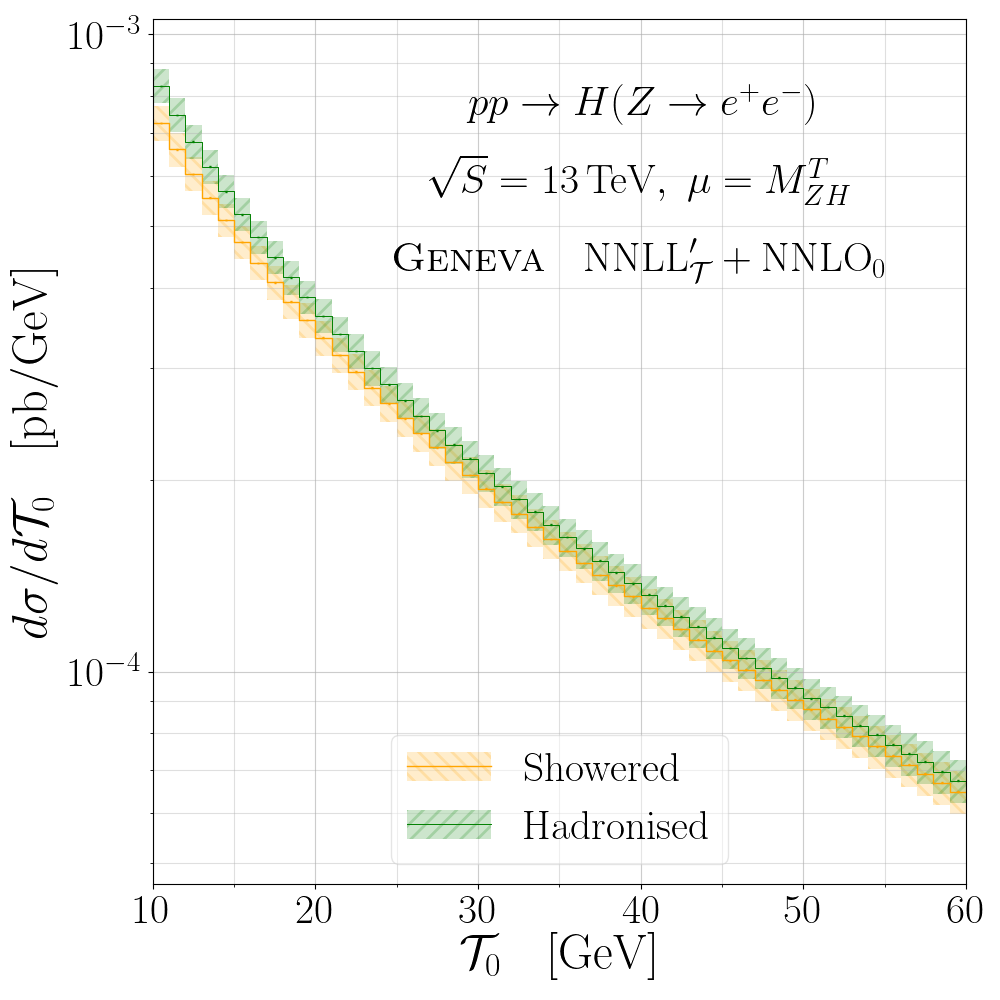}%
    \label{fig:Tau0transhad}
  \end{subfigure}
  \hspace*{\hspacebetweenthreeplots}
  \begin{subfigure}[b]{\rescalethreeplots}
    \includegraphics[width=\textwidth]{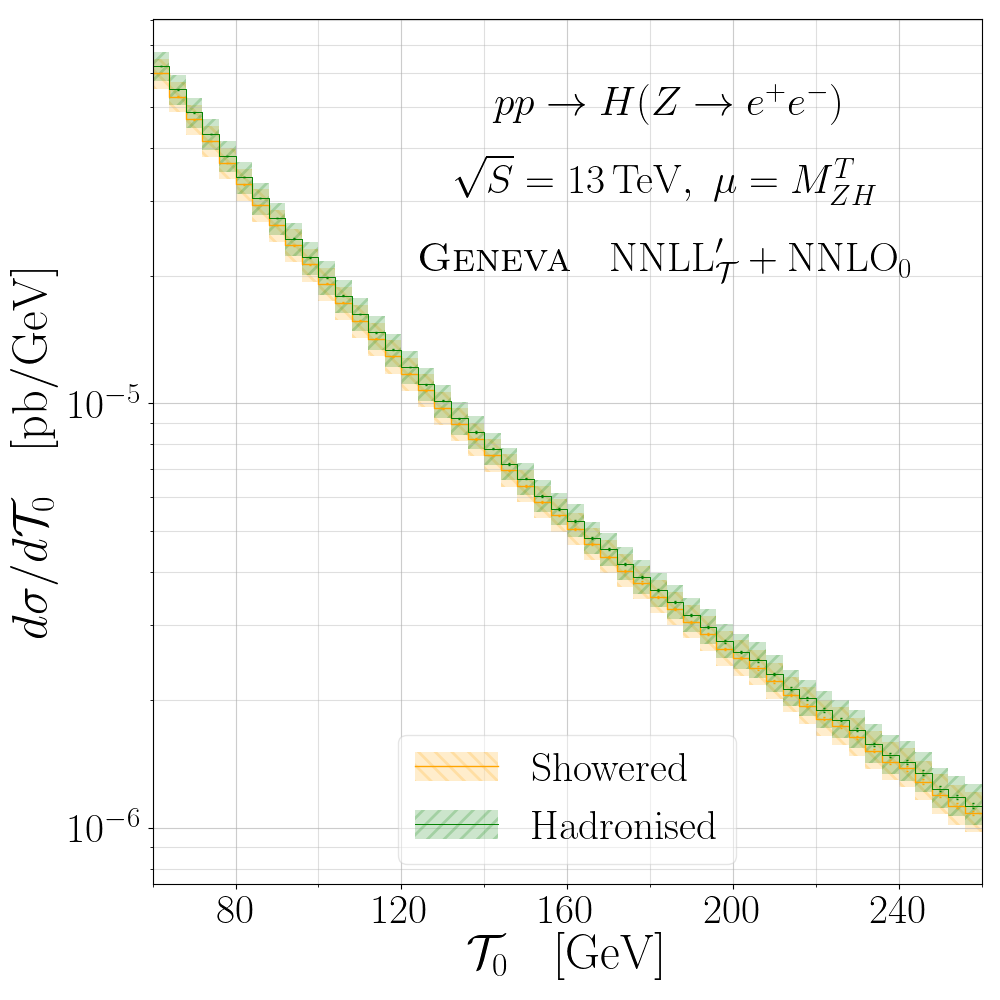}%
    \label{fig:Tau0tailhad}
  \end{subfigure}
  \vspace{\spacebeforefigurecaption}
  \caption{Comparison of the showered and hadronised $\Tau_0$ spectra in \geneva.}
\label{fig:tau0hadvalidation}
\end{figure*}

From here on we leave behind the $W^{\pm}H$ process and focus only on the $ZH$ case, which displays broadly similar behaviour but exhibits a few more interesting subtleties due to the presence of the aforementioned $gg$ production channel at NNLO. In \figs{ZHMatrix1}{ZHMatrix2} we show comparisons of transverse-momentum and rapidity distributions between \geneva and \Matrix, again neglecting the $gg$ channel for the time being. At the value of $\taucut$ chosen, we observe very good agreement with the \Matrix results with the exception of the case of the Higgs boson transverse momentum. We find that agreement for this distribution is improved at a lower value of the cut at the expense of a higher statistical uncertainty (as previously discussed in \subsec{Powerterms}).

We now consider the inclusion of the $gg$ channel at FO.
In \fig{ggeff} we show the impact at the partonic level in \geneva, focusing on the Higgs boson transverse momentum and the  invariant mass of the $VH$ system. We observe an effect of up to $\sim20\%$ on the differential distributions, demonstrating the importance of including this channel. We see also an increase in the scale uncertainties, related to the fact that the process \mbox{$gg\to ZH$} is included in effect only at leading order (albeit $\ord{\as^2}$). In \fig{matrixgg} we compare the \geneva predictions including the $gg$ channel with those of \Matrix for the Higgs boson transverse momentum and rapidity and again find good agreement.

Unfortunately, we were unable to compare with the results of similar calculations presented in \citerefs{Astill:2016hpa,Astill:2018ivh}. The reason for this was that in the case of $W^{\pm}H$ production, the authors in \citeref{Astill:2016hpa} neglected all top-quark effects while using a $3\times3$ CKM matrix. While each of these options can be set separately in the current public release of \openloops~\cite{Buccioni:2019sur} which we rely on to provide amplitudes, the combination of both is not possible. In the case of $ZH$ production, the same authors in \citerefs{Astill:2018ivh} included the decay of the Higgs boson into a $b\bar{b}$ pair at NLO, which at present is omitted in our calculation. We intend to include the higher-order corrections to the decay in a future release of the program which will enable us to make a detailed comparison of the two results.

\begin{figure*}[t]
  \begin{subfigure}[b]{\rescaletwoplots}
    \includegraphics[width=\textwidth]{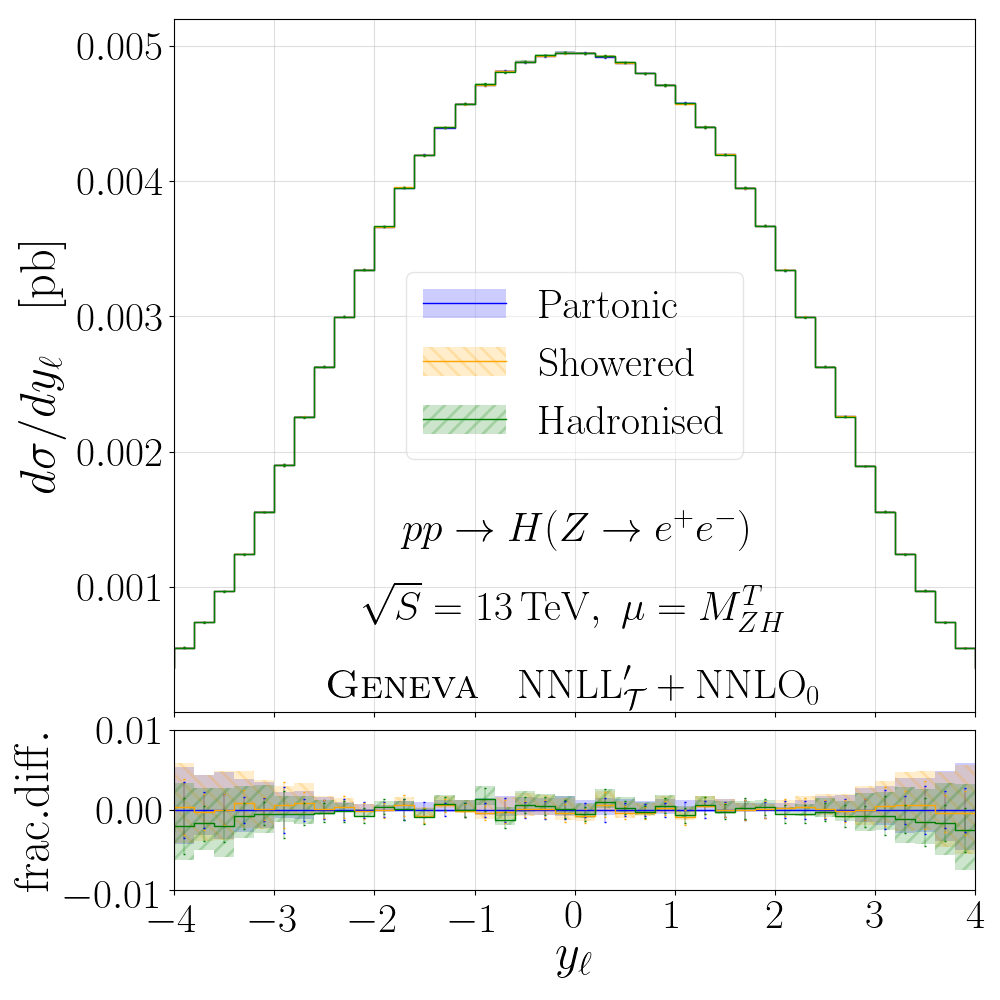}%
    \label{fig:cmpleprap}
  \end{subfigure}
  \hspace*{\hspacebetweentwoplots}
  \begin{subfigure}[b]{\rescaletwoplots}
    \includegraphics[width=\textwidth]{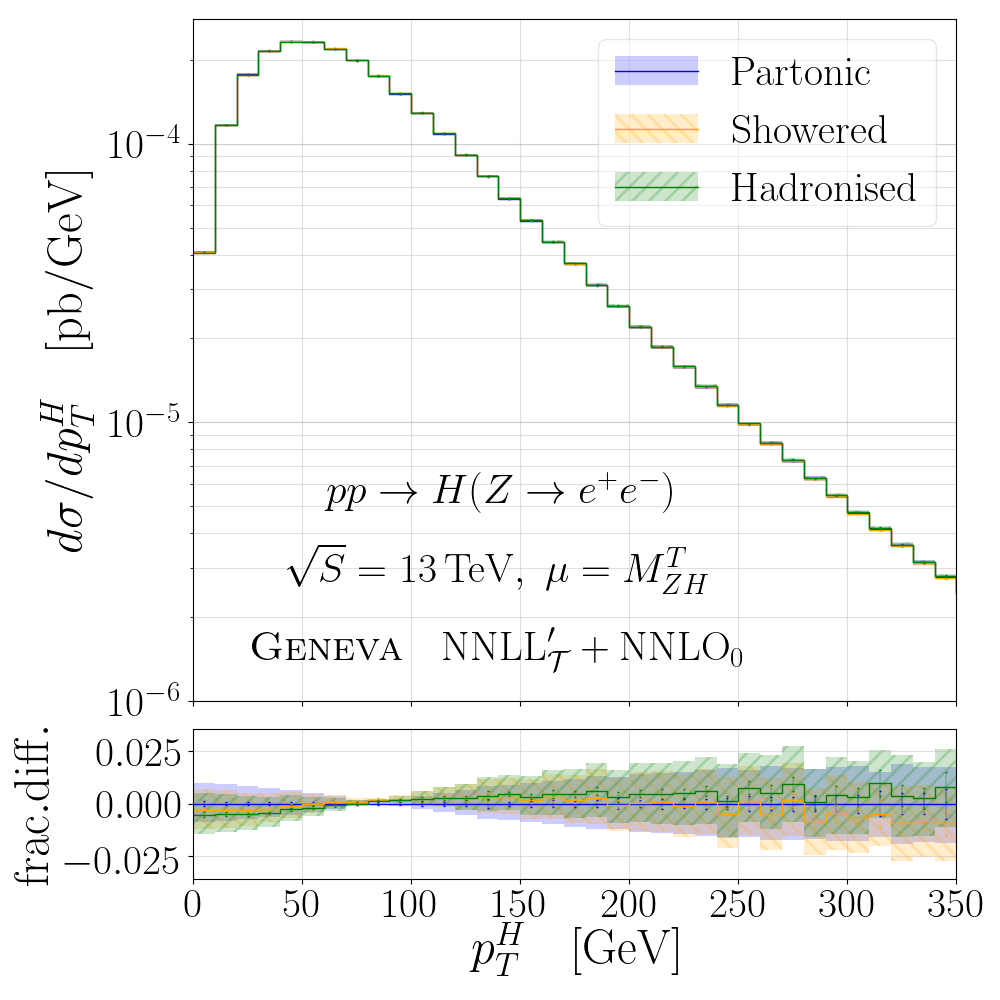}%
    \label{fig:cmpHpT}
  \end{subfigure}
  \vspace{\spacebeforefigurecaption}
  \caption{Comparison of showered and hadronised spectra for inclusive quantities in \geneva.}
\label{fig:inclusivevalidation}
\end{figure*}
\begin{figure*}[t]
  \begin{subfigure}[b]{\rescaletwoplots}
    \includegraphics[width=\textwidth]{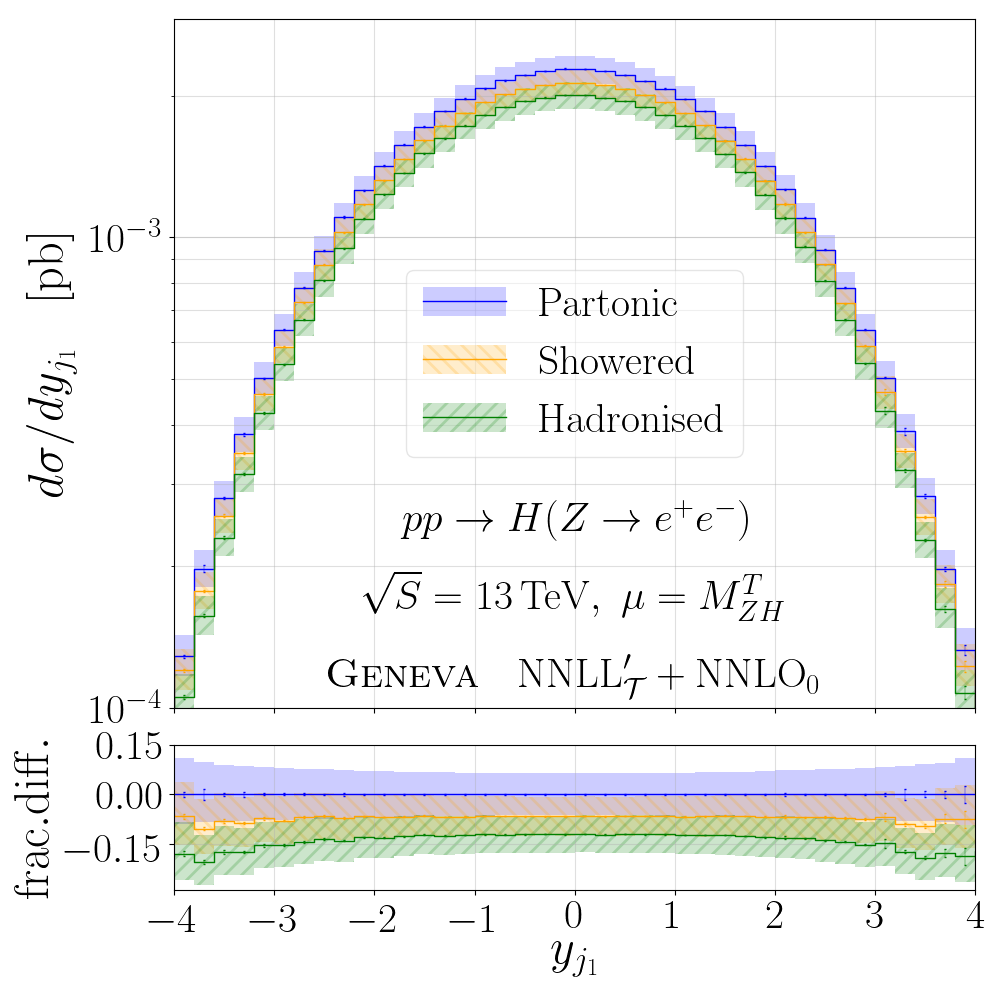}%
    \label{fig:cmpjetrap}
  \end{subfigure}
  \hspace*{\hspacebetweentwoplots}
  \begin{subfigure}[b]{\rescaletwoplots}
    \includegraphics[width=\textwidth]{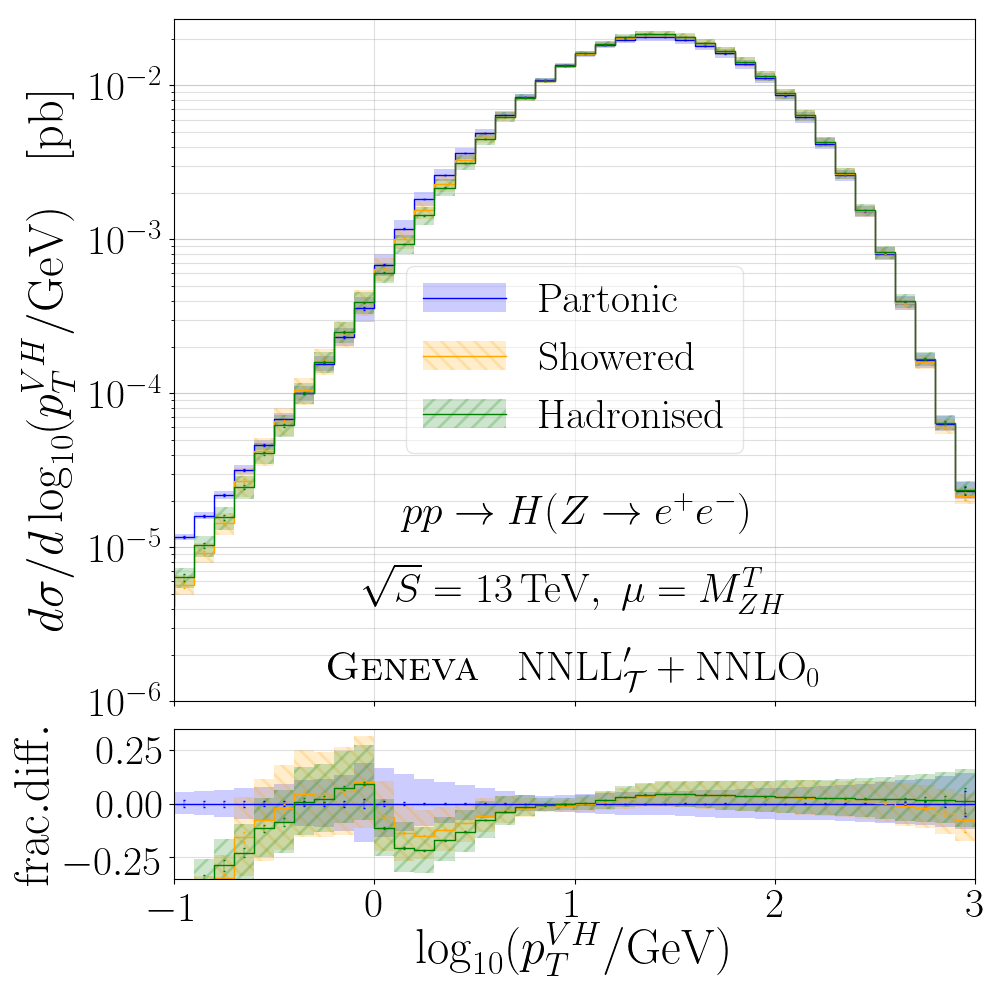}%
    \label{fig:cmpVHpT}
  \end{subfigure}
  \vspace{\spacebeforefigurecaption}
  \caption{Comparison of showered and hadronised spectra for exclusive quantities in \geneva.}
\label{fig:exclusivevalidation}
\end{figure*}

\section{Results}
\label{sec:Results}

We now present our predictions for various spectra after interfacing with the parton shower provided by \pythiaEight v8.235~\cite{Sjostrand:2006za, Sjostrand:2007gs}. For definiteness, we have chosen \pythiaEight's tune 18, we have set \mbox{$p_{T_0}^{\rm ref} = 2.4\GeV$}, and have run with all matrix element corrections switched off, since now the radiative effects entering at higher order are provided by \geneva.
In order to keep the analysis as simple as possible, we have also switched off all  QED effects in the showering.
In the following, we adopt the scale choice \mbox{$\mu_{\rm FO}=M_{VH}^T$} but otherwise use the same values of the parameters as in \sec{Validation}. Our scale uncertainty bands are calculated using a different procedure depending on whether quantities are either exclusive or inclusive in the additional radiation, as described in \subsec{Scales}. We reconstruct jets using the \FastJet algorithm~\cite{Cacciari:2005hq,Cacciari:2011ma} with a jet radius \mbox{$R=0.4$} and a minimum \mbox{$p_T^{j,\cut}=30\GeV$}.

In \fig{tau0validation} we show the $\Tau_0$ distribution at the partonic and showered level for $ZH$ production with the $gg$ channel switched off in three different regions: the peak (left pane), where the resummation effects are expected to be dominant; the transition region (centre pane) where the resummed and FO contributions should be on the same footing; and the tail (right pane), where the resummation is switched off and the FO perturbative expansion is a valid approximation.  We confirm that, as expected, the NNLL$'$ accuracy of the $\Tau_0$ distribution is preserved by the shower above $\Tau_0^\cut$. The shape below $\Tau_0^\cut$ is determined entirely by \pythiaEight, but the cross section falling below the cut is preserved as required (apart from the small spillover discussed in \subsec{PSmatching}). The small contribution appearing between \mbox{$0 < \Tau_0 < \Tau_0^\cut$} at the partonic level is due to the nonsingular $\Phi_1$ events which cannot be projected on a valid Born-like configuration and are therefore included only at fixed order.

In \fig{tau0hadvalidation} we turn on the hadronisation and show its impact on the showered distribution: we observe a large difference only in the peak region, as expected,  with the corrections at larger values of $\Tau_0$  being suppressed as $\ord{\lqcd/Q}$.

\begin{figure*}[ht!]
  \begin{subfigure}[b]{\rescaletwoplots}
    \includegraphics[width=\textwidth]{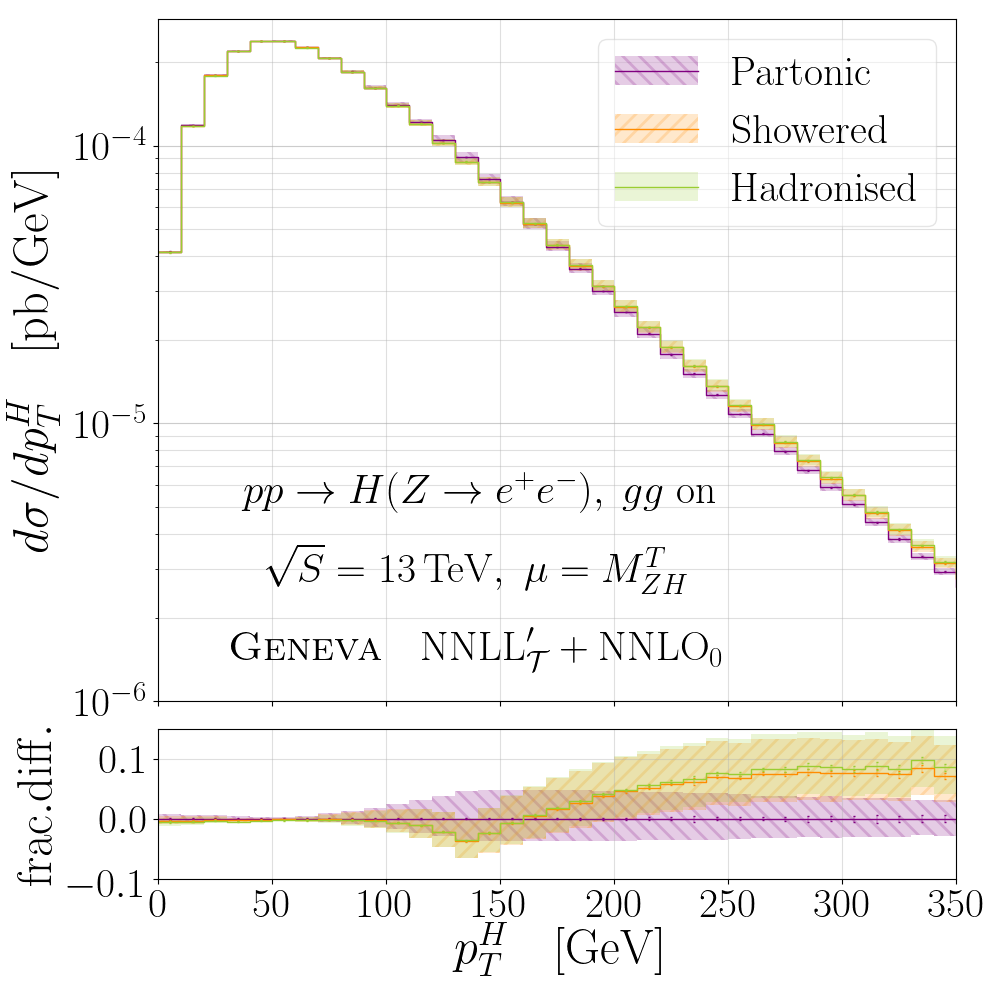}%
    \label{fig:cmpHpTgg}
  \end{subfigure}
  \hspace*{\hspacebetweentwoplots}
  \begin{subfigure}[b]{\rescaletwoplots}
    \includegraphics[width=\textwidth]{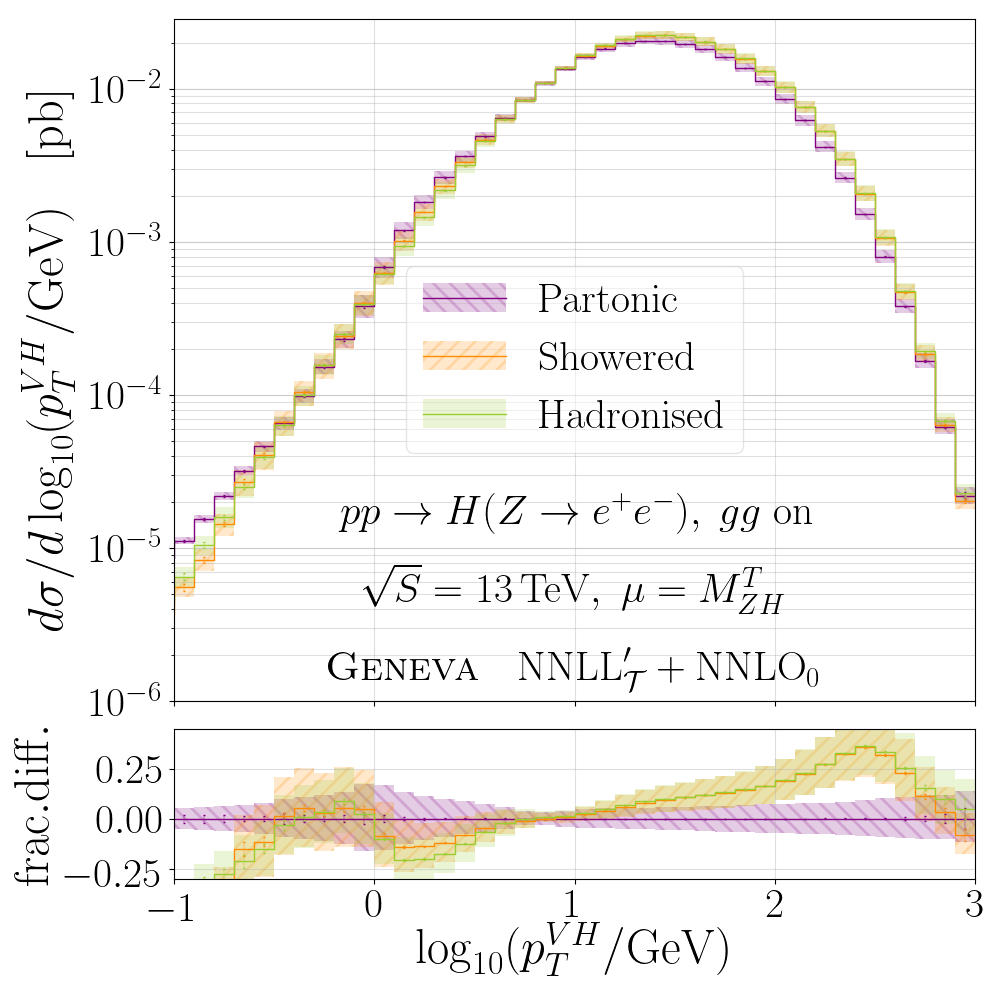}%
    \label{fig:cmpVHpTgg}
  \end{subfigure}
  \vspace{\spacebeforefigurecaption}
  \caption{Comparison of showered and hadronised spectra for quantities in \geneva with the inclusion of the gluon-fusion contributions to $ZH$ production.}
\label{fig:validationwgg}
\end{figure*}

\begin{figure*}[ht!]
  \begin{subfigure}[b]{\rescaletwoplots}
    \includegraphics[width=\textwidth]{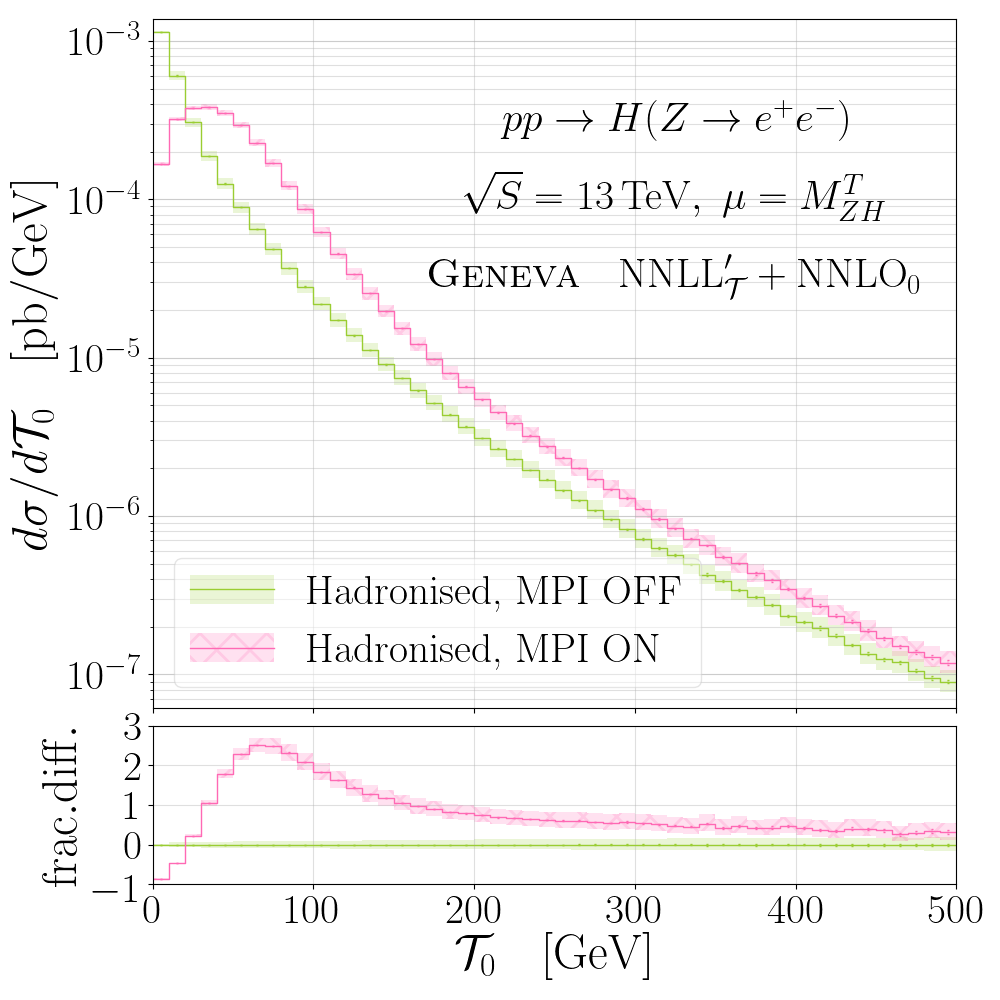}%
    \label{fig:Tau0MPIeff}
  \end{subfigure}
  \hspace*{\hspacebetweentwoplots}
  \begin{subfigure}[b]{\rescaletwoplots}
    \includegraphics[width=\textwidth]{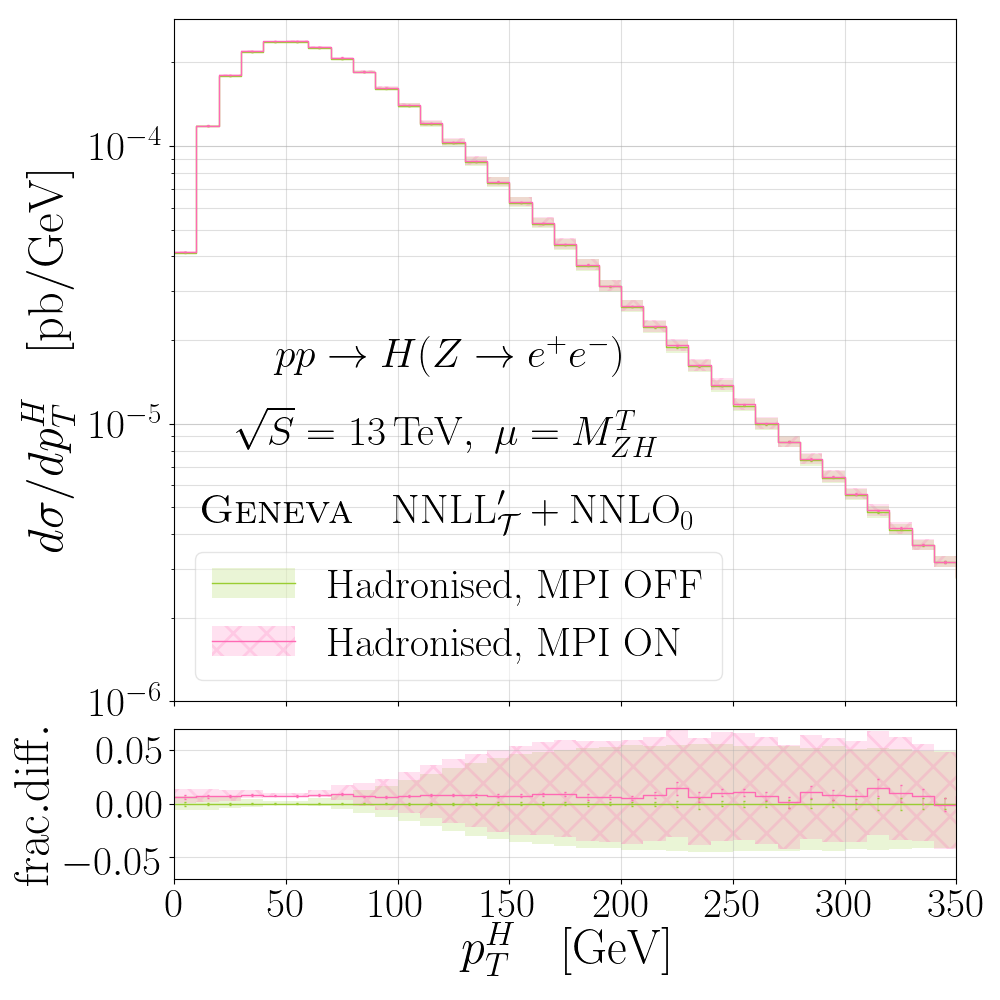}%
    \label{fig:HpTMPIeff}
  \end{subfigure}
  \vspace{\spacebeforefigurecaption}
  \caption{Impact of the inclusion of MPI effects to $ZH$ production at the hadronised level in \geneva. The $gg$ channel is included.}
\label{fig:MPIeff}
\end{figure*}

We continue with an examination of the effects of the shower on distributions other than $\Tau_0$. In \fig{inclusivevalidation}  we consider the transverse momentum of the Higgs boson and the rapidity of the hardest lepton, both quantities which are inclusive over any additional radiation. We note that showering does not significantly change the normalisation or the shape of the distributions, demonstrating that the NNLO accuracy is maintained even after showering and hadronisation. Additionally, we note that the scale variation uncertainties are unaffected by the shower and hadronisation stages. This is to be expected and a consequence of the fact that in this simple analysis we have neglected any uncertainty originating from the interface of our partonic predictions to the shower and from the hadronisation model. One could explore these effects in more detail by studying, for example, the variation in our predictions after modification of the shower starting scale for the former or explore different tuning parameters for the latter. Such investigation is beyond the scope of the current study.

In \fig{exclusivevalidation} we show instead quantities exclusive in the additional radiation. Although we cannot claim NNLL$'$ accuracy in the resummation for these observables, we may anticipate that a certain amount of the accuracy from the prediction of $\Tau_0$ may be inherited by other quantities. In the case of the rapidity of the hardest jet we see that the shower causes an overall shift of the distribution downwards by $\mathcal{O}(10\%)$, most likely due to the acceptance cut on the jet above $p_T^{j,\cut}$. Considering instead the transverse momentum of the $VH$ system, we see that the shape of the distribution in the resummation region is significantly modified by the shower.

We now proceed to study the effect of the shower after including the gluon-fusion channel. The majority of distributions show similar effects, with the inclusive quantities surviving the shower stage unmodified. Two distributions for which more significant differences are seen, however, are shown in \fig{validationwgg}, where we plot the transverse momenta of the Higgs boson and of the $VH$ system. We notice a significant deviation after the inclusion of the shower in the hard region. This is most likely to be a consequence of our choice of starting the shower at a very high scale for these contributions, and is in accordance with previous observations for similar $gg$-initiated processes~\cite{Alioli:2016xab,Heinrich:2017kxx}. Since the showering of these contributions starts at $\ord{\as^3}$, in the present calculation we lack any further means by which we may constrain its effects. This motivates a future inclusion of the $gg$-initiated process at NLO.

Finally, in \fig{MPIeff} we examine the difference in the spectra when multi-parton interaction~(MPI) effects are included in predictions for $ZH$ production at the hadronised level and with the $gg$ channel switched on. We observe that the $\Tau_0$ distribution is significantly modified by inclusion of MPI, as already seen in \citeref{Alioli:2016wqt} -- this follows from the definition of the beam thrust at the analysis level, which involves a sum over all final particles including those arising from secondary collisions. In the case of an inclusive quantity however (for example the Higgs boson transverse momentum), the shape of the distribution changes very little.

\section{Conclusions}
\label{sec:Conclusions}
We have implemented the Higgsstrahlung process in the \geneva
framework, which provides resummed predictions matched to the
fixed-order calculation and a parton shower at NNLL$'$+NNLO accuracy.  In
order to make a consistent choice for the profile functions used for
the determination of the resummation scales, we have studied the
interplay between the singular and nonsingular contributions in
different regions of the Born-like phase space. As expected, we find
that the region in which the resummation is applicable depends mostly
on the value of the invariant mass of the $VH$ system and it can be
defined in a similar way for all classes of events showing only a
mild dependence on the other kinematical variables.

We have confirmed the fixed-order accuracy of our results by
comparison with the program \Matrix and found very good agreement
at \mbox{$\Tau_0^\cut = 1\GeV$}. Only for the Higgs boson transverse-momentum distribution
do we find that an improved agreement can be reached by lowering 
the \zerojet resolution cutoff to \mbox{$\Tau_0^\cut = 0.1\GeV$}, at the price of an increased statistical
uncertainty. This leads us to conclude that the discrepancy is most
likely due to the missing $\ord{\as^2}$ power corrections
in \geneva as compared to \Matrix. 

We have provided predictions at the showered and hadronised levels by
interfacing with the parton shower program \pythiaEight. We first confirmed
that the accuracy of the $\Tau_0$ distribution is unaffected by the
showering and then showed that the inclusive distributions retain
their NNLO accuracy and are mostly unchanged by the shower. The
shower effects were found to be more significant for more exclusive distributions.

We have also included the gluon-fusion channel and studied the
differential distributions which are affected most by its
inclusion. We observe larger shower effects connected with these configurations,
possibly related to the higher starting scale used for the shower.

Finally we were also able to include MPI effects and showed their impact on inclusive and
MPI-sensitive distributions.

The code used for this study is available upon request to the authors and will be made public
in a future \geneva release at

\begin{center}
\url{http://geneva.physics.lbl.gov}.\\~\\
\end{center}

\vspace{-1em}
There are some clear directions in which this work could be furthered. At present, the
decay of the Higgs boson in our implementation can only be provided at LO by \pythiaEight.
A full NNLO calculation of the Higgs boson decay to a $b\bar{b}$ pair would be desirable
in order to improve the description of the final states which are experimentally accessible.
The combination of the production and decay processes in the narrow-width approximation is
in principle feasible within the \geneva framework, and we plan to study this in a future
work. Another avenue worth pursuing is the inclusion of subleading power
corrections at the fully differential level, which would reduce the size of the neglected
terms and thus improve the predictions for distributions even when a larger value of the
resolution cutoff is used. These effects are likely to become more important as processes
with more complex final-state phase spaces are considered.

Given the difficulty of discovering New Physics at the LHC, it is now more important than ever
to be able to make precise predictions of the SM backgrounds both at the fiducial
cross section level and when extrapolated over the full phase space. Since Monte Carlo event generators
are the primary tool used to provide these predictions, it is vital that they are made as accurate as possible.
This allows state-of-the-art theoretical calculations to be made available to experimental collaborations
so that they can be used directly in analyses.

\section{Acknowledgements}
\label{sec:Acknowledgements}
We are grateful to Christian Bauer, Frank Tackmann and Emanuele Re for
useful discussions and for providing comments on the manuscript.  We
also thank Jonas Lindert for his help with the \openloops2
implementation.  The work of SA, AB, SK and LR is supported by the ERC
Starting Grant REINVENT-714788.  SA and ML acknowledge funding from
Fondazione Cariplo and Regione Lombardia, grant 2017-2070.  We
acknowledge the CINECA award under the ISCRA initiative for the
availability of the high performance computing resources needed for
this work.  This research also used resources of the National Energy
Research Scientific Computing Center (NERSC), a U.S. Department of
Energy Office of Science User Facility operated under Contract
No. DE-AC02-05CH11231.


\bibliography{geneva}

\begin{thebibliography}{79}%
\makeatletter
\providecommand \@ifxundefined [1]{%
 \@ifx{#1\undefined}
}%
\providecommand \@ifnum [1]{%
 \ifnum #1\expandafter \@firstoftwo
 \else \expandafter \@secondoftwo
 \fi
}%
\providecommand \@ifx [1]{%
 \ifx #1\expandafter \@firstoftwo
 \else \expandafter \@secondoftwo
 \fi
}%
\providecommand \natexlab [1]{#1}%
\providecommand \enquote  [1]{``#1''}%
\providecommand \bibnamefont  [1]{#1}%
\providecommand \bibfnamefont [1]{#1}%
\providecommand \citenamefont [1]{#1}%
\providecommand \href@noop [0]{\@secondoftwo}%
\providecommand \href [0]{\begingroup \@sanitize@url \@href}%
\providecommand \@href[1]{\@@startlink{#1}\@@href}%
\providecommand \@@href[1]{\endgroup#1\@@endlink}%
\providecommand \@sanitize@url [0]{\catcode `\\12\catcode `\$12\catcode
  `\&12\catcode `\#12\catcode `\^12\catcode `\_12\catcode `\%12\relax}%
\providecommand \@@startlink[1]{}%
\providecommand \@@endlink[0]{}%
\providecommand \url  [0]{\begingroup\@sanitize@url \@url }%
\providecommand \@url [1]{\endgroup\@href {#1}{\urlprefix }}%
\providecommand \urlprefix  [0]{URL }%
\providecommand \Eprint [0]{\href }%
\providecommand \doibase [0]{http://dx.doi.org/}%
\providecommand \selectlanguage [0]{\@gobble}%
\providecommand \bibinfo  [0]{\@secondoftwo}%
\providecommand \bibfield  [0]{\@secondoftwo}%
\providecommand \translation [1]{[#1]}%
\providecommand \BibitemOpen [0]{}%
\providecommand \bibitemStop [0]{}%
\providecommand \bibitemNoStop [0]{.\EOS\space}%
\providecommand \EOS [0]{\spacefactor3000\relax}%
\providecommand \BibitemShut  [1]{\csname bibitem#1\endcsname}%
\let\auto@bib@innerbib\@empty
\bibitem [{\citenamefont {Sirunyan}\ \emph {et~al.}(2018)\citenamefont
  {Sirunyan} \emph {et~al.}}]{Sirunyan:2018kst}%
  \BibitemOpen
  \bibfield  {author} {\bibinfo {author} {\bibfnamefont {A.~M.}\ \bibnamefont
  {Sirunyan}} \emph {et~al.} (\bibinfo {collaboration} {CMS}),\ }\href
  {\doibase 10.1103/PhysRevLett.121.121801} {\bibfield  {journal} {\bibinfo
  {journal} {Phys. Rev. Lett.}\ }\textbf {\bibinfo {volume} {121}},\ \bibinfo
  {pages} {121801} (\bibinfo {year} {2018})},\ \Eprint
  {http://arxiv.org/abs/1808.08242} {arXiv:1808.08242 [hep-ex]} \BibitemShut
  {NoStop}%
\bibitem [{\citenamefont {Aaboud}\ \emph {et~al.}(2018)\citenamefont {Aaboud}
  \emph {et~al.}}]{Aaboud:2018zhk}%
  \BibitemOpen
  \bibfield  {author} {\bibinfo {author} {\bibfnamefont {M.}~\bibnamefont
  {Aaboud}} \emph {et~al.} (\bibinfo {collaboration} {ATLAS}),\ }\href
  {\doibase 10.1016/j.physletb.2018.09.013} {\bibfield  {journal} {\bibinfo
  {journal} {Phys. Lett.}\ }\textbf {\bibinfo {volume} {B786}},\ \bibinfo
  {pages} {59} (\bibinfo {year} {2018})},\ \Eprint
  {http://arxiv.org/abs/1808.08238} {arXiv:1808.08238 [hep-ex]} \BibitemShut
  {NoStop}%
\bibitem [{\citenamefont {Brein}\ \emph {et~al.}(2012)\citenamefont {Brein},
  \citenamefont {Harlander}, \citenamefont {Wiesemann},\ and\ \citenamefont
  {Zirke}}]{Brein:2011vx}%
  \BibitemOpen
  \bibfield  {author} {\bibinfo {author} {\bibfnamefont {O.}~\bibnamefont
  {Brein}}, \bibinfo {author} {\bibfnamefont {R.}~\bibnamefont {Harlander}},
  \bibinfo {author} {\bibfnamefont {M.}~\bibnamefont {Wiesemann}}, \ and\
  \bibinfo {author} {\bibfnamefont {T.}~\bibnamefont {Zirke}},\ }\href
  {\doibase 10.1140/epjc/s10052-012-1868-6} {\bibfield  {journal} {\bibinfo
  {journal} {Eur. Phys. J.}\ }\textbf {\bibinfo {volume} {C72}},\ \bibinfo
  {pages} {1868} (\bibinfo {year} {2012})},\ \Eprint
  {http://arxiv.org/abs/1111.0761} {arXiv:1111.0761 [hep-ph]} \BibitemShut
  {NoStop}%
\bibitem [{\citenamefont {Brein}\ \emph {et~al.}(2004)\citenamefont {Brein},
  \citenamefont {Djouadi},\ and\ \citenamefont {Harlander}}]{Brein:2003wg}%
  \BibitemOpen
  \bibfield  {author} {\bibinfo {author} {\bibfnamefont {O.}~\bibnamefont
  {Brein}}, \bibinfo {author} {\bibfnamefont {A.}~\bibnamefont {Djouadi}}, \
  and\ \bibinfo {author} {\bibfnamefont {R.}~\bibnamefont {Harlander}},\ }\href
  {\doibase 10.1016/j.physletb.2003.10.112} {\bibfield  {journal} {\bibinfo
  {journal} {Phys. Lett.}\ }\textbf {\bibinfo {volume} {B579}},\ \bibinfo
  {pages} {149} (\bibinfo {year} {2004})},\ \Eprint
  {http://arxiv.org/abs/hep-ph/0307206} {arXiv:hep-ph/0307206 [hep-ph]}
  \BibitemShut {NoStop}%
\bibitem [{\citenamefont {Brein}\ \emph {et~al.}(2013)\citenamefont {Brein},
  \citenamefont {Harlander},\ and\ \citenamefont {Zirke}}]{Brein:2012ne}%
  \BibitemOpen
  \bibfield  {author} {\bibinfo {author} {\bibfnamefont {O.}~\bibnamefont
  {Brein}}, \bibinfo {author} {\bibfnamefont {R.~V.}\ \bibnamefont
  {Harlander}}, \ and\ \bibinfo {author} {\bibfnamefont {T.~J.~E.}\
  \bibnamefont {Zirke}},\ }\href {\doibase 10.1016/j.cpc.2012.11.002}
  {\bibfield  {journal} {\bibinfo  {journal} {Comput. Phys. Commun.}\ }\textbf
  {\bibinfo {volume} {184}},\ \bibinfo {pages} {998} (\bibinfo {year}
  {2013})},\ \Eprint {http://arxiv.org/abs/1210.5347} {arXiv:1210.5347
  [hep-ph]} \BibitemShut {NoStop}%
\bibitem [{\citenamefont {Ferrera}\ \emph {et~al.}(2011)\citenamefont
  {Ferrera}, \citenamefont {Grazzini},\ and\ \citenamefont
  {Tramontano}}]{Ferrera:2011bk}%
  \BibitemOpen
  \bibfield  {author} {\bibinfo {author} {\bibfnamefont {G.}~\bibnamefont
  {Ferrera}}, \bibinfo {author} {\bibfnamefont {M.}~\bibnamefont {Grazzini}}, \
  and\ \bibinfo {author} {\bibfnamefont {F.}~\bibnamefont {Tramontano}},\
  }\href {\doibase 10.1103/PhysRevLett.107.152003} {\bibfield  {journal}
  {\bibinfo  {journal} {Phys. Rev. Lett.}\ }\textbf {\bibinfo {volume} {107}},\
  \bibinfo {pages} {152003} (\bibinfo {year} {2011})},\ \Eprint
  {http://arxiv.org/abs/1107.1164} {arXiv:1107.1164 [hep-ph]} \BibitemShut
  {NoStop}%
\bibitem [{\citenamefont {Ferrera}\ \emph {et~al.}(2014)\citenamefont
  {Ferrera}, \citenamefont {Grazzini},\ and\ \citenamefont
  {Tramontano}}]{Ferrera:2013yga}%
  \BibitemOpen
  \bibfield  {author} {\bibinfo {author} {\bibfnamefont {G.}~\bibnamefont
  {Ferrera}}, \bibinfo {author} {\bibfnamefont {M.}~\bibnamefont {Grazzini}}, \
  and\ \bibinfo {author} {\bibfnamefont {F.}~\bibnamefont {Tramontano}},\
  }\href {\doibase 10.1007/JHEP04(2014)039} {\bibfield  {journal} {\bibinfo
  {journal} {JHEP}\ }\textbf {\bibinfo {volume} {04}},\ \bibinfo {pages} {039}
  (\bibinfo {year} {2014})},\ \Eprint {http://arxiv.org/abs/1312.1669}
  {arXiv:1312.1669 [hep-ph]} \BibitemShut {NoStop}%
\bibitem [{\citenamefont {Ferrera}\ \emph {et~al.}(2015)\citenamefont
  {Ferrera}, \citenamefont {Grazzini},\ and\ \citenamefont
  {Tramontano}}]{Ferrera:2014lca}%
  \BibitemOpen
  \bibfield  {author} {\bibinfo {author} {\bibfnamefont {G.}~\bibnamefont
  {Ferrera}}, \bibinfo {author} {\bibfnamefont {M.}~\bibnamefont {Grazzini}}, \
  and\ \bibinfo {author} {\bibfnamefont {F.}~\bibnamefont {Tramontano}},\
  }\href {\doibase 10.1016/j.physletb.2014.11.040} {\bibfield  {journal}
  {\bibinfo  {journal} {Phys. Lett.}\ }\textbf {\bibinfo {volume} {B740}},\
  \bibinfo {pages} {51} (\bibinfo {year} {2015})},\ \Eprint
  {http://arxiv.org/abs/1407.4747} {arXiv:1407.4747 [hep-ph]} \BibitemShut
  {NoStop}%
\bibitem [{\citenamefont {Campbell}\ \emph {et~al.}(2016)\citenamefont
  {Campbell}, \citenamefont {Ellis},\ and\ \citenamefont
  {Williams}}]{Campbell:2016jau}%
  \BibitemOpen
  \bibfield  {author} {\bibinfo {author} {\bibfnamefont {J.~M.}\ \bibnamefont
  {Campbell}}, \bibinfo {author} {\bibfnamefont {R.~K.}\ \bibnamefont {Ellis}},
  \ and\ \bibinfo {author} {\bibfnamefont {C.}~\bibnamefont {Williams}},\
  }\href {\doibase 10.1007/JHEP06(2016)179} {\bibfield  {journal} {\bibinfo
  {journal} {JHEP}\ }\textbf {\bibinfo {volume} {06}},\ \bibinfo {pages} {179}
  (\bibinfo {year} {2016})},\ \Eprint {http://arxiv.org/abs/1601.00658}
  {arXiv:1601.00658 [hep-ph]} \BibitemShut {NoStop}%
\bibitem [{\citenamefont {Ferrera}\ \emph {et~al.}(2018)\citenamefont
  {Ferrera}, \citenamefont {Somogyi},\ and\ \citenamefont
  {Tramontano}}]{Ferrera:2017zex}%
  \BibitemOpen
  \bibfield  {author} {\bibinfo {author} {\bibfnamefont {G.}~\bibnamefont
  {Ferrera}}, \bibinfo {author} {\bibfnamefont {G.}~\bibnamefont {Somogyi}}, \
  and\ \bibinfo {author} {\bibfnamefont {F.}~\bibnamefont {Tramontano}},\
  }\href {\doibase 10.1016/j.physletb.2018.03.021} {\bibfield  {journal}
  {\bibinfo  {journal} {Phys. Lett.}\ }\textbf {\bibinfo {volume} {B780}},\
  \bibinfo {pages} {346} (\bibinfo {year} {2018})},\ \Eprint
  {http://arxiv.org/abs/1705.10304} {arXiv:1705.10304 [hep-ph]} \BibitemShut
  {NoStop}%
\bibitem [{\citenamefont {Caola}\ \emph {et~al.}(2018)\citenamefont {Caola},
  \citenamefont {Luisoni}, \citenamefont {Melnikov},\ and\ \citenamefont
  {R{\"o}ntsch}}]{Caola:2017xuq}%
  \BibitemOpen
  \bibfield  {author} {\bibinfo {author} {\bibfnamefont {F.}~\bibnamefont
  {Caola}}, \bibinfo {author} {\bibfnamefont {G.}~\bibnamefont {Luisoni}},
  \bibinfo {author} {\bibfnamefont {K.}~\bibnamefont {Melnikov}}, \ and\
  \bibinfo {author} {\bibfnamefont {R.}~\bibnamefont {R{\"o}ntsch}},\ }\href
  {\doibase 10.1103/PhysRevD.97.074022} {\bibfield  {journal} {\bibinfo
  {journal} {Phys. Rev.}\ }\textbf {\bibinfo {volume} {D97}},\ \bibinfo {pages}
  {074022} (\bibinfo {year} {2018})},\ \Eprint
  {http://arxiv.org/abs/1712.06954} {arXiv:1712.06954 [hep-ph]} \BibitemShut
  {NoStop}%
\bibitem [{\citenamefont {Gauld}\ \emph {et~al.}(2019)\citenamefont {Gauld},
  \citenamefont {Gehrmann-De~Ridder}, \citenamefont {Glover}, \citenamefont
  {Huss},\ and\ \citenamefont {Majer}}]{Gauld:2019yng}%
  \BibitemOpen
  \bibfield  {author} {\bibinfo {author} {\bibfnamefont {R.}~\bibnamefont
  {Gauld}}, \bibinfo {author} {\bibfnamefont {A.}~\bibnamefont
  {Gehrmann-De~Ridder}}, \bibinfo {author} {\bibfnamefont {E.~W.~N.}\
  \bibnamefont {Glover}}, \bibinfo {author} {\bibfnamefont {A.}~\bibnamefont
  {Huss}}, \ and\ \bibinfo {author} {\bibfnamefont {I.}~\bibnamefont {Majer}},\
  }\href@noop {} {\  (\bibinfo {year} {2019})},\ \Eprint
  {http://arxiv.org/abs/1907.05836} {arXiv:1907.05836 [hep-ph]} \BibitemShut
  {NoStop}%
\bibitem [{\citenamefont {Dawson}\ \emph {et~al.}(2012)\citenamefont {Dawson},
  \citenamefont {Han}, \citenamefont {Lai}, \citenamefont {Leibovich},\ and\
  \citenamefont {Lewis}}]{Dawson:2012gs}%
  \BibitemOpen
  \bibfield  {author} {\bibinfo {author} {\bibfnamefont {S.}~\bibnamefont
  {Dawson}}, \bibinfo {author} {\bibfnamefont {T.}~\bibnamefont {Han}},
  \bibinfo {author} {\bibfnamefont {W.~K.}\ \bibnamefont {Lai}}, \bibinfo
  {author} {\bibfnamefont {A.~K.}\ \bibnamefont {Leibovich}}, \ and\ \bibinfo
  {author} {\bibfnamefont {I.}~\bibnamefont {Lewis}},\ }\href {\doibase
  10.1103/PhysRevD.86.074007} {\bibfield  {journal} {\bibinfo  {journal} {Phys.
  Rev.}\ }\textbf {\bibinfo {volume} {D86}},\ \bibinfo {pages} {074007}
  (\bibinfo {year} {2012})},\ \Eprint {http://arxiv.org/abs/1207.4207}
  {arXiv:1207.4207 [hep-ph]} \BibitemShut {NoStop}%
\bibitem [{\citenamefont {Shao}\ \emph {et~al.}(2014)\citenamefont {Shao},
  \citenamefont {Li},\ and\ \citenamefont {Li}}]{Shao:2013uba}%
  \BibitemOpen
  \bibfield  {author} {\bibinfo {author} {\bibfnamefont {D.~Y.}\ \bibnamefont
  {Shao}}, \bibinfo {author} {\bibfnamefont {C.~S.}\ \bibnamefont {Li}}, \ and\
  \bibinfo {author} {\bibfnamefont {H.~T.}\ \bibnamefont {Li}},\ }\href
  {\doibase 10.1007/JHEP02(2014)117} {\bibfield  {journal} {\bibinfo  {journal}
  {JHEP}\ }\textbf {\bibinfo {volume} {02}},\ \bibinfo {pages} {117} (\bibinfo
  {year} {2014})},\ \Eprint {http://arxiv.org/abs/1309.5015} {arXiv:1309.5015
  [hep-ph]} \BibitemShut {NoStop}%
\bibitem [{\citenamefont {Li}\ and\ \citenamefont {Liu}(2014)}]{Li:2014ria}%
  \BibitemOpen
  \bibfield  {author} {\bibinfo {author} {\bibfnamefont {Y.}~\bibnamefont
  {Li}}\ and\ \bibinfo {author} {\bibfnamefont {X.}~\bibnamefont {Liu}},\
  }\href {\doibase 10.1007/JHEP06(2014)028} {\bibfield  {journal} {\bibinfo
  {journal} {JHEP}\ }\textbf {\bibinfo {volume} {06}},\ \bibinfo {pages} {028}
  (\bibinfo {year} {2014})},\ \Eprint {http://arxiv.org/abs/1401.2149}
  {arXiv:1401.2149 [hep-ph]} \BibitemShut {NoStop}%
\bibitem [{\citenamefont {Harlander}\ \emph {et~al.}(2014)\citenamefont
  {Harlander}, \citenamefont {Kulesza}, \citenamefont {Theeuwes},\ and\
  \citenamefont {Zirke}}]{Harlander:2014wda}%
  \BibitemOpen
  \bibfield  {author} {\bibinfo {author} {\bibfnamefont {R.~V.}\ \bibnamefont
  {Harlander}}, \bibinfo {author} {\bibfnamefont {A.}~\bibnamefont {Kulesza}},
  \bibinfo {author} {\bibfnamefont {V.}~\bibnamefont {Theeuwes}}, \ and\
  \bibinfo {author} {\bibfnamefont {T.}~\bibnamefont {Zirke}},\ }\href
  {\doibase 10.1007/JHEP11(2014)082} {\bibfield  {journal} {\bibinfo  {journal}
  {JHEP}\ }\textbf {\bibinfo {volume} {11}},\ \bibinfo {pages} {082} (\bibinfo
  {year} {2014})},\ \Eprint {http://arxiv.org/abs/1410.0217} {arXiv:1410.0217
  [hep-ph]} \BibitemShut {NoStop}%
\bibitem [{\citenamefont {Denner}\ \emph {et~al.}(2012)\citenamefont {Denner},
  \citenamefont {Dittmaier}, \citenamefont {Kallweit},\ and\ \citenamefont
  {M{\"u}ck}}]{Denner:2011id}%
  \BibitemOpen
  \bibfield  {author} {\bibinfo {author} {\bibfnamefont {A.}~\bibnamefont
  {Denner}}, \bibinfo {author} {\bibfnamefont {S.}~\bibnamefont {Dittmaier}},
  \bibinfo {author} {\bibfnamefont {S.}~\bibnamefont {Kallweit}}, \ and\
  \bibinfo {author} {\bibfnamefont {A.}~\bibnamefont {M{\"u}ck}},\ }\href
  {\doibase 10.1007/JHEP03(2012)075} {\bibfield  {journal} {\bibinfo  {journal}
  {JHEP}\ }\textbf {\bibinfo {volume} {03}},\ \bibinfo {pages} {075} (\bibinfo
  {year} {2012})},\ \Eprint {http://arxiv.org/abs/1112.5142} {arXiv:1112.5142
  [hep-ph]} \BibitemShut {NoStop}%
\bibitem [{\citenamefont {Denner}\ \emph {et~al.}(2015)\citenamefont {Denner},
  \citenamefont {Dittmaier}, \citenamefont {Kallweit},\ and\ \citenamefont
  {M{\"u}ck}}]{Denner:2014cla}%
  \BibitemOpen
  \bibfield  {author} {\bibinfo {author} {\bibfnamefont {A.}~\bibnamefont
  {Denner}}, \bibinfo {author} {\bibfnamefont {S.}~\bibnamefont {Dittmaier}},
  \bibinfo {author} {\bibfnamefont {S.}~\bibnamefont {Kallweit}}, \ and\
  \bibinfo {author} {\bibfnamefont {A.}~\bibnamefont {M{\"u}ck}},\ }\href
  {\doibase 10.1016/j.cpc.2015.04.021} {\bibfield  {journal} {\bibinfo
  {journal} {Comput. Phys. Commun.}\ }\textbf {\bibinfo {volume} {195}},\
  \bibinfo {pages} {161} (\bibinfo {year} {2015})},\ \Eprint
  {http://arxiv.org/abs/1412.5390} {arXiv:1412.5390 [hep-ph]} \BibitemShut
  {NoStop}%
\bibitem [{\citenamefont {Luisoni}\ \emph {et~al.}(2013)\citenamefont
  {Luisoni}, \citenamefont {Nason}, \citenamefont {Oleari},\ and\ \citenamefont
  {Tramontano}}]{Luisoni:2013kna}%
  \BibitemOpen
  \bibfield  {author} {\bibinfo {author} {\bibfnamefont {G.}~\bibnamefont
  {Luisoni}}, \bibinfo {author} {\bibfnamefont {P.}~\bibnamefont {Nason}},
  \bibinfo {author} {\bibfnamefont {C.}~\bibnamefont {Oleari}}, \ and\ \bibinfo
  {author} {\bibfnamefont {F.}~\bibnamefont {Tramontano}},\ }\href {\doibase
  10.1007/JHEP10(2013)083} {\bibfield  {journal} {\bibinfo  {journal} {JHEP}\
  }\textbf {\bibinfo {volume} {10}},\ \bibinfo {pages} {083} (\bibinfo {year}
  {2013})},\ \Eprint {http://arxiv.org/abs/1306.2542} {arXiv:1306.2542
  [hep-ph]} \BibitemShut {NoStop}%
\bibitem [{\citenamefont {Hamilton}\ \emph {et~al.}(2012)\citenamefont
  {Hamilton}, \citenamefont {Nason},\ and\ \citenamefont
  {Zanderighi}}]{Hamilton:2012np}%
  \BibitemOpen
  \bibfield  {author} {\bibinfo {author} {\bibfnamefont {K.}~\bibnamefont
  {Hamilton}}, \bibinfo {author} {\bibfnamefont {P.}~\bibnamefont {Nason}}, \
  and\ \bibinfo {author} {\bibfnamefont {G.}~\bibnamefont {Zanderighi}},\
  }\href {\doibase 10.1007/JHEP10(2012)155} {\bibfield  {journal} {\bibinfo
  {journal} {JHEP}\ }\textbf {\bibinfo {volume} {10}},\ \bibinfo {pages} {155}
  (\bibinfo {year} {2012})},\ \Eprint {http://arxiv.org/abs/1206.3572}
  {arXiv:1206.3572 [hep-ph]} \BibitemShut {NoStop}%
\bibitem [{\citenamefont {Hamilton}\ \emph {et~al.}(2013)\citenamefont
  {Hamilton}, \citenamefont {Nason}, \citenamefont {Oleari},\ and\
  \citenamefont {Zanderighi}}]{Hamilton:2012rf}%
  \BibitemOpen
  \bibfield  {author} {\bibinfo {author} {\bibfnamefont {K.}~\bibnamefont
  {Hamilton}}, \bibinfo {author} {\bibfnamefont {P.}~\bibnamefont {Nason}},
  \bibinfo {author} {\bibfnamefont {C.}~\bibnamefont {Oleari}}, \ and\ \bibinfo
  {author} {\bibfnamefont {G.}~\bibnamefont {Zanderighi}},\ }\href {\doibase
  10.1007/JHEP05(2013)082} {\bibfield  {journal} {\bibinfo  {journal} {JHEP}\
  }\textbf {\bibinfo {volume} {05}},\ \bibinfo {pages} {082} (\bibinfo {year}
  {2013})},\ \Eprint {http://arxiv.org/abs/1212.4504} {arXiv:1212.4504
  [hep-ph]} \BibitemShut {NoStop}%
\bibitem [{\citenamefont {{H\"oche, Stefan and Li, Ye and Prestel,
  Stefan}}(2015)}]{Hoeche:2014aia}%
  \BibitemOpen
  \bibfield  {author} {\bibinfo {author} {\bibnamefont {{H\"oche, Stefan and
  Li, Ye and Prestel, Stefan}}},\ }\href {\doibase 10.1103/PhysRevD.91.074015}
  {\bibfield  {journal} {\bibinfo  {journal} {Phys. Rev. D}\ }\textbf {\bibinfo
  {volume} {91}},\ \bibinfo {pages} {074015} (\bibinfo {year} {2015})},\
  \Eprint {http://arxiv.org/abs/1405.3607} {arXiv:1405.3607 [hep-ph]}
  \BibitemShut {NoStop}%
\bibitem [{\citenamefont {Alioli}\ \emph {et~al.}(2015)\citenamefont {Alioli},
  \citenamefont {Bauer}, \citenamefont {Berggren}, \citenamefont {Tackmann},\
  and\ \citenamefont {Walsh}}]{Alioli:2015toa}%
  \BibitemOpen
  \bibfield  {author} {\bibinfo {author} {\bibfnamefont {S.}~\bibnamefont
  {Alioli}}, \bibinfo {author} {\bibfnamefont {C.~W.}\ \bibnamefont {Bauer}},
  \bibinfo {author} {\bibfnamefont {C.}~\bibnamefont {Berggren}}, \bibinfo
  {author} {\bibfnamefont {F.~J.}\ \bibnamefont {Tackmann}}, \ and\ \bibinfo
  {author} {\bibfnamefont {J.~R.}\ \bibnamefont {Walsh}},\ }\href {\doibase
  10.1103/PhysRevD.92.094020} {\bibfield  {journal} {\bibinfo  {journal} {Phys.
  Rev.}\ }\textbf {\bibinfo {volume} {D92}},\ \bibinfo {pages} {094020}
  (\bibinfo {year} {2015})},\ \Eprint {http://arxiv.org/abs/1508.01475}
  {arXiv:1508.01475 [hep-ph]} \BibitemShut {NoStop}%
\bibitem [{\citenamefont {Monni}\ \emph {et~al.}(2019)\citenamefont {Monni},
  \citenamefont {Nason}, \citenamefont {Re}, \citenamefont {Wiesemann},\ and\
  \citenamefont {Zanderighi}}]{Monni:2019whf}%
  \BibitemOpen
  \bibfield  {author} {\bibinfo {author} {\bibfnamefont {P.~F.}\ \bibnamefont
  {Monni}}, \bibinfo {author} {\bibfnamefont {P.}~\bibnamefont {Nason}},
  \bibinfo {author} {\bibfnamefont {E.}~\bibnamefont {Re}}, \bibinfo {author}
  {\bibfnamefont {M.}~\bibnamefont {Wiesemann}}, \ and\ \bibinfo {author}
  {\bibfnamefont {G.}~\bibnamefont {Zanderighi}},\ }\href@noop {} {\  (\bibinfo
  {year} {2019})},\ \Eprint {http://arxiv.org/abs/1908.06987} {arXiv:1908.06987
  [hep-ph]} \BibitemShut {NoStop}%
\bibitem [{\citenamefont {Astill}\ \emph {et~al.}(2016)\citenamefont {Astill},
  \citenamefont {Bizon}, \citenamefont {Re},\ and\ \citenamefont
  {Zanderighi}}]{Astill:2016hpa}%
  \BibitemOpen
  \bibfield  {author} {\bibinfo {author} {\bibfnamefont {W.}~\bibnamefont
  {Astill}}, \bibinfo {author} {\bibfnamefont {W.}~\bibnamefont {Bizon}},
  \bibinfo {author} {\bibfnamefont {E.}~\bibnamefont {Re}}, \ and\ \bibinfo
  {author} {\bibfnamefont {G.}~\bibnamefont {Zanderighi}},\ }\href {\doibase
  10.1007/JHEP06(2016)154} {\bibfield  {journal} {\bibinfo  {journal} {JHEP}\
  }\textbf {\bibinfo {volume} {06}},\ \bibinfo {pages} {154} (\bibinfo {year}
  {2016})},\ \Eprint {http://arxiv.org/abs/1603.01620} {arXiv:1603.01620
  [hep-ph]} \BibitemShut {NoStop}%
\bibitem [{\citenamefont {Astill}\ \emph {et~al.}(2018)\citenamefont {Astill},
  \citenamefont {Bizon"}, \citenamefont {Re},\ and\ \citenamefont
  {Zanderighi}}]{Astill:2018ivh}%
  \BibitemOpen
  \bibfield  {author} {\bibinfo {author} {\bibfnamefont {W.}~\bibnamefont
  {Astill}}, \bibinfo {author} {\bibfnamefont {W.}~\bibnamefont {Bizon"}},
  \bibinfo {author} {\bibfnamefont {E.}~\bibnamefont {Re}}, \ and\ \bibinfo
  {author} {\bibfnamefont {G.}~\bibnamefont {Zanderighi}},\ }\href {\doibase
  10.1007/JHEP11(2018)157} {\bibfield  {journal} {\bibinfo  {journal} {JHEP}\
  }\textbf {\bibinfo {volume} {11}},\ \bibinfo {pages} {157} (\bibinfo {year}
  {2018})},\ \Eprint {http://arxiv.org/abs/1804.08141} {arXiv:1804.08141
  [hep-ph]} \BibitemShut {NoStop}%
\bibitem [{\citenamefont {Granata}\ \emph {et~al.}(2017)\citenamefont
  {Granata}, \citenamefont {Lindert}, \citenamefont {Oleari},\ and\
  \citenamefont {Pozzorini}}]{Granata:2017iod}%
  \BibitemOpen
  \bibfield  {author} {\bibinfo {author} {\bibfnamefont {F.}~\bibnamefont
  {Granata}}, \bibinfo {author} {\bibfnamefont {J.~M.}\ \bibnamefont
  {Lindert}}, \bibinfo {author} {\bibfnamefont {C.}~\bibnamefont {Oleari}}, \
  and\ \bibinfo {author} {\bibfnamefont {S.}~\bibnamefont {Pozzorini}},\ }\href
  {\doibase 10.1007/JHEP09(2017)012} {\bibfield  {journal} {\bibinfo  {journal}
  {JHEP}\ }\textbf {\bibinfo {volume} {09}},\ \bibinfo {pages} {012} (\bibinfo
  {year} {2017})},\ \Eprint {http://arxiv.org/abs/1706.03522} {arXiv:1706.03522
  [hep-ph]} \BibitemShut {NoStop}%
\bibitem [{\citenamefont {Alioli}\ \emph {et~al.}(2013)\citenamefont {Alioli},
  \citenamefont {Bauer}, \citenamefont {Berggren}, \citenamefont {Hornig},
  \citenamefont {Tackmann} \emph {et~al.}}]{Alioli:2012fc}%
  \BibitemOpen
  \bibfield  {author} {\bibinfo {author} {\bibfnamefont {S.}~\bibnamefont
  {Alioli}}, \bibinfo {author} {\bibfnamefont {C.~W.}\ \bibnamefont {Bauer}},
  \bibinfo {author} {\bibfnamefont {C.}~\bibnamefont {Berggren}}, \bibinfo
  {author} {\bibfnamefont {A.}~\bibnamefont {Hornig}}, \bibinfo {author}
  {\bibfnamefont {F.~J.}\ \bibnamefont {Tackmann}},  \emph {et~al.},\ }\href
  {\doibase 10.1007/JHEP09(2013)120} {\bibfield  {journal} {\bibinfo  {journal}
  {JHEP}\ }\textbf {\bibinfo {volume} {1309}},\ \bibinfo {pages} {120}
  (\bibinfo {year} {2013})},\ \Eprint {http://arxiv.org/abs/1211.7049}
  {arXiv:1211.7049 [hep-ph]} \BibitemShut {NoStop}%
\bibitem [{\citenamefont {Alioli}\ \emph {et~al.}(2014)\citenamefont {Alioli},
  \citenamefont {Bauer}, \citenamefont {Berggren}, \citenamefont {Tackmann},
  \citenamefont {Walsh} \emph {et~al.}}]{Alioli:2013hqa}%
  \BibitemOpen
  \bibfield  {author} {\bibinfo {author} {\bibfnamefont {S.}~\bibnamefont
  {Alioli}}, \bibinfo {author} {\bibfnamefont {C.~W.}\ \bibnamefont {Bauer}},
  \bibinfo {author} {\bibfnamefont {C.}~\bibnamefont {Berggren}}, \bibinfo
  {author} {\bibfnamefont {F.~J.}\ \bibnamefont {Tackmann}}, \bibinfo {author}
  {\bibfnamefont {J.~R.}\ \bibnamefont {Walsh}},  \emph {et~al.},\ }\href
  {\doibase 10.1007/JHEP06(2014)089} {\bibfield  {journal} {\bibinfo  {journal}
  {JHEP}\ }\textbf {\bibinfo {volume} {1406}},\ \bibinfo {pages} {089}
  (\bibinfo {year} {2014})},\ \Eprint {http://arxiv.org/abs/1311.0286}
  {arXiv:1311.0286 [hep-ph]} \BibitemShut {NoStop}%
\bibitem [{\citenamefont {Alioli}\ \emph {et~al.}(2016)\citenamefont {Alioli},
  \citenamefont {Bauer}, \citenamefont {Guns},\ and\ \citenamefont
  {Tackmann}}]{Alioli:2016wqt}%
  \BibitemOpen
  \bibfield  {author} {\bibinfo {author} {\bibfnamefont {S.}~\bibnamefont
  {Alioli}}, \bibinfo {author} {\bibfnamefont {C.~W.}\ \bibnamefont {Bauer}},
  \bibinfo {author} {\bibfnamefont {S.}~\bibnamefont {Guns}}, \ and\ \bibinfo
  {author} {\bibfnamefont {F.~J.}\ \bibnamefont {Tackmann}},\ }\href {\doibase
  10.1140/epjc/s10052-016-4458-1} {\bibfield  {journal} {\bibinfo  {journal}
  {Eur. Phys. J.}\ }\textbf {\bibinfo {volume} {C76}},\ \bibinfo {pages} {614}
  (\bibinfo {year} {2016})},\ \Eprint {http://arxiv.org/abs/1605.07192}
  {arXiv:1605.07192 [hep-ph]} \BibitemShut {NoStop}%
\bibitem [{\citenamefont {Frixione}\ \emph {et~al.}(1996)\citenamefont
  {Frixione}, \citenamefont {Kunszt},\ and\ \citenamefont
  {Signer}}]{Frixione:1995ms}%
  \BibitemOpen
  \bibfield  {author} {\bibinfo {author} {\bibfnamefont {S.}~\bibnamefont
  {Frixione}}, \bibinfo {author} {\bibfnamefont {Z.}~\bibnamefont {Kunszt}}, \
  and\ \bibinfo {author} {\bibfnamefont {A.}~\bibnamefont {Signer}},\ }\href
  {\doibase 10.1016/0550-3213(96)00110-1} {\bibfield  {journal} {\bibinfo
  {journal} {Nucl.Phys.}\ }\textbf {\bibinfo {volume} {B467}},\ \bibinfo
  {pages} {399} (\bibinfo {year} {1996})},\ \Eprint
  {http://arxiv.org/abs/hep-ph/9512328} {hep-ph/9512328} \BibitemShut {NoStop}%
\bibitem [{\citenamefont {Stewart}\ \emph
  {et~al.}(2010{\natexlab{a}})\citenamefont {Stewart}, \citenamefont
  {Tackmann},\ and\ \citenamefont {Waalewijn}}]{Stewart:2010tn}%
  \BibitemOpen
  \bibfield  {author} {\bibinfo {author} {\bibfnamefont {I.~W.}\ \bibnamefont
  {Stewart}}, \bibinfo {author} {\bibfnamefont {F.~J.}\ \bibnamefont
  {Tackmann}}, \ and\ \bibinfo {author} {\bibfnamefont {W.~J.}\ \bibnamefont
  {Waalewijn}},\ }\href {\doibase 10.1103/PhysRevLett.105.092002} {\bibfield
  {journal} {\bibinfo  {journal} {Phys. Rev. Lett.}\ }\textbf {\bibinfo
  {volume} {105}},\ \bibinfo {pages} {092002} (\bibinfo {year}
  {2010}{\natexlab{a}})},\ \Eprint {http://arxiv.org/abs/1004.2489}
  {arXiv:1004.2489 [hep-ph]} \BibitemShut {NoStop}%
\bibitem [{\citenamefont {Pietrulewicz}\ \emph {et~al.}(2016)\citenamefont
  {Pietrulewicz}, \citenamefont {Tackmann},\ and\ \citenamefont
  {Waalewijn}}]{Pietrulewicz:2016nwo}%
  \BibitemOpen
  \bibfield  {author} {\bibinfo {author} {\bibfnamefont {P.}~\bibnamefont
  {Pietrulewicz}}, \bibinfo {author} {\bibfnamefont {F.~J.}\ \bibnamefont
  {Tackmann}}, \ and\ \bibinfo {author} {\bibfnamefont {W.~J.}\ \bibnamefont
  {Waalewijn}},\ }\href {\doibase 10.1007/JHEP08(2016)002} {\bibfield
  {journal} {\bibinfo  {journal} {JHEP}\ }\textbf {\bibinfo {volume} {08}},\
  \bibinfo {pages} {002} (\bibinfo {year} {2016})},\ \Eprint
  {http://arxiv.org/abs/1601.05088} {arXiv:1601.05088 [hep-ph]} \BibitemShut
  {NoStop}%
\bibitem [{\citenamefont {Stewart}\ \emph
  {et~al.}(2010{\natexlab{b}})\citenamefont {Stewart}, \citenamefont
  {Tackmann},\ and\ \citenamefont {Waalewijn}}]{Stewart:2009yx}%
  \BibitemOpen
  \bibfield  {author} {\bibinfo {author} {\bibfnamefont {I.~W.}\ \bibnamefont
  {Stewart}}, \bibinfo {author} {\bibfnamefont {F.~J.}\ \bibnamefont
  {Tackmann}}, \ and\ \bibinfo {author} {\bibfnamefont {W.~J.}\ \bibnamefont
  {Waalewijn}},\ }\href {\doibase 10.1103/PhysRevD.81.094035} {\bibfield
  {journal} {\bibinfo  {journal} {Phys. Rev. D}\ }\textbf {\bibinfo {volume}
  {81}},\ \bibinfo {pages} {094035} (\bibinfo {year} {2010}{\natexlab{b}})},\
  \Eprint {http://arxiv.org/abs/0910.0467} {arXiv:0910.0467 [hep-ph]}
  \BibitemShut {NoStop}%
\bibitem [{\citenamefont {Stewart}\ \emph {et~al.}(2011)\citenamefont
  {Stewart}, \citenamefont {Tackmann},\ and\ \citenamefont
  {Waalewijn}}]{Stewart:2010pd}%
  \BibitemOpen
  \bibfield  {author} {\bibinfo {author} {\bibfnamefont {I.~W.}\ \bibnamefont
  {Stewart}}, \bibinfo {author} {\bibfnamefont {F.~J.}\ \bibnamefont
  {Tackmann}}, \ and\ \bibinfo {author} {\bibfnamefont {W.~J.}\ \bibnamefont
  {Waalewijn}},\ }\href {\doibase 10.1103/PhysRevLett.106.032001} {\bibfield
  {journal} {\bibinfo  {journal} {Phys. Rev. Lett.}\ }\textbf {\bibinfo
  {volume} {106}},\ \bibinfo {pages} {032001} (\bibinfo {year} {2011})},\
  \Eprint {http://arxiv.org/abs/1005.4060} {arXiv:1005.4060 [hep-ph]}
  \BibitemShut {NoStop}%
\bibitem [{\citenamefont {Berger}\ \emph {et~al.}(2011)\citenamefont {Berger},
  \citenamefont {Marcantonini}, \citenamefont {Stewart}, \citenamefont
  {Tackmann},\ and\ \citenamefont {Waalewijn}}]{Berger:2010xi}%
  \BibitemOpen
  \bibfield  {author} {\bibinfo {author} {\bibfnamefont {C.~F.}\ \bibnamefont
  {Berger}}, \bibinfo {author} {\bibfnamefont {C.}~\bibnamefont
  {Marcantonini}}, \bibinfo {author} {\bibfnamefont {I.~W.}\ \bibnamefont
  {Stewart}}, \bibinfo {author} {\bibfnamefont {F.~J.}\ \bibnamefont
  {Tackmann}}, \ and\ \bibinfo {author} {\bibfnamefont {W.~J.}\ \bibnamefont
  {Waalewijn}},\ }\href {\doibase 10.1007/JHEP04(2011)092} {\bibfield
  {journal} {\bibinfo  {journal} {JHEP}\ }\textbf {\bibinfo {volume} {1104}},\
  \bibinfo {pages} {092} (\bibinfo {year} {2011})},\ \Eprint
  {http://arxiv.org/abs/1012.4480} {arXiv:1012.4480 [hep-ph]} \BibitemShut
  {NoStop}%
\bibitem [{\citenamefont {Bauer}\ \emph {et~al.}(2000)\citenamefont {Bauer},
  \citenamefont {Fleming},\ and\ \citenamefont {Luke}}]{Bauer:2000ew}%
  \BibitemOpen
  \bibfield  {author} {\bibinfo {author} {\bibfnamefont {C.~W.}\ \bibnamefont
  {Bauer}}, \bibinfo {author} {\bibfnamefont {S.}~\bibnamefont {Fleming}}, \
  and\ \bibinfo {author} {\bibfnamefont {M.~E.}\ \bibnamefont {Luke}},\ }\href
  {\doibase 10.1103/PhysRevD.63.014006} {\bibfield  {journal} {\bibinfo
  {journal} {Phys. Rev. D}\ }\textbf {\bibinfo {volume} {63}},\ \bibinfo
  {pages} {014006} (\bibinfo {year} {2000})},\ \Eprint
  {http://arxiv.org/abs/hep-ph/0005275} {hep-ph/0005275} \BibitemShut {NoStop}%
\bibitem [{\citenamefont {Bauer}\ \emph {et~al.}(2001)\citenamefont {Bauer},
  \citenamefont {Fleming}, \citenamefont {Pirjol},\ and\ \citenamefont
  {Stewart}}]{Bauer:2000yr}%
  \BibitemOpen
  \bibfield  {author} {\bibinfo {author} {\bibfnamefont {C.~W.}\ \bibnamefont
  {Bauer}}, \bibinfo {author} {\bibfnamefont {S.}~\bibnamefont {Fleming}},
  \bibinfo {author} {\bibfnamefont {D.}~\bibnamefont {Pirjol}}, \ and\ \bibinfo
  {author} {\bibfnamefont {I.~W.}\ \bibnamefont {Stewart}},\ }\href {\doibase
  10.1103/PhysRevD.63.114020} {\bibfield  {journal} {\bibinfo  {journal} {Phys.
  Rev. D}\ }\textbf {\bibinfo {volume} {63}},\ \bibinfo {pages} {114020}
  (\bibinfo {year} {2001})},\ \Eprint {http://arxiv.org/abs/hep-ph/0011336}
  {hep-ph/0011336} \BibitemShut {NoStop}%
\bibitem [{\citenamefont {Bauer}\ and\ \citenamefont
  {Stewart}(2001)}]{Bauer:2001ct}%
  \BibitemOpen
  \bibfield  {author} {\bibinfo {author} {\bibfnamefont {C.~W.}\ \bibnamefont
  {Bauer}}\ and\ \bibinfo {author} {\bibfnamefont {I.~W.}\ \bibnamefont
  {Stewart}},\ }\href {\doibase 10.1016/S0370-2693(01)00902-9} {\bibfield
  {journal} {\bibinfo  {journal} {Phys. Lett. B}\ }\textbf {\bibinfo {volume}
  {516}},\ \bibinfo {pages} {134} (\bibinfo {year} {2001})},\ \Eprint
  {http://arxiv.org/abs/hep-ph/0107001} {hep-ph/0107001} \BibitemShut {NoStop}%
\bibitem [{\citenamefont {Bauer}\ \emph
  {et~al.}(2002{\natexlab{a}})\citenamefont {Bauer}, \citenamefont {Pirjol},\
  and\ \citenamefont {Stewart}}]{Bauer:2001yt}%
  \BibitemOpen
  \bibfield  {author} {\bibinfo {author} {\bibfnamefont {C.~W.}\ \bibnamefont
  {Bauer}}, \bibinfo {author} {\bibfnamefont {D.}~\bibnamefont {Pirjol}}, \
  and\ \bibinfo {author} {\bibfnamefont {I.~W.}\ \bibnamefont {Stewart}},\
  }\href {\doibase 10.1103/PhysRevD.65.054022} {\bibfield  {journal} {\bibinfo
  {journal} {Phys. Rev. D}\ }\textbf {\bibinfo {volume} {65}},\ \bibinfo
  {pages} {054022} (\bibinfo {year} {2002}{\natexlab{a}})},\ \Eprint
  {http://arxiv.org/abs/hep-ph/0109045} {hep-ph/0109045} \BibitemShut {NoStop}%
\bibitem [{\citenamefont {Bauer}\ \emph
  {et~al.}(2002{\natexlab{b}})\citenamefont {Bauer}, \citenamefont {Fleming},
  \citenamefont {Pirjol}, \citenamefont {Rothstein},\ and\ \citenamefont
  {Stewart}}]{Bauer:2002nz}%
  \BibitemOpen
  \bibfield  {author} {\bibinfo {author} {\bibfnamefont {C.~W.}\ \bibnamefont
  {Bauer}}, \bibinfo {author} {\bibfnamefont {S.}~\bibnamefont {Fleming}},
  \bibinfo {author} {\bibfnamefont {D.}~\bibnamefont {Pirjol}}, \bibinfo
  {author} {\bibfnamefont {I.~Z.}\ \bibnamefont {Rothstein}}, \ and\ \bibinfo
  {author} {\bibfnamefont {I.~W.}\ \bibnamefont {Stewart}},\ }\href {\doibase
  10.1103/PhysRevD.66.014017} {\bibfield  {journal} {\bibinfo  {journal} {Phys.
  Rev. D}\ }\textbf {\bibinfo {volume} {66}},\ \bibinfo {pages} {014017}
  (\bibinfo {year} {2002}{\natexlab{b}})},\ \Eprint
  {http://arxiv.org/abs/hep-ph/0202088} {hep-ph/0202088} \BibitemShut {NoStop}%
\bibitem [{\citenamefont {Beneke}\ and\ \citenamefont
  {Feldmann}(2003)}]{Beneke:2002ni}%
  \BibitemOpen
  \bibfield  {author} {\bibinfo {author} {\bibfnamefont {M.}~\bibnamefont
  {Beneke}}\ and\ \bibinfo {author} {\bibfnamefont {T.}~\bibnamefont
  {Feldmann}},\ }\href {\doibase 10.1016/S0370-2693(02)03204-5} {\bibfield
  {journal} {\bibinfo  {journal} {Phys. Lett.}\ }\textbf {\bibinfo {volume}
  {B553}},\ \bibinfo {pages} {267} (\bibinfo {year} {2003})},\ \Eprint
  {http://arxiv.org/abs/hep-ph/0211358} {arXiv:hep-ph/0211358 [hep-ph]}
  \BibitemShut {NoStop}%
\bibitem [{\citenamefont {Beneke}\ \emph {et~al.}(2002)\citenamefont {Beneke},
  \citenamefont {Chapovsky}, \citenamefont {Diehl},\ and\ \citenamefont
  {Feldmann}}]{Beneke:2002ph}%
  \BibitemOpen
  \bibfield  {author} {\bibinfo {author} {\bibfnamefont {M.}~\bibnamefont
  {Beneke}}, \bibinfo {author} {\bibfnamefont {A.~P.}\ \bibnamefont
  {Chapovsky}}, \bibinfo {author} {\bibfnamefont {M.}~\bibnamefont {Diehl}}, \
  and\ \bibinfo {author} {\bibfnamefont {T.}~\bibnamefont {Feldmann}},\ }\href
  {\doibase 10.1016/S0550-3213(02)00687-9} {\bibfield  {journal} {\bibinfo
  {journal} {Nucl. Phys.}\ }\textbf {\bibinfo {volume} {B643}},\ \bibinfo
  {pages} {431} (\bibinfo {year} {2002})},\ \Eprint
  {http://arxiv.org/abs/hep-ph/0206152} {arXiv:hep-ph/0206152 [hep-ph]}
  \BibitemShut {NoStop}%
\bibitem [{\citenamefont {Idilbi}\ \emph {et~al.}(2006)\citenamefont {Idilbi},
  \citenamefont {Ji},\ and\ \citenamefont {Yuan}}]{Idilbi:2006dg}%
  \BibitemOpen
  \bibfield  {author} {\bibinfo {author} {\bibfnamefont {A.}~\bibnamefont
  {Idilbi}}, \bibinfo {author} {\bibfnamefont {X.-d.}\ \bibnamefont {Ji}}, \
  and\ \bibinfo {author} {\bibfnamefont {F.}~\bibnamefont {Yuan}},\ }\href
  {\doibase 10.1016/j.nuclphysb.2006.07.002} {\bibfield  {journal} {\bibinfo
  {journal} {Nucl. Phys.}\ }\textbf {\bibinfo {volume} {B753}},\ \bibinfo
  {pages} {42} (\bibinfo {year} {2006})},\ \Eprint
  {http://arxiv.org/abs/hep-ph/0605068} {hep-ph/0605068} \BibitemShut {NoStop}%
\bibitem [{\citenamefont {Becher}\ \emph {et~al.}(2007)\citenamefont {Becher},
  \citenamefont {Neubert},\ and\ \citenamefont {Pecjak}}]{Becher:2006mr}%
  \BibitemOpen
  \bibfield  {author} {\bibinfo {author} {\bibfnamefont {T.}~\bibnamefont
  {Becher}}, \bibinfo {author} {\bibfnamefont {M.}~\bibnamefont {Neubert}}, \
  and\ \bibinfo {author} {\bibfnamefont {B.~D.}\ \bibnamefont {Pecjak}},\
  }\href {\doibase 10.1088/1126-6708/2007/01/076} {\bibfield  {journal}
  {\bibinfo  {journal} {JHEP}\ }\textbf {\bibinfo {volume} {01}},\ \bibinfo
  {pages} {076} (\bibinfo {year} {2007})},\ \Eprint
  {http://arxiv.org/abs/hep-ph/0607228} {hep-ph/0607228} \BibitemShut {NoStop}%
\bibitem [{\citenamefont {Stewart}\ \emph
  {et~al.}(2010{\natexlab{c}})\citenamefont {Stewart}, \citenamefont
  {Tackmann},\ and\ \citenamefont {Waalewijn}}]{Stewart:2010qs}%
  \BibitemOpen
  \bibfield  {author} {\bibinfo {author} {\bibfnamefont {I.~W.}\ \bibnamefont
  {Stewart}}, \bibinfo {author} {\bibfnamefont {F.~J.}\ \bibnamefont
  {Tackmann}}, \ and\ \bibinfo {author} {\bibfnamefont {W.~J.}\ \bibnamefont
  {Waalewijn}},\ }\href {\doibase 10.1007/JHEP09(2010)005} {\bibfield
  {journal} {\bibinfo  {journal} {JHEP}\ }\textbf {\bibinfo {volume} {1009}},\
  \bibinfo {pages} {005} (\bibinfo {year} {2010}{\natexlab{c}})},\ \Eprint
  {http://arxiv.org/abs/1002.2213} {arXiv:1002.2213 [hep-ph]} \BibitemShut
  {NoStop}%
\bibitem [{\citenamefont {Monni}\ \emph {et~al.}(2011)\citenamefont {Monni},
  \citenamefont {Gehrmann},\ and\ \citenamefont {Luisoni}}]{Monni:2011gb}%
  \BibitemOpen
  \bibfield  {author} {\bibinfo {author} {\bibfnamefont {P.~F.}\ \bibnamefont
  {Monni}}, \bibinfo {author} {\bibfnamefont {T.}~\bibnamefont {Gehrmann}}, \
  and\ \bibinfo {author} {\bibfnamefont {G.}~\bibnamefont {Luisoni}},\ }\href
  {\doibase 10.1007/JHEP08(2011)010} {\bibfield  {journal} {\bibinfo  {journal}
  {JHEP}\ }\textbf {\bibinfo {volume} {08}},\ \bibinfo {pages} {010} (\bibinfo
  {year} {2011})},\ \Eprint {http://arxiv.org/abs/1105.4560} {arXiv:1105.4560
  [hep-ph]} \BibitemShut {NoStop}%
\bibitem [{\citenamefont {Kelley}\ \emph {et~al.}(2011)\citenamefont {Kelley},
  \citenamefont {Schwartz}, \citenamefont {Schabinger},\ and\ \citenamefont
  {Zhu}}]{Kelley:2011ng}%
  \BibitemOpen
  \bibfield  {author} {\bibinfo {author} {\bibfnamefont {R.}~\bibnamefont
  {Kelley}}, \bibinfo {author} {\bibfnamefont {M.~D.}\ \bibnamefont
  {Schwartz}}, \bibinfo {author} {\bibfnamefont {R.~M.}\ \bibnamefont
  {Schabinger}}, \ and\ \bibinfo {author} {\bibfnamefont {H.~X.}\ \bibnamefont
  {Zhu}},\ }\href {\doibase 10.1103/PhysRevD.84.045022} {\bibfield  {journal}
  {\bibinfo  {journal} {Phys. Rev.}\ }\textbf {\bibinfo {volume} {D84}},\
  \bibinfo {pages} {045022} (\bibinfo {year} {2011})},\ \Eprint
  {http://arxiv.org/abs/1105.3676} {arXiv:1105.3676 [hep-ph]} \BibitemShut
  {NoStop}%
\bibitem [{\citenamefont {Hornig}\ \emph {et~al.}(2011)\citenamefont {Hornig},
  \citenamefont {Lee}, \citenamefont {Stewart}, \citenamefont {Walsh},\ and\
  \citenamefont {Zuberi}}]{Hornig:2011iu}%
  \BibitemOpen
  \bibfield  {author} {\bibinfo {author} {\bibfnamefont {A.}~\bibnamefont
  {Hornig}}, \bibinfo {author} {\bibfnamefont {C.}~\bibnamefont {Lee}},
  \bibinfo {author} {\bibfnamefont {I.~W.}\ \bibnamefont {Stewart}}, \bibinfo
  {author} {\bibfnamefont {J.~R.}\ \bibnamefont {Walsh}}, \ and\ \bibinfo
  {author} {\bibfnamefont {S.}~\bibnamefont {Zuberi}},\ }\href {\doibase
  10.1007/JHEP08(2011)054} {\bibfield  {journal} {\bibinfo  {journal} {JHEP}\
  }\textbf {\bibinfo {volume} {08}},\ \bibinfo {pages} {054} (\bibinfo {year}
  {2011})},\ \Eprint {http://arxiv.org/abs/1105.4628} {arXiv:1105.4628
  [hep-ph]} \BibitemShut {NoStop}%
\bibitem [{\citenamefont {Gaunt}\ \emph
  {et~al.}(2014{\natexlab{a}})\citenamefont {Gaunt}, \citenamefont
  {Stahlhofen},\ and\ \citenamefont {Tackmann}}]{Gaunt:2014xga}%
  \BibitemOpen
  \bibfield  {author} {\bibinfo {author} {\bibfnamefont {J.~R.}\ \bibnamefont
  {Gaunt}}, \bibinfo {author} {\bibfnamefont {M.}~\bibnamefont {Stahlhofen}}, \
  and\ \bibinfo {author} {\bibfnamefont {F.~J.}\ \bibnamefont {Tackmann}},\
  }\href {\doibase 10.1007/JHEP04(2014)113} {\bibfield  {journal} {\bibinfo
  {journal} {JHEP}\ }\textbf {\bibinfo {volume} {04}},\ \bibinfo {pages} {113}
  (\bibinfo {year} {2014}{\natexlab{a}})},\ \Eprint
  {http://arxiv.org/abs/1401.5478} {arXiv:1401.5478 [hep-ph]} \BibitemShut
  {NoStop}%
\bibitem [{\citenamefont {Gaunt}\ \emph
  {et~al.}(2014{\natexlab{b}})\citenamefont {Gaunt}, \citenamefont
  {Stahlhofen},\ and\ \citenamefont {Tackmann}}]{Gaunt:2014cfa}%
  \BibitemOpen
  \bibfield  {author} {\bibinfo {author} {\bibfnamefont {J.}~\bibnamefont
  {Gaunt}}, \bibinfo {author} {\bibfnamefont {M.}~\bibnamefont {Stahlhofen}}, \
  and\ \bibinfo {author} {\bibfnamefont {F.~J.}\ \bibnamefont {Tackmann}},\
  }\href {\doibase 10.1007/JHEP08(2014)020} {\bibfield  {journal} {\bibinfo
  {journal} {JHEP}\ }\textbf {\bibinfo {volume} {08}},\ \bibinfo {pages} {020}
  (\bibinfo {year} {2014}{\natexlab{b}})},\ \Eprint
  {http://arxiv.org/abs/1405.1044} {arXiv:1405.1044 [hep-ph]} \BibitemShut
  {NoStop}%
\bibitem [{\citenamefont {Kang}\ \emph {et~al.}(2015)\citenamefont {Kang},
  \citenamefont {Labun},\ and\ \citenamefont {Lee}}]{Kang:2015moa}%
  \BibitemOpen
  \bibfield  {author} {\bibinfo {author} {\bibfnamefont {D.}~\bibnamefont
  {Kang}}, \bibinfo {author} {\bibfnamefont {O.~Z.}\ \bibnamefont {Labun}}, \
  and\ \bibinfo {author} {\bibfnamefont {C.}~\bibnamefont {Lee}},\ }\href
  {\doibase 10.1016/j.physletb.2015.06.057} {\bibfield  {journal} {\bibinfo
  {journal} {Phys. Lett. B}\ }\textbf {\bibinfo {volume} {748}},\ \bibinfo
  {pages} {45} (\bibinfo {year} {2015})},\ \Eprint
  {http://arxiv.org/abs/1504.04006} {arXiv:1504.04006 [hep-ph]} \BibitemShut
  {NoStop}%
\bibitem [{\citenamefont {Gaunt}\ \emph {et~al.}(2015)\citenamefont {Gaunt},
  \citenamefont {Stahlhofen}, \citenamefont {Tackmann},\ and\ \citenamefont
  {Walsh}}]{Gaunt:2015pea}%
  \BibitemOpen
  \bibfield  {author} {\bibinfo {author} {\bibfnamefont {J.}~\bibnamefont
  {Gaunt}}, \bibinfo {author} {\bibfnamefont {M.}~\bibnamefont {Stahlhofen}},
  \bibinfo {author} {\bibfnamefont {F.~J.}\ \bibnamefont {Tackmann}}, \ and\
  \bibinfo {author} {\bibfnamefont {J.~R.}\ \bibnamefont {Walsh}},\ }\href
  {\doibase 10.1007/JHEP09(2015)058} {\bibfield  {journal} {\bibinfo  {journal}
  {JHEP}\ }\textbf {\bibinfo {volume} {09}},\ \bibinfo {pages} {058} (\bibinfo
  {year} {2015})},\ \Eprint {http://arxiv.org/abs/1505.04794} {arXiv:1505.04794
  [hep-ph]} \BibitemShut {NoStop}%
\bibitem [{\citenamefont {Ligeti}\ \emph {et~al.}(2008)\citenamefont {Ligeti},
  \citenamefont {Stewart},\ and\ \citenamefont {Tackmann}}]{Ligeti:2008ac}%
  \BibitemOpen
  \bibfield  {author} {\bibinfo {author} {\bibfnamefont {Z.}~\bibnamefont
  {Ligeti}}, \bibinfo {author} {\bibfnamefont {I.~W.}\ \bibnamefont {Stewart}},
  \ and\ \bibinfo {author} {\bibfnamefont {F.~J.}\ \bibnamefont {Tackmann}},\
  }\href {\doibase 10.1103/PhysRevD.78.114014} {\bibfield  {journal} {\bibinfo
  {journal} {Phys. Rev. D}\ }\textbf {\bibinfo {volume} {78}},\ \bibinfo
  {pages} {114014} (\bibinfo {year} {2008})},\ \Eprint
  {http://arxiv.org/abs/0807.1926} {arXiv:0807.1926 [hep-ph]} \BibitemShut
  {NoStop}%
\bibitem [{\citenamefont {Abbate}\ \emph {et~al.}(2011)\citenamefont {Abbate},
  \citenamefont {Fickinger}, \citenamefont {Hoang}, \citenamefont {Mateu},\
  and\ \citenamefont {Stewart}}]{Abbate:2010xh}%
  \BibitemOpen
  \bibfield  {author} {\bibinfo {author} {\bibfnamefont {R.}~\bibnamefont
  {Abbate}}, \bibinfo {author} {\bibfnamefont {M.}~\bibnamefont {Fickinger}},
  \bibinfo {author} {\bibfnamefont {A.~H.}\ \bibnamefont {Hoang}}, \bibinfo
  {author} {\bibfnamefont {V.}~\bibnamefont {Mateu}}, \ and\ \bibinfo {author}
  {\bibfnamefont {I.~W.}\ \bibnamefont {Stewart}},\ }\href {\doibase
  10.1103/PhysRevD.83.074021} {\bibfield  {journal} {\bibinfo  {journal} {Phys.
  Rev. D}\ }\textbf {\bibinfo {volume} {83}},\ \bibinfo {pages} {074021}
  (\bibinfo {year} {2011})},\ \Eprint {http://arxiv.org/abs/1006.3080}
  {arXiv:1006.3080 [hep-ph]} \BibitemShut {NoStop}%
\bibitem [{\citenamefont {Stewart}\ \emph {et~al.}(2014)\citenamefont
  {Stewart}, \citenamefont {Tackmann}, \citenamefont {Walsh},\ and\
  \citenamefont {Zuberi}}]{Stewart:2013faa}%
  \BibitemOpen
  \bibfield  {author} {\bibinfo {author} {\bibfnamefont {I.~W.}\ \bibnamefont
  {Stewart}}, \bibinfo {author} {\bibfnamefont {F.~J.}\ \bibnamefont
  {Tackmann}}, \bibinfo {author} {\bibfnamefont {J.~R.}\ \bibnamefont {Walsh}},
  \ and\ \bibinfo {author} {\bibfnamefont {S.}~\bibnamefont {Zuberi}},\ }\href
  {\doibase 10.1103/PhysRevD.89.054001} {\bibfield  {journal} {\bibinfo
  {journal} {Phys. Rev. D}\ }\textbf {\bibinfo {volume} {89}},\ \bibinfo
  {pages} {054001} (\bibinfo {year} {2014})},\ \Eprint
  {http://arxiv.org/abs/1307.1808} {arXiv:1307.1808} \BibitemShut {NoStop}%
\bibitem [{\citenamefont {Gangal}\ \emph {et~al.}(2015)\citenamefont {Gangal},
  \citenamefont {Stahlhofen},\ and\ \citenamefont {Tackmann}}]{Gangal:2014qda}%
  \BibitemOpen
  \bibfield  {author} {\bibinfo {author} {\bibfnamefont {S.}~\bibnamefont
  {Gangal}}, \bibinfo {author} {\bibfnamefont {M.}~\bibnamefont {Stahlhofen}},
  \ and\ \bibinfo {author} {\bibfnamefont {F.~J.}\ \bibnamefont {Tackmann}},\
  }\href {\doibase 10.1103/PhysRevD.91.054023} {\bibfield  {journal} {\bibinfo
  {journal} {Phys. Rev. D}\ }\textbf {\bibinfo {volume} {91}},\ \bibinfo
  {pages} {054023} (\bibinfo {year} {2015})},\ \Eprint
  {http://arxiv.org/abs/1412.4792} {arXiv:1412.4792 [hep-ph]} \BibitemShut
  {NoStop}%
\bibitem [{\citenamefont {Catani}\ and\ \citenamefont
  {Seymour}(1996)}]{Catani:1996jh}%
  \BibitemOpen
  \bibfield  {author} {\bibinfo {author} {\bibfnamefont {S.}~\bibnamefont
  {Catani}}\ and\ \bibinfo {author} {\bibfnamefont {M.~H.}\ \bibnamefont
  {Seymour}},\ }\href {\doibase 10.1016/0370-2693(96)00425-X} {\bibfield
  {journal} {\bibinfo  {journal} {Phys. Lett. B}\ }\textbf {\bibinfo {volume}
  {378}},\ \bibinfo {pages} {287} (\bibinfo {year} {1996})},\ \Eprint
  {http://arxiv.org/abs/hep-ph/9602277} {hep-ph/9602277} \BibitemShut {NoStop}%
\bibitem [{\citenamefont {Catani}\ and\ \citenamefont
  {Seymour}(1997)}]{Catani:1996vz}%
  \BibitemOpen
  \bibfield  {author} {\bibinfo {author} {\bibfnamefont {S.}~\bibnamefont
  {Catani}}\ and\ \bibinfo {author} {\bibfnamefont {M.~H.}\ \bibnamefont
  {Seymour}},\ }\href {\doibase 10.1016/S0550-3213(96)00589-5} {\bibfield
  {journal} {\bibinfo  {journal} {Nucl. Phys. B}\ }\textbf {\bibinfo {volume}
  {485}},\ \bibinfo {pages} {291} (\bibinfo {year} {1997})},\ \bibinfo {note}
  {[Erratum-ibid. {\bf B510}, 503 (1998)]},\ \Eprint
  {http://arxiv.org/abs/hep-ph/9605323} {hep-ph/9605323} \BibitemShut {NoStop}%
\bibitem [{\citenamefont {Caola}\ \emph {et~al.}(2017)\citenamefont {Caola},
  \citenamefont {Melnikov},\ and\ \citenamefont {R{\"o}ntsch}}]{Caola:2017dug}%
  \BibitemOpen
  \bibfield  {author} {\bibinfo {author} {\bibfnamefont {F.}~\bibnamefont
  {Caola}}, \bibinfo {author} {\bibfnamefont {K.}~\bibnamefont {Melnikov}}, \
  and\ \bibinfo {author} {\bibfnamefont {R.}~\bibnamefont {R{\"o}ntsch}},\
  }\href {\doibase 10.1140/epjc/s10052-017-4774-0} {\bibfield  {journal}
  {\bibinfo  {journal} {Eur. Phys. J.}\ }\textbf {\bibinfo {volume} {C77}},\
  \bibinfo {pages} {248} (\bibinfo {year} {2017})},\ \Eprint
  {http://arxiv.org/abs/1702.01352} {arXiv:1702.01352 [hep-ph]} \BibitemShut
  {NoStop}%
\bibitem [{\citenamefont {Czakon}(2010)}]{Czakon:2010td}%
  \BibitemOpen
  \bibfield  {author} {\bibinfo {author} {\bibfnamefont {M.}~\bibnamefont
  {Czakon}},\ }\href {\doibase 10.1016/j.physletb.2010.08.036} {\bibfield
  {journal} {\bibinfo  {journal} {Phys. Lett.}\ }\textbf {\bibinfo {volume}
  {B693}},\ \bibinfo {pages} {259} (\bibinfo {year} {2010})},\ \Eprint
  {http://arxiv.org/abs/1005.0274} {arXiv:1005.0274 [hep-ph]} \BibitemShut
  {NoStop}%
\bibitem [{\citenamefont {Czakon}\ and\ \citenamefont
  {Heymes}(2014)}]{Czakon:2014oma}%
  \BibitemOpen
  \bibfield  {author} {\bibinfo {author} {\bibfnamefont {M.}~\bibnamefont
  {Czakon}}\ and\ \bibinfo {author} {\bibfnamefont {D.}~\bibnamefont
  {Heymes}},\ }\href {\doibase 10.1016/j.nuclphysb.2014.11.006} {\bibfield
  {journal} {\bibinfo  {journal} {Nucl. Phys.}\ }\textbf {\bibinfo {volume}
  {B890}},\ \bibinfo {pages} {152} (\bibinfo {year} {2014})},\ \Eprint
  {http://arxiv.org/abs/1408.2500} {arXiv:1408.2500 [hep-ph]} \BibitemShut
  {NoStop}%
\bibitem [{\citenamefont {Gehrmann-De~Ridder}\ \emph
  {et~al.}(2005)\citenamefont {Gehrmann-De~Ridder}, \citenamefont {Gehrmann},\
  and\ \citenamefont {Glover}}]{GehrmannDeRidder:2005cm}%
  \BibitemOpen
  \bibfield  {author} {\bibinfo {author} {\bibfnamefont {A.}~\bibnamefont
  {Gehrmann-De~Ridder}}, \bibinfo {author} {\bibfnamefont {T.}~\bibnamefont
  {Gehrmann}}, \ and\ \bibinfo {author} {\bibfnamefont {E.~W.~N.}\ \bibnamefont
  {Glover}},\ }\href {\doibase 10.1088/1126-6708/2005/09/056} {\bibfield
  {journal} {\bibinfo  {journal} {JHEP}\ }\textbf {\bibinfo {volume} {09}},\
  \bibinfo {pages} {056} (\bibinfo {year} {2005})},\ \Eprint
  {http://arxiv.org/abs/hep-ph/0505111} {arXiv:hep-ph/0505111 [hep-ph]}
  \BibitemShut {NoStop}%
\bibitem [{\citenamefont {Magnea}\ \emph {et~al.}(2018)\citenamefont {Magnea},
  \citenamefont {Maina}, \citenamefont {Pelliccioli}, \citenamefont
  {Signorile-Signorile}, \citenamefont {Torrielli},\ and\ \citenamefont
  {Uccirati}}]{Magnea:2018hab}%
  \BibitemOpen
  \bibfield  {author} {\bibinfo {author} {\bibfnamefont {L.}~\bibnamefont
  {Magnea}}, \bibinfo {author} {\bibfnamefont {E.}~\bibnamefont {Maina}},
  \bibinfo {author} {\bibfnamefont {G.}~\bibnamefont {Pelliccioli}}, \bibinfo
  {author} {\bibfnamefont {C.}~\bibnamefont {Signorile-Signorile}}, \bibinfo
  {author} {\bibfnamefont {P.}~\bibnamefont {Torrielli}}, \ and\ \bibinfo
  {author} {\bibfnamefont {S.}~\bibnamefont {Uccirati}},\ }\href {\doibase
  10.1007/JHEP06(2019)013, 10.1007/JHEP12(2018)107} {\bibfield  {journal}
  {\bibinfo  {journal} {JHEP}\ }\textbf {\bibinfo {volume} {12}},\ \bibinfo
  {pages} {107} (\bibinfo {year} {2018})},\ \bibinfo {note} {[Erratum:
  JHEP06,013(2019)]},\ \Eprint {http://arxiv.org/abs/1806.09570}
  {arXiv:1806.09570 [hep-ph]} \BibitemShut {NoStop}%
\bibitem [{\citenamefont {Grazzini}\ \emph {et~al.}(2018)\citenamefont
  {Grazzini}, \citenamefont {Kallweit},\ and\ \citenamefont
  {Wiesemann}}]{Grazzini:2017mhc}%
  \BibitemOpen
  \bibfield  {author} {\bibinfo {author} {\bibfnamefont {M.}~\bibnamefont
  {Grazzini}}, \bibinfo {author} {\bibfnamefont {S.}~\bibnamefont {Kallweit}},
  \ and\ \bibinfo {author} {\bibfnamefont {M.}~\bibnamefont {Wiesemann}},\
  }\href {\doibase 10.1140/epjc/s10052-018-5771-7} {\bibfield  {journal}
  {\bibinfo  {journal} {Eur. Phys. J.}\ }\textbf {\bibinfo {volume} {C78}},\
  \bibinfo {pages} {537} (\bibinfo {year} {2018})},\ \Eprint
  {http://arxiv.org/abs/1711.06631} {arXiv:1711.06631 [hep-ph]} \BibitemShut
  {NoStop}%
\bibitem [{\citenamefont {Catani}\ and\ \citenamefont
  {Grazzini}(2007)}]{Catani:2007vq}%
  \BibitemOpen
  \bibfield  {author} {\bibinfo {author} {\bibfnamefont {S.}~\bibnamefont
  {Catani}}\ and\ \bibinfo {author} {\bibfnamefont {M.}~\bibnamefont
  {Grazzini}},\ }\href {\doibase 10.1103/PhysRevLett.98.222002} {\bibfield
  {journal} {\bibinfo  {journal} {Phys. Rev. Lett.}\ }\textbf {\bibinfo
  {volume} {98}},\ \bibinfo {pages} {222002} (\bibinfo {year} {2007})},\
  \Eprint {http://arxiv.org/abs/hep-ph/0703012} {hep-ph/0703012} \BibitemShut
  {NoStop}%
\bibitem [{\citenamefont {Bozzi}\ \emph {et~al.}(2006)\citenamefont {Bozzi},
  \citenamefont {Catani}, \citenamefont {de~Florian},\ and\ \citenamefont
  {Grazzini}}]{Bozzi:2005wk}%
  \BibitemOpen
  \bibfield  {author} {\bibinfo {author} {\bibfnamefont {G.}~\bibnamefont
  {Bozzi}}, \bibinfo {author} {\bibfnamefont {S.}~\bibnamefont {Catani}},
  \bibinfo {author} {\bibfnamefont {D.}~\bibnamefont {de~Florian}}, \ and\
  \bibinfo {author} {\bibfnamefont {M.}~\bibnamefont {Grazzini}},\ }\href
  {\doibase 10.1016/j.nuclphysb.2005.12.022} {\bibfield  {journal} {\bibinfo
  {journal} {Nucl. Phys.}\ }\textbf {\bibinfo {volume} {B737}},\ \bibinfo
  {pages} {73} (\bibinfo {year} {2006})},\ \Eprint
  {http://arxiv.org/abs/hep-ph/0508068} {arXiv:hep-ph/0508068 [hep-ph]}
  \BibitemShut {NoStop}%
\bibitem [{\citenamefont {Catani}\ \emph {et~al.}(2014)\citenamefont {Catani},
  \citenamefont {Cieri}, \citenamefont {de~Florian}, \citenamefont {Ferrera},\
  and\ \citenamefont {Grazzini}}]{Catani:2013tia}%
  \BibitemOpen
  \bibfield  {author} {\bibinfo {author} {\bibfnamefont {S.}~\bibnamefont
  {Catani}}, \bibinfo {author} {\bibfnamefont {L.}~\bibnamefont {Cieri}},
  \bibinfo {author} {\bibfnamefont {D.}~\bibnamefont {de~Florian}}, \bibinfo
  {author} {\bibfnamefont {G.}~\bibnamefont {Ferrera}}, \ and\ \bibinfo
  {author} {\bibfnamefont {M.}~\bibnamefont {Grazzini}},\ }\href {\doibase
  10.1016/j.nuclphysb.2014.02.011} {\bibfield  {journal} {\bibinfo  {journal}
  {Nucl. Phys.}\ }\textbf {\bibinfo {volume} {B881}},\ \bibinfo {pages} {414}
  (\bibinfo {year} {2014})},\ \Eprint {http://arxiv.org/abs/1311.1654}
  {arXiv:1311.1654 [hep-ph]} \BibitemShut {NoStop}%
\bibitem [{\citenamefont {Catani}\ \emph {et~al.}(2019)\citenamefont {Catani},
  \citenamefont {Devoto}, \citenamefont {Grazzini}, \citenamefont {Kallweit},\
  and\ \citenamefont {Mazzitelli}}]{Catani:2019hip}%
  \BibitemOpen
  \bibfield  {author} {\bibinfo {author} {\bibfnamefont {S.}~\bibnamefont
  {Catani}}, \bibinfo {author} {\bibfnamefont {S.}~\bibnamefont {Devoto}},
  \bibinfo {author} {\bibfnamefont {M.}~\bibnamefont {Grazzini}}, \bibinfo
  {author} {\bibfnamefont {S.}~\bibnamefont {Kallweit}}, \ and\ \bibinfo
  {author} {\bibfnamefont {J.}~\bibnamefont {Mazzitelli}},\ }\href {\doibase
  10.1007/JHEP07(2019)100} {\bibfield  {journal} {\bibinfo  {journal} {JHEP}\
  }\textbf {\bibinfo {volume} {07}},\ \bibinfo {pages} {100} (\bibinfo {year}
  {2019})},\ \Eprint {http://arxiv.org/abs/1906.06535} {arXiv:1906.06535
  [hep-ph]} \BibitemShut {NoStop}%
\bibitem [{\citenamefont {Butterworth}\ \emph {et~al.}(2016)\citenamefont
  {Butterworth} \emph {et~al.}}]{Butterworth:2015oua}%
  \BibitemOpen
  \bibfield  {author} {\bibinfo {author} {\bibfnamefont {J.}~\bibnamefont
  {Butterworth}} \emph {et~al.},\ }\href {\doibase
  10.1088/0954-3899/43/2/023001} {\bibfield  {journal} {\bibinfo  {journal} {J.
  Phys.}\ }\textbf {\bibinfo {volume} {G43}},\ \bibinfo {pages} {023001}
  (\bibinfo {year} {2016})},\ \Eprint {http://arxiv.org/abs/1510.03865}
  {arXiv:1510.03865 [hep-ph]} \BibitemShut {NoStop}%
\bibitem [{\citenamefont {Buckley}\ \emph {et~al.}(2015)\citenamefont
  {Buckley}, \citenamefont {Ferrando}, \citenamefont {Lloyd}, \citenamefont
  {Nordstr{\"o}m}, \citenamefont {Page}, \citenamefont {R{\"u}fenacht},
  \citenamefont {Sch{\"o}nherr},\ and\ \citenamefont {Watt}}]{Buckley:2014ana}%
  \BibitemOpen
  \bibfield  {author} {\bibinfo {author} {\bibfnamefont {A.}~\bibnamefont
  {Buckley}}, \bibinfo {author} {\bibfnamefont {J.}~\bibnamefont {Ferrando}},
  \bibinfo {author} {\bibfnamefont {S.}~\bibnamefont {Lloyd}}, \bibinfo
  {author} {\bibfnamefont {K.}~\bibnamefont {Nordstr{\"o}m}}, \bibinfo {author}
  {\bibfnamefont {B.}~\bibnamefont {Page}}, \bibinfo {author} {\bibfnamefont
  {M.}~\bibnamefont {R{\"u}fenacht}}, \bibinfo {author} {\bibfnamefont
  {M.}~\bibnamefont {Sch{\"o}nherr}}, \ and\ \bibinfo {author} {\bibfnamefont
  {G.}~\bibnamefont {Watt}},\ }\href {\doibase 10.1140/epjc/s10052-015-3318-8}
  {\bibfield  {journal} {\bibinfo  {journal} {Eur. Phys. J.}\ }\textbf
  {\bibinfo {volume} {C75}},\ \bibinfo {pages} {132} (\bibinfo {year}
  {2015})},\ \Eprint {http://arxiv.org/abs/1412.7420} {arXiv:1412.7420
  [hep-ph]} \BibitemShut {NoStop}%
\bibitem [{\citenamefont {Tanabashi}\ \emph {et~al.}(2018)\citenamefont
  {Tanabashi} \emph {et~al.}}]{Tanabashi:2018oca}%
  \BibitemOpen
  \bibfield  {author} {\bibinfo {author} {\bibfnamefont {M.}~\bibnamefont
  {Tanabashi}} \emph {et~al.} (\bibinfo {collaboration} {Particle Data
  Group}),\ }\href {\doibase 10.1103/PhysRevD.98.030001} {\bibfield  {journal}
  {\bibinfo  {journal} {Phys. Rev.}\ }\textbf {\bibinfo {volume} {D98}},\
  \bibinfo {pages} {030001} (\bibinfo {year} {2018})}\BibitemShut {NoStop}%
\bibitem [{\citenamefont {Buccioni}\ \emph {et~al.}(2019)\citenamefont
  {Buccioni}, \citenamefont {Lang}, \citenamefont {Lindert}, \citenamefont
  {Maierh{\"o}fer}, \citenamefont {Pozzorini}, \citenamefont {Zhang},\ and\
  \citenamefont {Zoller}}]{Buccioni:2019sur}%
  \BibitemOpen
  \bibfield  {author} {\bibinfo {author} {\bibfnamefont {F.}~\bibnamefont
  {Buccioni}}, \bibinfo {author} {\bibfnamefont {J.-N.}\ \bibnamefont {Lang}},
  \bibinfo {author} {\bibfnamefont {J.~M.}\ \bibnamefont {Lindert}}, \bibinfo
  {author} {\bibfnamefont {P.}~\bibnamefont {Maierh{\"o}fer}}, \bibinfo
  {author} {\bibfnamefont {S.}~\bibnamefont {Pozzorini}}, \bibinfo {author}
  {\bibfnamefont {H.}~\bibnamefont {Zhang}}, \ and\ \bibinfo {author}
  {\bibfnamefont {M.~F.}\ \bibnamefont {Zoller}},\ }\href@noop {} {\  (\bibinfo
  {year} {2019})},\ \Eprint {http://arxiv.org/abs/1907.13071} {arXiv:1907.13071
  [hep-ph]} \BibitemShut {NoStop}%
\bibitem [{\citenamefont {Sj{\"o}strand}\ \emph {et~al.}(2006)\citenamefont
  {Sj{\"o}strand}, \citenamefont {Mrenna},\ and\ \citenamefont
  {Skands}}]{Sjostrand:2006za}%
  \BibitemOpen
  \bibfield  {author} {\bibinfo {author} {\bibfnamefont {T.}~\bibnamefont
  {Sj{\"o}strand}}, \bibinfo {author} {\bibfnamefont {S.}~\bibnamefont
  {Mrenna}}, \ and\ \bibinfo {author} {\bibfnamefont {P.}~\bibnamefont
  {Skands}},\ }\href@noop {} {\bibfield  {journal} {\bibinfo  {journal} {JHEP}\
  }\textbf {\bibinfo {volume} {05}},\ \bibinfo {pages} {026} (\bibinfo {year}
  {2006})},\ \Eprint {http://arxiv.org/abs/hep-ph/0603175} {hep-ph/0603175}
  \BibitemShut {NoStop}%
\bibitem [{\citenamefont {Sj{\"o}strand}\ \emph {et~al.}(2008)\citenamefont
  {Sj{\"o}strand}, \citenamefont {Mrenna},\ and\ \citenamefont
  {Skands}}]{Sjostrand:2007gs}%
  \BibitemOpen
  \bibfield  {author} {\bibinfo {author} {\bibfnamefont {T.}~\bibnamefont
  {Sj{\"o}strand}}, \bibinfo {author} {\bibfnamefont {S.}~\bibnamefont
  {Mrenna}}, \ and\ \bibinfo {author} {\bibfnamefont {P.~Z.}\ \bibnamefont
  {Skands}},\ }\href {\doibase 10.1016/j.cpc.2008.01.036} {\bibfield  {journal}
  {\bibinfo  {journal} {Comput. Phys. Commun.}\ }\textbf {\bibinfo {volume}
  {178}},\ \bibinfo {pages} {852} (\bibinfo {year} {2008})},\ \Eprint
  {http://arxiv.org/abs/0710.3820} {arXiv:0710.3820 [hep-ph]} \BibitemShut
  {NoStop}%
\bibitem [{\citenamefont {Cacciari}\ and\ \citenamefont
  {Salam}(2006)}]{Cacciari:2005hq}%
  \BibitemOpen
  \bibfield  {author} {\bibinfo {author} {\bibfnamefont {M.}~\bibnamefont
  {Cacciari}}\ and\ \bibinfo {author} {\bibfnamefont {G.~P.}\ \bibnamefont
  {Salam}},\ }\href {\doibase 10.1016/j.physletb.2006.08.037} {\bibfield
  {journal} {\bibinfo  {journal} {Phys. Lett.}\ }\textbf {\bibinfo {volume}
  {B641}},\ \bibinfo {pages} {57} (\bibinfo {year} {2006})},\ \Eprint
  {http://arxiv.org/abs/hep-ph/0512210} {arXiv:hep-ph/0512210 [hep-ph]}
  \BibitemShut {NoStop}%
\bibitem [{\citenamefont {Cacciari}\ \emph {et~al.}(2012)\citenamefont
  {Cacciari}, \citenamefont {Salam},\ and\ \citenamefont
  {Soyez}}]{Cacciari:2011ma}%
  \BibitemOpen
  \bibfield  {author} {\bibinfo {author} {\bibfnamefont {M.}~\bibnamefont
  {Cacciari}}, \bibinfo {author} {\bibfnamefont {G.~P.}\ \bibnamefont {Salam}},
  \ and\ \bibinfo {author} {\bibfnamefont {G.}~\bibnamefont {Soyez}},\ }\href
  {\doibase 10.1140/epjc/s10052-012-1896-2} {\bibfield  {journal} {\bibinfo
  {journal} {Eur. Phys. J.}\ }\textbf {\bibinfo {volume} {C72}},\ \bibinfo
  {pages} {1896} (\bibinfo {year} {2012})},\ \Eprint
  {http://arxiv.org/abs/1111.6097} {arXiv:1111.6097 [hep-ph]} \BibitemShut
  {NoStop}%
\bibitem [{\citenamefont {Alioli}\ \emph {et~al.}(2017)\citenamefont {Alioli},
  \citenamefont {Caola}, \citenamefont {Luisoni},\ and\ \citenamefont
  {R{\"o}ntsch}}]{Alioli:2016xab}%
  \BibitemOpen
  \bibfield  {author} {\bibinfo {author} {\bibfnamefont {S.}~\bibnamefont
  {Alioli}}, \bibinfo {author} {\bibfnamefont {F.}~\bibnamefont {Caola}},
  \bibinfo {author} {\bibfnamefont {G.}~\bibnamefont {Luisoni}}, \ and\
  \bibinfo {author} {\bibfnamefont {R.}~\bibnamefont {R{\"o}ntsch}},\ }\href
  {\doibase 10.1103/PhysRevD.95.034042} {\bibfield  {journal} {\bibinfo
  {journal} {Phys. Rev.}\ }\textbf {\bibinfo {volume} {D95}},\ \bibinfo {pages}
  {034042} (\bibinfo {year} {2017})},\ \Eprint
  {http://arxiv.org/abs/1609.09719} {arXiv:1609.09719 [hep-ph]} \BibitemShut
  {NoStop}%
\bibitem [{\citenamefont {Heinrich}\ \emph {et~al.}(2017)\citenamefont
  {Heinrich}, \citenamefont {Jones}, \citenamefont {Kerner}, \citenamefont
  {Luisoni},\ and\ \citenamefont {Vryonidou}}]{Heinrich:2017kxx}%
  \BibitemOpen
  \bibfield  {author} {\bibinfo {author} {\bibfnamefont {G.}~\bibnamefont
  {Heinrich}}, \bibinfo {author} {\bibfnamefont {S.~P.}\ \bibnamefont {Jones}},
  \bibinfo {author} {\bibfnamefont {M.}~\bibnamefont {Kerner}}, \bibinfo
  {author} {\bibfnamefont {G.}~\bibnamefont {Luisoni}}, \ and\ \bibinfo
  {author} {\bibfnamefont {E.}~\bibnamefont {Vryonidou}},\ }\href {\doibase
  10.1007/JHEP08(2017)088} {\bibfield  {journal} {\bibinfo  {journal} {JHEP}\
  }\textbf {\bibinfo {volume} {08}},\ \bibinfo {pages} {088} (\bibinfo {year}
  {2017})},\ \Eprint {http://arxiv.org/abs/1703.09252} {arXiv:1703.09252
  [hep-ph]} \BibitemShut {NoStop}%
\end{thebibliography}%

\end{document}